\PassOptionsToPackage{unicode}{hyperref}
\PassOptionsToPackage{hyphens}{url}
\documentclass[
  11pt,
]{article}
\usepackage{xcolor}
\usepackage[margin=1in]{geometry}
\usepackage{amsmath,amssymb}
\usepackage{iftex}
\ifPDFTeX
  \usepackage[T1]{fontenc}
  \usepackage[utf8]{inputenc}
  \usepackage{textcomp} 
\else 
  \usepackage{unicode-math} 
  \defaultfontfeatures{Scale=MatchLowercase}
  \defaultfontfeatures[\rmfamily]{Ligatures=TeX,Scale=1}
\fi
\usepackage{lmodern}
\ifPDFTeX\else
\fi
\IfFileExists{upquote.sty}{\usepackage{upquote}}{}
\IfFileExists{microtype.sty}{
  \usepackage[]{microtype}
  \UseMicrotypeSet[protrusion]{basicmath} 
}{}
\makeatletter
\@ifundefined{KOMAClassName}{
  \IfFileExists{parskip.sty}{%
    \usepackage{parskip}
  }{
    \setlength{\parindent}{0pt}
    \setlength{\parskip}{6pt plus 2pt minus 1pt}}
}{
  \KOMAoptions{parskip=half}}
\makeatother
\usepackage{longtable,booktabs,array}
\usepackage{calc} 
\usepackage{etoolbox}
\makeatletter
\patchcmd\longtable{\par}{\if@noskipsec\mbox{}\fi\par}{}{}
\makeatother
\IfFileExists{footnotehyper.sty}{\usepackage{footnotehyper}}{\usepackage{footnote}}
\makesavenoteenv{longtable}
\providecommand{\tightlist}{%
  \setlength{\itemsep}{0pt}\setlength{\parskip}{0pt}}

\usepackage{colortbl}
\usepackage{pdflscape}
\usepackage{array}
\usepackage{booktabs}
\usepackage{longtable}
\usepackage{tikz}
\usepackage{pgfplots}
\pgfplotsset{compat=1.18}
\usetikzlibrary{positioning, arrows.meta, fit, backgrounds, shapes.geometric}

\definecolor{tableheaderbg}{RGB}{223, 234, 209}
\definecolor{tablerowalt}{RGB}{248, 250, 244}

\renewcommand{\arraystretch}{1.25}

\arrayrulecolor{black!55}
\usepackage{bookmark}
\IfFileExists{xurl.sty}{\usepackage{xurl}}{} 
\hypersetup{
  hidelinks,
  pdfcreator={LaTeX via pandoc}}

\author{}
\date{}

\begin{document}

\section{Proof of Useful Attestation}\label{proof-of-useful-attestation}

\subsection{A Consensus Primitive for Attestation-Native
Chains}\label{a-consensus-primitive-for-attestation-native-chains}

\textbf{Stefan Stefanović}

\emph{Ligate Labs}

\textbf{Working Paper v0.9.2}

\textbf{Date:} 2026-05-25

\textbf{Contact:} hello@ligate.io

\newpage

\tableofcontents

\newpage

\subsection{Abstract}\label{abstract}

Consensus mechanisms for chains whose primary economic activity is
attestation production - content provenance, AI-output attribution,
threshold-signed credentials, supply-chain traceability - are misaligned
with the workload they secure. Validators on a generic Proof of Stake
chain earn the same fees regardless of whether they handle attestation
work correctly or selectively censor it. \textbf{Proof of Useful
Attestation (PoUA)} changes that.

In PoUA, validator influence is computed as bonded stake multiplied by a
non-transferable reputation score that grows with valid attestation
processing and shrinks under detected misbehavior. The primitive is
designed for chains whose runtime, fee market, and economic model are
purpose-built for attestations - Ligate Chain is the worked example
throughout. We give the protocol specification, a threat model under
standard partial-synchrony, an incentive analysis under a
profit-maximizing validator model, and a concrete integration with the
Sovereign SDK rollup framework. PoUA inherits the safety and liveness
properties of its underlying BFT primitive (Tendermint-style optimistic
finality, in deployment) and constructs a multiplicative cost-to-attack
premium of \(4\times\) to \(10\times\) over equivalent pure-stake chains
(Figure \ref{fig:cost-to-attack}; reproduced empirically by
\texttt{prototypes/poua-sim/scripts/run\_capital\_scan.py}).

The contribution is not a new cryptographic primitive. It is a
synthesis: reputation-weighted consensus (Yu et al., 2019; Eyal, 2015),
proof-of-useful-work (Helium 2018; Filecoin 2017), and restaking with
non-transferable bonds (EigenLayer, 2023), recombined to fit
attestation-native chains, and given the specific mechanism choices,
Sybil-resistance argument, and engineering integration that prior work
does not. The hard part - defending against compound
capital-and-grinding adversaries who control validators, attestor sets,
and submitter addresses simultaneously - is treated through a layered
defense whose load-bearing piece is a formal cost-to-grind bound (Lemma
1).

\begin{center}\rule{0.5\linewidth}{0.5pt}\end{center}

\subsection{1. Introduction}\label{introduction}

\subsubsection{1.1 The Attestation-Native Chain
Thesis}\label{the-attestation-native-chain-thesis}

A chain whose primary economic activity is the production and
verification of cryptographic attestations against typed schemas - call
it an \emph{attestation-native chain} - should not be built on consensus
mechanisms designed for general-purpose state transition. Ethereum
displaced computing-on-Bitcoin by recognizing that smart contracts
needed their own runtime. Filecoin and Helium displaced
storage-on-Ethereum and wireless-on-Ethereum by recognizing that storage
and coverage needed their own consensus. Attestation work is the next
case. A general-purpose chain hosting attestation contracts can serve
the workload, but cannot defend it.

Ligate Chain is the worked example of an attestation-native chain. The
runtime is built around schemas, attestor sets, and threshold-signed
attestations as first-class primitives. The fee market, the validator
economic model, and the consensus mechanism follow from that choice.
This paper covers the consensus component: \textbf{Proof of Useful
Attestation (PoUA)}, a weighting primitive in which validator influence
is tied to the validator's history of producing valid attestation work.

\textbf{What PoUA is and is not.} PoUA is a consensus weighting
primitive layered on top of a data availability substrate (Celestia in
our reference implementation, DA-agnostic in principle). It is not an
alternative to DA; immutability and ordering guarantees still come from
the substrate. What PoUA adds is reputation-weighted signer integrity:
validators are weighted by stake \(\times\) non-transferable reputation,
where reputation evolves from the validator's history of valid
attestation work and slashed misbehavior. The chain enforces this
weighting at consensus time, not as application-level computation. The
result is a security model where the cost-to-grind for an attacker is
bounded below by an intrinsic protocol parameter
(\(\tau_{\text{burn}}\)), independent of the underlying DA layer's
economics. Section 3.8 works the distinction out in terms of the system
model; §11 Q11 answers the ``why not just use Celestia raw'' framing
directly.

The argument is economic, not aesthetic. A chain whose security budget
is sized to its attestation workload, and whose security mechanism pays
the validators most useful to that workload, has a defensibility profile
- a moat - that a generic chain hosting attestation contracts does not.
Section 5 quantifies the moat: a \(4\times\) to \(10\times\)
multiplicative premium on cost-to-attack over equivalent pure-stake
Proof of Stake chains.

\subsubsection{1.2 Why Now: The 2026
Inflection}\label{why-now-the-2026-inflection}

Three shifts in late 2025 and early 2026 make this work timely.

\textbf{The provenance crisis.} Generative AI now produces audio, video,
and prose indistinguishable from human-made content, at commodity
prices. A substantial fraction of newly published online content in 2026
has some form of generative-AI involvement, with provenance
documentation usually missing. The demand for on-chain attestation -
that a piece of content is human-produced, that a piece of evidence is
AI-augmented, that an AI output came from a specific model at a specific
time - is no longer speculative. News organizations are defending
against synthetic-evidence lawsuits. Regulators are implementing the EU
AI Act's transparency clauses. Consumer-AI products are being required
to disclose model involvement. The attestation workload is a
volume-and-pricing problem with an actual customer base.

\textbf{The restaking maturity.} EigenLayer launched on Ethereum mainnet
in 2023 and showed that consensus security can be reused, rebonded, and
slashed across multiple application surfaces at scale. That conceptual
breakthrough opened space for further specialization. PoUA is one such
specialization: not a layer atop an existing chain, but a primitive
built into a chain whose attestation workload \emph{is} the application
surface, with bonding and slashing tied directly to that workload's
correctness.

\textbf{The Sovereign SDK production-readiness.} The Sovereign SDK
rollup framework, which provides modular consensus, data availability,
and execution layers with hooks for custom kernels, hit its first
production release window in 2025-2026. This is the substrate Ligate
Chain is built on, and the substrate any attestation-native chain
following can be built on without reimplementing the lower stack.
Section 7 details the PoUA integration as a kernel extension.

These three together - a near-term validated demand surface, a maturity
in the restaking-and-specialization paradigm, and a substrate that
admits the proposed mechanism - give PoUA a design window that did not
exist eighteen months ago.

\subsubsection{1.3 The Misalignment
Problem}\label{the-misalignment-problem}

A growing class of decentralized applications - content provenance for
AI-generated media, sponsorship attestation in autonomous-agent
transactions, regulatory time-locks, threshold-signed credentials,
supply-chain traceability - has a common structural feature. Its
on-chain footprint is dominated not by general-purpose state transitions
but by \emph{attestations}: cryptographically signed statements of the
form ``set of authorities \(\mathcal{A}\) attests that statement \(s\)
holds against schema \(\sigma\) at time \(t\).''

Built on general-purpose blockchains (Ethereum, Solana, Cosmos
application chains), this workload has three problems.

The first is a \textbf{composability tax}. Each attestation pays the
cost of a generic smart-contract state write, despite the underlying
operation being simple - verify \(k\)-of-\(n\) signatures and write a
hash. On Ethereum mainnet a single attestation costs \$0.50 to \$5.00 in
gas; on most Layer-2 networks, \$0.01 to \$0.10. For applications
producing thousands of attestations per second, which is the design
envelope for mainstream content-provenance products, this pricing is
prohibitive even on the cheapest host chains.

The second is \textbf{schema fragmentation}. Attestation schemas live in
independently deployed contracts. There is no global registry, no typed
composition primitive, no protocol-level guarantee that two schemas
claiming the same name refer to the same underlying contract.
Cross-schema dependencies become ad-hoc external calls without
compile-time guarantees. Consumers of attestation data either solve
discovery off-chain or trust a centralized registry.

The third is \textbf{misaligned consensus incentives} - the problem this
paper concerns. Validators on general-purpose chains earn fees from any
state transition. They have no economic reason to specialize in
attestation workloads, and no penalty for behaviors specifically harmful
to attestation integrity: selectively excluding attestations from
certain schemas, accepting invalid threshold signatures, extracting MEV
from attestation reordering. The chain's economic security is
indifferent to the application-layer correctness of its dominant
workload.

An attestation-native chain whose runtime, fee market, and consensus
mechanism are built for attestation production addresses all three
problems. The remaining components of that architecture (per-schema fee
markets, native delegation, cross-schema composition typing, time-locked
/ commit-reveal schemas) are the subject of companion papers; this one
is about consensus.

\subsubsection{1.4 The Central Question}\label{the-central-question}

In a Proof of Stake chain with attestation as its primary workload, a
validator's stake is fungible with stake on any other Proof of Stake
chain. Nothing ties consensus security to attestation correctness beyond
the indirect channel of slashing for consensus-layer double-signing. An
adversary with capital can buy stake, perform attestation work badly -
censor schemas they disfavor, accept invalid attestations from corrupt
attestor sets they control, extract MEV from attestation reordering -
and suffer no consequence. The standard PoS slashing conditions do not
trigger on attestation-specific misbehavior.

The question of this paper:

\begin{quote}
\textbf{Can a consensus mechanism be designed in which a validator's
influence is causally linked to their history of producing valid
attestation work, in a Sybil-resistant manner that cannot be replicated
by stake-only chains?}
\end{quote}

PoUA answers yes. The rest of the paper specifies the mechanism,
analyzes its security and incentive properties, demonstrates the answer
is implementable on a production rollup framework, and quantifies the
resulting moat.

\subsubsection{1.5 Approach in Brief}\label{approach-in-brief}

The mechanism, before the formal specification, in three points.

First: validator influence is computed as bonded stake times a
non-transferable reputation score. Where standard PoS uses
\(w_v = s_v\), PoUA uses \(w_v = s_v \cdot r_v\) with \(r_v\) a
multiplier in a bounded interval \([r_{\min}, r_{\max}]\).

Second: reputation accumulates through validator-side participation in
valid attestation processing, weighted by the economic value of the
attestations included, and decays through detected misbehavior. The
``useful'' in \emph{Proof of Useful Attestation} lives here. Reputation
rewards work the chain's economy values, not arbitrary on-chain
activity.

Third: the reputation interval is bounded both ways. \(r_{\min} > 0\)
ensures new validators have non-zero consensus weight from stake alone,
eliminating cold-start lockout. \(r_{\max} < \infty\) prevents runaway
concentration on long-running validators.

The mechanism inherits the safety and liveness of its underlying BFT
primitive (Theorems 1 and 2 in §5.2). It does not weaken anything a
chain operator currently relies on. It strengthens cost-to-attack
against capital-only adversaries by a multiplicative factor related to
the honest validator set's average reputation (§5.3), and it ties the
chain's economic security to the chain's productive workload in a way
pure-stake PoS chains cannot replicate without changing consensus.

The key formal result, derived in Section 5.3, is that the
cost-to-attack premium against a capital adversary is:

\[\kappa = \frac{\bar{r}_H}{r_{\min}}\]

where \(\bar{r}_H\) is the mean reputation of honest validators at
attack time. With recommended parameters
(\(r_{\max}/r_{\min} \in [4, 10]\)), a healthy steady-state chain
achieves a cost-to-attack premium of \(4\times\) to \(10\times\) over
equivalent pure-stake PoS. This is the formal moat referenced in Section
1.1, and the §5.3 derivation matches empirical Monte Carlo from the
reference simulator (Figure \ref{fig:cost-to-attack};
\texttt{run\_capital\_scan.py}) to within binomial variance.

\subsubsection{1.6 Contributions}\label{contributions}

The paper contributes five things.

A \textbf{mechanism specification} in §4.1-4.4 gives the validator
weighting formula, the reputation update function, the slashing
conditions, and the bootstrap procedure at enough detail for an
implementer to build PoUA into a production rollup.

A \textbf{threat model and security analysis} in §5 articulates three
adversary archetypes - capital, reputation, and compound
capital-and-grinding - establishes that PoUA inherits BFT safety and
liveness under partial-synchrony with \(f < n/3\) Byzantine validators
(Theorems 1 and 2), and derives the multiplicative cost-to-attack
premium \(\kappa\) that constitutes PoUA's formal moat over pure-stake
PoS.

An \textbf{incentive analysis} in §6 shows that under a
profit-maximizing validator model, the unique equilibrium has all
validators performing valid attestation work, with quantitative bounds
on the cost of deviation. Reputation, being non-transferable and having
a bounded forward-revenue value, acts as a time-locked incentive
alignment that pure-stake PoS lacks.

An \textbf{implementation specification} in §7 covers the integration of
PoUA into the Sovereign SDK rollup framework: reputation state, slashing
surfaces, v0 parameters, storage cost, migration from a stake-only
bootstrapping phase. Every integration point is identified against an
existing Sovereign SDK module surface; the implementation is engineering
work, not research.

A \textbf{comparative analysis} in §8 positions PoUA against
reputation-weighted consensus (RepuCoin, EigenTrust),
proof-of-useful-work systems (Helium, Filecoin), restaking (EigenLayer),
and pure-stake Proof of Stake (Tendermint, Algorand) across six axes.
PoUA is novel as a synthesis, not in any single component, and §8
identifies the specific synthesis point.

\subsubsection{1.6.1 Status of Claims: Proven, Bounded, and
Empirical}\label{status-of-claims-proven-bounded-and-empirical}

Reviewers asking the right questions converge on the same separation. We
name it explicitly so that readers know which claims rest on formal
proof, which rest on stated assumptions, and which require devnet or
simulator validation we have not yet completed.

\textbf{Proven, in the sense of formal mathematical argument under
standard cryptographic and BFT assumptions:}

\begin{itemize}
\tightlist
\item
  BFT safety and liveness inheritance under \(f < n/3\) Byzantine
  \emph{weight} (Theorems 1 and 2, §5.2, via the weighted
  quorum-intersection Lemma 2).
\item
  The cost-to-attack premium algebra \(\kappa = \bar{r}_H / r_{\min}\)
  for a pure capital adversary (§5.3).
\item
  The cost-to-grind lower bound under Layer 3 with stated burn
  destination:
  \(F^{\text{net, per member}}_{\mathcal{CR}} \geq \tau_{\text{burn}} \cdot \Delta r / [\eta \cdot \alpha_{\text{eff}}(m, k)]\)
  (Lemma 1, §5.5.3), with explicit cartel-size and burn-destination
  parameter dependence.
\item
  Boundedness, monotonicity, and stability of the reputation update
  function (§4.3, by clip plus linearity of the additive update).
\end{itemize}

\textbf{Bounded under stated assumptions, where the assumptions are
non-trivial and named:}

\begin{itemize}
\tightlist
\item
  The ``up to \(4\times\) to \(10\times\) moat'' headline is a
  \emph{steady-state ceiling}, depressed during the warmup window, the
  validator-set ramp, and post-slash recovery. §5.3.1 quantifies the
  transition envelope; the realized \(\kappa\) is a stake-weighted
  average of validator-specific reputation values, which approaches the
  ceiling only when warmup is complete, churn is low, and no recent
  major slash has occurred.
\item
  The cartel cost-to-grind bound holds as stated under the recommended
  pure-burn destination. The treasury and redistribution alternatives in
  §5.5.3 carry weaker bounds explicitly derived per variant.
\item
  The honest equilibrium argument (§6.2) assumes profit-maximizing
  validators with full information about protocol rules and other
  validators' strategies. Validators with non-economic motives sit
  outside the model.
\item
  Reputation as forward-revenue (§6.3) assumes attestation fee flow
  \(R_f\) is positive and approximately stationary across the
  validator's discount horizon. In low-volume periods the slash
  deterrent attenuates.
\end{itemize}

\textbf{Empirically validated via the reference simulator}
(\href{https://github.com/ligate-io/ligate-research/tree/main/prototypes/poua-sim}{\texttt{prototypes/poua-sim/}},
M1-M7):

\begin{itemize}
\tightlist
\item
  The cost-to-attack premium \(\kappa\) algebra (§5.3) reproduces under
  Monte Carlo to within binomial sampling variance (Figure 2). The
  transition-state \(\kappa\) envelope (§5.3.1) reproduces empirically
  across warmup / ramp / steady / post-slash phases (Figure 3).
\item
  Lemma 1's cartel cost-to-grind bound (§5.5.3) reproduces under all
  three Layer 3 burn destinations across \(m \in \{1, 2, 3, 4\}\) in
  \(k = 12\) to floating-point precision (Figure 6).
\item
  The full named-deviation strategy space (§6.2) collapses to
  \(r_{\min}\) under the full layered defense (Panel C of Figure 8),
  validating the honest-equilibrium claim against the M6 strategy-search
  heatmap including GRIND\_VIA\_STAGED\_SUBMITTERS at small / medium /
  large pool sizes.
\item
  Scale invariance of \(\kappa\) holds across
  \(|V| \in \{50, 100, 250, 500, 1000\}\) (§5.3.2, Figure 4); the §5.3
  small-set Lemma 1 example generalizes to mainnet-scale validator sets
  without parameter retuning.
\item
  Eclipse recovery follows the analytical §4.3 update rate (§5.5.6.1,
  Figure 7); \(\kappa\) is insensitive to adversarial latency in the
  single-canonical-chain model (§5.3.2.1, Figure 5).
\item
  The three-rebase auto-adjuster (§4.4.2 + §4.4.3) Lyapunov function
  \(V = D_\tau^2 + D_\eta^2 + D_\lambda^2\) is non-increasing under
  correlated drift
  (\texttt{test\_three\_rebases\_concurrent\_no\_amplification}).
\end{itemize}

\textbf{Empirical or heuristic, named as such, still requiring devnet
validation (deferred to v0.9 and beyond):}

\begin{itemize}
\tightlist
\item
  A2 (censorship) and A3 (grinding) detection \emph{power} against
  realistic adversaries. Appendix A gives analytical false-positive
  bounds under stated null hypotheses (\(\chi^2\) for A2,
  Erdős-Rényi-style for A3); the synthetic-attestor TPR scan in §A.4
  saturates and does not provide a calibrated true-positive curve.
  Devnet attestation traffic is required to position the detectors
  against realistic graph structure.
\item
  The Erdős-Rényi null hypothesis for A3 detection does not match real
  chain transaction graphs, which are typically scale-free. §A.4's
  Chung-Lu comparison figure (Figure 10) quantifies the gap; production
  deployment should re-derive the threshold against an empirical
  chain-graph baseline (post-devnet calibration).
\item
  Full mechanism-design-grade incentive compatibility. §6 plus the M6
  strategy-search heatmap give a layered economic argument against the
  named rational-deviation strategies; a Hurwicz-style proof of full
  strategy-proofness across an unconstrained strategy space remains open
  (v1.0+ research).
\end{itemize}

The honest one-line takeaway: \textbf{PoUA is a mechanism design
proposal with a formal economic floor (Lemma 1) plus inherited BFT
safety and liveness (Theorems 1-2), empirically validated through M1-M7
of the reference simulator across the named adversary models, with the
remaining heuristic surfaces (A2 / A3 detector power, full
strategy-proofness) explicitly deferred to devnet calibration and
post-arXiv research.} It is not a complete cryptographic security proof.
Where the paper makes ``if-then'' arguments, the ``if'' is named and
bounded; where the argument is heuristic, the heuristic is labeled and
the limitation acknowledged.

\subsubsection{1.7 Scope and Non-Goals}\label{scope-and-non-goals}

\textbf{In scope:}

\begin{itemize}
\tightlist
\item
  Validator selection and weighting in a single attestation-native
  chain.
\item
  Sybil resistance under economic attackers with stake-and-reputation
  acquisition costs.
\item
  Slashing conditions specific to attestation workloads.
\item
  Behavior under partial-synchrony with \(f < n/3\) Byzantine
  validators.
\item
  Concrete implementation atop the Sovereign SDK rollup framework.
\end{itemize}

\textbf{Explicitly out of scope:}

\begin{itemize}
\tightlist
\item
  Cross-chain reputation portability. Reputation is local to a single
  chain; portability across chains is a future work direction (Section
  9).
\item
  Cross-shard reputation, in the event Ligate Chain is sharded in the
  future. We treat the chain as monolithic throughout.
\item
  Privacy-preserving reputation. The reputation score is fully public;
  private-reputation extensions are future work.
\item
  Quantum-secure variants. PoUA inherits whatever post-quantum stance
  the underlying BFT primitive adopts; no quantum-resistance claims are
  made or required by the mechanism.
\item
  Mechanism design for the chain's broader fee market (per-schema fees,
  sponsored gas) and other attestation-native primitives (cross-schema
  composition, native delegation, time-locked schemas). These are
  companion concerns addressed in separate working papers.
\end{itemize}

\subsubsection{1.8 Document Structure}\label{document-structure}

Section 1.6.1 separates the paper's claims into proven,
bounded-under-stated-assumptions, and empirical-or-heuristic; readers in
a hurry may want to start there. Section 2 surveys background and prior
art across Proof of Stake, Proof of Useful Work, reputation-weighted
consensus, restaking, and Proof of Authority families. Section 3 fixes
notation and the system model. Section 4 specifies the PoUA protocol in
full, including the §4.4.1 \(\alpha\)-\(\beta\) Pareto frontier and the
§4.4.2 adaptive \(\tau_{\text{burn}}\) rebase mechanism. Section 5
analyzes security, including the transition-state \(\kappa\) envelope
(§5.3.1) and the layered defense against compound capital-plus-grinding
adversaries (§5.5). Section 6 analyzes incentives, including the §6.3.1
volume-dependence of the slash deterrent. Section 7 describes the Ligate
Chain implementation, including concrete v0 parameter recommendations.
Section 8 compares PoUA with prior systems across six analytical axes.
Section 9 lists limitations and future work. Section 10 concludes.
Section 11 collects frequently asked questions and addresses common
misunderstandings raised in early review. References follow. Appendix A
specifies the statistical detection procedures for heuristic slashing
conditions, with analytical false-positive bounds and an empirical FPR
comparison (ER vs.~scale-free null) in §A.4. Appendix B collects formal
definitions used throughout.

\textbf{On figures and empirical validation.} v0.7 incorporates five
empirical figures from the reference simulator at
\texttt{prototypes/poua-sim/} alongside the analytical derivations:
cost-to-attack with Monte Carlo overlay (Figure
\ref{fig:cost-to-attack}), realized \(\kappa\) across the chain
lifecycle (Figure \ref{fig:kappa-trajectory}), Lemma 1 cartel + burn
destinations (Figure \ref{fig:lemma1-burn-destinations}),
volume-deterrent ratio (Figure \ref{fig:volume-deterrent}), and A3
detector FPR under realistic null hypotheses (Figure
\ref{fig:a3-fpr-comparison}). Cross-language test vectors at
\texttt{prototypes/poua-sim/test\_vectors/} encode the analytical truths
so that a future Rust implementation in \texttt{ligate-chain} can
re-validate the same algebra. See the simulator README for the
regeneration workflow.

\begin{center}\rule{0.5\linewidth}{0.5pt}\end{center}

\subsection{2. Background and Related
Work}\label{background-and-related-work}

\subsubsection{2.1 Proof of Stake}\label{proof-of-stake}

The dominant family of permissionless consensus mechanisms in production
today, PoS protocols (Buchman 2016; Buterin \& Griffith 2017; Gilad et
al.~2017; Yin et al.~2019) select block proposers and finalizers as a
function of bonded capital. Validators deposit a token bond, propose and
vote on blocks, and earn protocol-specified rewards. Misbehavior -
specifically, equivocation (signing two conflicting blocks at the same
height) and surround voting - is detectable on-chain and punished by
\emph{slashing}: forfeiture of a fraction of the bond.

PoS is well-suited to chains whose validators' primary economic activity
is consensus itself. It is poorly suited to chains where consensus is a
means to an end and the chain's distinctive value lies elsewhere: PoS
validators are paid the same regardless of whether the application-layer
workload is processed correctly or selectively censored.

\subsubsection{2.2 Proof of Useful Work}\label{proof-of-useful-work}

A line of work originating with Helium's Proof of Coverage (Haleem et
al., 2018) and including Filecoin's Proof-of-Spacetime (Benet et al.,
2017) and Chia's Proof of Space-and-Time (Cohen, 2019) replaces (or
augments) traditional Proof-of-Work computation with proofs that the
validator is performing some socially or economically useful task:
providing wireless coverage, storing data, persisting capacity over
time. Validator influence is gated on observable performance of the
task.

PoUA is structurally analogous: validator influence is gated on
observable performance of attestation processing. The
proof-of-useful-work tradition typically requires hardware-attested or
cryptographically committed measurements (storage challenges, coverage
beacons); PoUA's ``useful work'' is verifiable at protocol level
(attestation transactions either pass quorum verification or do not) and
requires no external measurement infrastructure.

\subsubsection{2.3 Reputation-Weighted
Consensus}\label{reputation-weighted-consensus}

A line of academic work explores augmenting traditional consensus with
reputation scores derived from observable validator behavior. RepuCoin
(Yu et al., 2019) builds reputation from PoW mining history and uses it
to weight BFT votes, achieving Sybil resistance with sub-50\%
honest-stake assumptions. The earlier EigenTrust algorithm (Kamvar et
al., 2003) for peer-to-peer networks established the broader pattern of
using transitive interaction history to weight network influence in a
decentralized setting. The general ``trust and reputation'' literature
in distributed systems (Resnick et al., 2000; Hoffman et al., 2009)
provides the theoretical underpinning for both lines.

These mechanisms are well-explored in research and yet remain thin on
production deployment. The reasons are real: Sybil resistance is hard to
formalize when reputation is observable on-chain (since adversaries can
inject behavior into the observation channel), heuristic detection of
grinding patterns is brittle, and formal proofs of incentive
compatibility for reputation-weighted BFT are sparse. PoUA inherits the
structural pattern - observable behavior reputation augmenting consensus
weighting - and specializes the reputation-update function to
attestation work specifically. To our knowledge, no prior
reputation-weighted consensus mechanism couples reputation accumulation
to a chain's application-layer productive workload as PoUA does, and
none constructs a defense against compound capital-plus-grinding
adversaries with the layered-defense plus formal cost-to-grind argument
we develop in §5.5.

\subsubsection{2.4 Restaking and Reused
Stake}\label{restaking-and-reused-stake}

EigenLayer (2023) introduced the abstraction of \emph{restaking}: a
staked validator on a primary chain (Ethereum) opts to additionally
stake - and submit to slashing on - a secondary protocol's correctness
conditions. The validator's bond is reused as economic security across
multiple protocols.

PoUA can be viewed as a \emph{single-chain restaking} mechanism in which
the secondary ``protocol'' being restaked is the chain's own attestation
workload. The novelty relative to restaking is that the reputation
component is \emph{non-transferable} and \emph{intrinsic to the chain}:
it cannot be unbundled from a validator's identity, and cannot be reused
on other protocols.

\subsubsection{2.5 Proof of Authority and Permissioned
Variants}\label{proof-of-authority-and-permissioned-variants}

PoA (Aura, Clique, IBFT) restricts validator set membership to a fixed,
identity-bound roster. It is widely deployed in enterprise and
consortium chains. PoUA shares with PoA the property that identity
matters beyond just stake; it differs in that PoUA does not require
permissioned admission and produces reputation on-chain, while PoA
admits validators by off-chain governance.

\subsubsection{2.6 Hybrid Stake-Plus-Reputation
Mechanisms}\label{hybrid-stake-plus-reputation-mechanisms}

The Algorand consensus committee selection is technically
pure-stake-weighted (Gilad et al., 2017), but operational deployments
augment with reputation-like signals (relay-node uptime). Snowman
(Avalanche) introduces metastability properties that depend implicitly
on validator response latencies, a behavior-dependent component. Cosmos
Hub validators are subject to ``tombstoning'' for repeated offenses, a
coarse reputation signal.

PoUA is, to our knowledge, the first proposal to systematically couple
consensus weighting to a productive application-layer workload in a
non-mining, non-storage chain context.

\begin{center}\rule{0.5\linewidth}{0.5pt}\end{center}

\subsection{3. System Model}\label{system-model}

\subsubsection{3.1 Network and Adversary}\label{network-and-adversary}

We assume a partially synchronous network model (Dwork, Lynch,
Stockmeyer, 1988): there exists an unknown but finite Global
Stabilization Time (GST) after which message delays are bounded by a
known constant \(\Delta\). Before GST, the adversary may delay messages
arbitrarily.

The validator set at epoch \(t\) is \(V(t) = \{v_1, \ldots, v_n\}\) of
size \(n\). Up to \(f < n/3\) validators may be Byzantine: they may
deviate from the protocol arbitrarily, including coordinating among
themselves, withholding messages, equivocating, and arbitrary deviations
consistent with their cryptographic credentials. The remaining \(n - f\)
validators are \emph{honest} and follow the protocol.

The Byzantine bound \(f < n/3\) is the weakest assumption under which
BFT safety and liveness can be guaranteed in partial synchrony (Dwork et
al., 1988).

\textbf{Reference simulator coverage of network adversity.} The
reference simulator at
\href{https://github.com/ligate-io/ligate-research/tree/main/prototypes/poua-sim}{\texttt{prototypes/poua-sim/}}
implements four \texttt{NetworkScheduler} protocols that empirically
exercise the partial-synchrony model above under adversarial conditions:
uniform / adversarial latency, partition with drops, target-validator
eclipse, and scale benchmarking across
\(|V| \in \{50, 100, 250, 500, 1000\}\). Together these cover the four
adversarial-network categories from §5.2's safety / liveness inheritance
argument. The simulator's per-validator delivery queue lets blocks
propagate at scheduler-determined slot offsets while preserving the §4.3
voter-share semantics: the per-block \(g_{\text{vote}}\) denominator is
fixed at block creation, so late-arriving voters use the same
denominator as on-time ones. Empirical results land in §5.3.2 (scale
invariance), §5.3.2.1 (adversarial latency), and §5.5.6.1 (eclipse
recovery).

\subsubsection{3.2 Cryptographic
Assumptions}\label{cryptographic-assumptions}

We assume the existence of:

\begin{itemize}
\tightlist
\item
  An EUF-CMA-secure digital signature scheme with public verification.
\item
  A collision-resistant hash function
  \(H : \{0,1\}^* \to \{0,1\}^{256}\) used for transaction hashing,
  attestation payload commitment, and Merkle accumulation.
\item
  A pseudorandom function family suitable for committee selection
  (instantiated in deployment as VRF; see Section 7).
\end{itemize}

\subsubsection{3.3 Validators, Attestors, and Their
Distinction}\label{validators-attestors-and-their-distinction}

PoUA distinguishes two roles, both economically active on-chain:

\begin{itemize}
\tightlist
\item
  \textbf{Validators} (\(v \in V\)): order, propose, and vote on blocks.
  Bonded with stake \(s_v\). Subject to slashing for consensus-layer
  misbehavior. \emph{Hold reputation.}
\item
  \textbf{Attestors} (\(a\)): sign attestation payloads against
  registered schemas. Members of \emph{attestor sets} registered
  on-chain. Bonded with separate (typically smaller) stake. Subject to
  slashing for attestation-layer misbehavior (signing a payload that
  fails to verify).
\end{itemize}

Validators and attestors are distinct sets. A single party may operate
both validator and attestor nodes, but their roles do not commingle:
validator reputation is built from validator-side processing of
attestations (inclusion, ordering, vote), not from being an attestor.
This separation prevents ``I attested to my own attestation'' reputation
farming.

\subsubsection{3.4 Schemas and Attestor
Sets}\label{schemas-and-attestor-sets}

The chain's runtime maintains:

\begin{itemize}
\tightlist
\item
  \textbf{Schemas} \(\sigma \in \Sigma\): typed attestation contracts.
  Each \(\sigma\) specifies a payload type, a designated \emph{attestor
  set} \(\mathcal{A}_\sigma\), a signature threshold \(k_\sigma\), and
  fee parameters. Schemas are registered on-chain.
\item
  \textbf{Attestor sets} \(\mathcal{A} = \{a_1, \ldots, a_m\}\):
  enumerated public-key sets registered on-chain. Multiple schemas may
  share a single attestor set.
\item
  \textbf{Attestations} \(\alpha = (\sigma, p, \Sigma_{k_\sigma})\): a
  tuple of schema id, payload hash \(p\), and a
  \(k_\sigma\)-of-\(|\mathcal{A}_\sigma|\) threshold signature
  \(\Sigma_{k_\sigma}\) over \((p, \sigma, \text{submitter})\). An
  attestation is \emph{valid} if \(\Sigma_{k_\sigma}\) verifies under
  the public keys in \(\mathcal{A}_\sigma\) at the schema's threshold.
\end{itemize}

\subsubsection{3.5 Stake, Reputation, and
Weight}\label{stake-reputation-and-weight}

Each validator \(v\) at time \(t\) has:

\begin{itemize}
\tightlist
\item
  \textbf{Stake} \(s_v(t) \in \mathbb{R}_{\geq 0}\): tokens bonded to
  \(v\)'s consensus identity.
\item
  \textbf{Reputation} \(r_v(t) \in [r_{\min}, r_{\max}]\): a
  non-transferable scalar in a bounded interval, with
  \(0 < r_{\min} < r_{\max}\).
\item
  \textbf{Weight} \(w_v(t) = s_v(t) \cdot r_v(t)\): the quantity used in
  validator selection and BFT vote tallying.
\end{itemize}

We require \(r_{\min} > 0\) to ensure newly registered validators with
no reputation history retain non-zero consensus weight from stake alone,
eliminating cold-start lockout.

We require \(r_{\max} < \infty\) to prevent reputation runaway from
concentrating consensus weight in a small validator subset. The specific
values of \(r_{\min}, r_{\max}\) are protocol parameters subject to
governance; see Section 4.4 for design guidance.

\subsubsection{3.6 Time}\label{time}

Time is discretized into slots of fixed duration \(\tau\). Slot \(t\)
produces (or fails to produce) a block \(B_t\). We assume the BFT
primitive achieves block finality within \(O(1)\) slots after GST
(instantiated as Tendermint-style two-round optimistic finality in
deployment).

\subsubsection{3.7 System Diagram}\label{system-diagram}

Figure 1 collects the entities and their relationships. The validator
role and the attestor role are distinct (§3.3): validators order blocks
and accumulate reputation through processing; attestors sign payloads
against schemas they have been registered to. Both bond stake; only
validators carry reputation.

\begin{figure}[h]
\centering
\begin{tikzpicture}[
  node distance=2.0cm,
  every node/.style={font=\small},
  entity/.style={rectangle, draw, rounded corners=2pt, minimum height=1.0cm, minimum width=2.6cm, align=center, fill=tablerowalt},
  consensus/.style={rectangle, draw, rounded corners=2pt, minimum height=1.0cm, minimum width=2.6cm, align=center, fill=tableheaderbg},
  arrow/.style={-Stealth, thick, draw=black!65}
]

  \node[consensus] (validator) {Validator $v$ \\ stake $s_v$, rep $r_v$ \\ weight $w_v = s_v r_v$};
  \node[consensus, right=of validator] (bft) {BFT vote tally \\ (commit if $\sum_v w_v > 2/3 \, S$)};

  \node[entity, below=of validator] (block) {Block $B$ \\ contains attestations};

  \node[entity, below=of block, xshift=-2.6cm] (attestation) {Attestation $\alpha$ \\ $(\sigma, p, \Sigma_k)$};
  \node[entity, below=of block, xshift=2.6cm] (schema) {Schema $\sigma$ \\ binds attestor set};
  \node[entity, right=of schema] (attestorset) {Attestor set $\mathcal{A}_\sigma$ \\ $k$-of-$n$ keys};

  \draw[arrow] (validator.east) -- node[above, midway, font=\scriptsize] {weighted vote} (bft.west);
  \draw[arrow] (validator.south) -- node[right, midway, font=\scriptsize] {proposes / votes} (block.north);
  \draw[arrow] (attestation.north) -- node[left, midway, font=\scriptsize] {included in} (block.south west);
  \draw[arrow] (attestation.east) -- node[above, midway, font=\scriptsize] {against} (schema.west);
  \draw[arrow] (schema.east) -- node[above, midway, font=\scriptsize] {uses} (attestorset.west);
  \draw[arrow, dashed] (block.south east) -- node[right, midway, font=\scriptsize] {fee + valid attestation $\to$ $g_v$} ([xshift=-2pt] schema.north east);

\end{tikzpicture}
\caption{System diagram. Solid arrows are protocol-level relationships; the dashed arrow shows the reputation-accumulation channel: a valid attestation, included in a block proposed (or voted on) by validator $v$, contributes to $v$'s good-behavior score $g_v(t)$ via §4.3, weighted by the attestation's fee.}
\label{fig:system}
\end{figure}
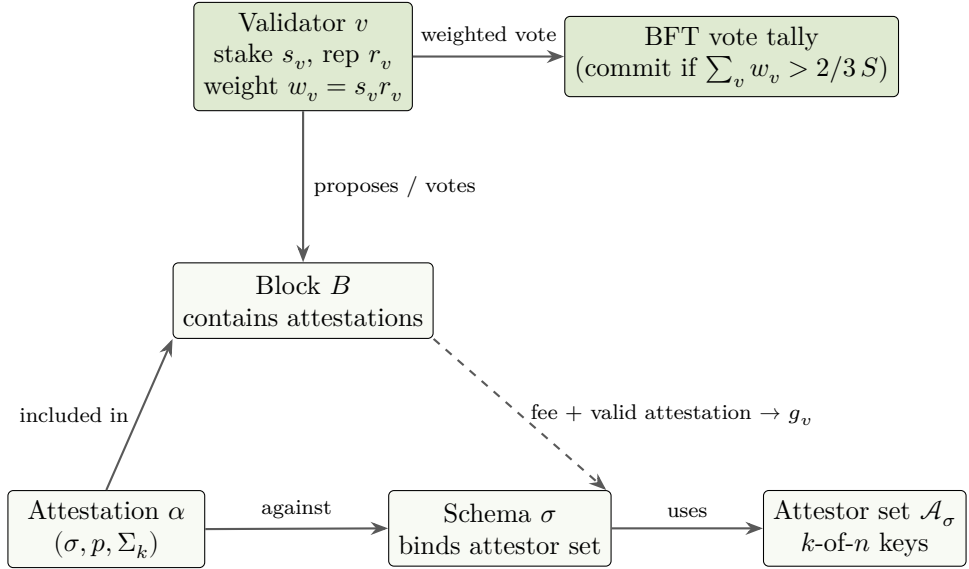

\subsubsection{3.8 Why Data Availability Alone Is
Insufficient}\label{why-data-availability-alone-is-insufficient}

A natural simplification of any attestation system is ``use a data
availability layer and call it done.'' DA gives immutable byte storage
and verifiable ordering. For pure attestation logs (e.g., timestamped
commitments to off-chain content), this is enough.

PoUA is not a pure attestation log. It is a reputation-weighted
attestation system, which requires three things DA cannot provide.

\textbf{Signer reputation as queryable state.} Section 4.3's reputation
update is a stateful computation. Light clients must verify ``validator
\(v\)'s reputation is \(r\) at epoch \(t\)'' without re-executing the
chain history. This requires consensus on per-epoch reputation, exposed
as committed state. A DA layer does not produce committed state; it
produces ordered, available data.

\textbf{Schema-scoped attestor sets.} Each schema (§3.4) declares an
attestor set with minimum reputation requirements. Validators not in the
set, or below the threshold, cannot produce valid attestations for that
schema. Enforcement requires consensus to evaluate set membership and
reputation predicates at attestation time. DA layers accept all
submitted bytes regardless of predicate.

\textbf{Protocol-enforced burn.} Lemma 1 (§5.5.3) bounds the adversary's
cost-to-grind by
\(F_{\text{net}} \geq \tau_{\text{burn}} \cdot \Delta r / (\eta \cdot \alpha_{\text{eff}})\).
The bound holds only if the chain enforces the \(\tau_{\text{burn}}\)
fraction at fee-distribution time. DA layers charge per-byte; they do
not redirect fees to provable destruction. Without protocol-enforced
burn, \(F_{\text{net}} \to 0\) and security collapses.

These three requirements collectively define the consensus surface PoUA
needs above DA. The surface is small (per-validator reputation,
per-schema policy, fee economics) but nonempty. Pure DA cannot host it.
This is why PoUA runs as a Sovereign SDK rollup on Celestia rather than
as a Celestia namespace alone.

The two layers are complementary: Celestia secures the bytes; PoUA
secures the signer. Both are required, and the security argument in §5
holds only when both are in place. §11 Q11 answers the same question
directly for FAQ readers.

\begin{center}\rule{0.5\linewidth}{0.5pt}\end{center}

\subsection{4. The PoUA Protocol}\label{the-poua-protocol}

\subsubsection{4.1 Validator Selection}\label{validator-selection}

At each slot \(t\), a block proposer is selected pseudorandomly weighted
by validator weight:

\[\Pr[\text{proposer}(t) = v] = \frac{w_v(t)}{\sum_{u \in V(t)} w_u(t)}\]

Selection is computed deterministically from a VRF output committed at
the previous block, so all honest validators agree on the proposer for
slot \(t\) at the moment slot \(t-1\) is finalized.

The validator set \(V(t)\) is updated at epoch boundaries (every \(E\)
slots, where \(E\) is a protocol parameter typically chosen as
\(2^{14}\) slots \(\approx\) 4 hours at \(\tau = 1\,\text{s}\)). Within
an epoch, the validator set is fixed; reputation updates are deferred to
epoch boundaries to amortize state cost (see Section 7 for storage
analysis).

\subsubsection{4.2 Vote Weighting}\label{vote-weighting}

When the BFT primitive collects pre-commits and commits, votes are
weighted by \(w_v(t)\):

\[\text{commit}(B_t) \iff \sum_{v : v \text{ commits } B_t} w_v(t) > \frac{2}{3} \cdot \sum_{u \in V(t)} w_u(t)\]

This is the sole point at which reputation enters the BFT vote tally.
All other BFT properties (proposer selection, view changes, equivocation
detection) are inherited unmodified from the underlying primitive.

\subsubsection{4.3 The Reputation Update
Function}\label{the-reputation-update-function}

At each epoch boundary \(t = E \cdot k\) for \(k \in \mathbb{N}\),
reputation is updated for each \(v \in V(t)\) based on the validator's
attestation-processing performance during the epoch:

\[r_v(t + E) = \text{clip}_{[r_{\min}, r_{\max}]}\left(r_v(t) + \eta \cdot g_v(t) - \lambda \cdot b_v(t)\right)\]

where:

\begin{itemize}
\tightlist
\item
  \(g_v(t) \in \mathbb{R}_{\geq 0}\) is the \emph{good behavior score}:
  the fee-weighted measure of \(v\)'s contribution to processing valid
  attestations during the epoch, both as a proposer and as a voting
  validator. To prevent reputation accumulation from concentrating
  exclusively on the small subset of validators selected as proposer
  (which would create a positive-feedback loop in which
  already-high-reputation validators win more proposer slots and
  accumulate more reputation, entrenching an early-rich set), \(g_v(t)\)
  has both a proposer and a voter component:
\end{itemize}

\[g_v(t) = \min\bigl(G_{\max},\; \alpha \cdot G_v^{\text{prop}}(t) + \beta \cdot G_v^{\text{vote}}(t)\bigr)\]

where:

\begin{itemize}
\item
  \(G_v^{\text{prop}}(t) = \sum_{B \in \text{Proposed}_v(t, t+E)} \sum_{\alpha \in B} \mathbb{1}[\alpha \text{ valid}] \cdot \text{fee}(\alpha)\):
  the fee-weighted count of valid attestations \(v\) included in blocks
  \(v\) proposed.
\item
  \(G_v^{\text{vote}}(t) = \sum_{B \in \text{VotedOn}_v(t, t+E)} \frac{\sum_{\alpha \in B} \mathbb{1}[\alpha \text{ valid}] \cdot \text{fee}(\alpha)}{|\text{voters}(B)|}\):
  the per-voter share of valid-attestation work in blocks \(v\) voted on
  (committed) but did not propose. Dividing by \(|\text{voters}(B)|\)
  keeps the total reputation injection per block constant: a single
  block contributes the same total reputation regardless of how many
  voters participated.
\item
  \(\alpha, \beta \geq 0\) with \(\alpha + \beta = 1\) are protocol
  parameters splitting reputation between proposer and voter pools.
  Recommended \(\alpha = 0.7\), \(\beta = 0.3\) (proposer earns the
  majority share, voters share the rest).
\item
  \(G_{\max}\) is a per-epoch growth cap, calibrated so that no
  validator can move reputation by more than
  \((r_{\max} - r_{\min}) / T_{\text{ramp}}\) in a single epoch (i.e.,
  the fastest possible ramp from \(r_{\min}\) to \(r_{\max}\) takes at
  least \(T_{\text{ramp}}\) epochs of full participation). This both
  throttles grinding and bounds the speed at which any one validator can
  pull ahead of the field.
\end{itemize}

\(\text{fee}(\alpha)\) is the protocol-paid fee for attestation
\(\alpha\). Weighting by fee aligns reputation accumulation with the
chain's revenue and prevents reputation grinding via low-value
attestations.

\begin{itemize}
\item
  \(b_v(t) \geq 0\) is the \emph{bad behavior score}: the count of
  slashable infractions detected for \(v\) in the epoch, weighted by
  severity (see Section 4.5).
\item
  \(\eta > 0, \lambda > 0\) are protocol parameters tuning growth and
  decay rates. \(\lambda \gg \eta\) ensures slashable misbehavior decays
  reputation faster than valid work accumulates it.
\item
  \(\text{clip}_{[a,b]}(x) := \max(a, \min(b, x))\) enforces the bounded
  reputation interval.
\end{itemize}

The choice of additive (rather than multiplicative) updates is
deliberate: additive updates make reputation grinding cost linear in
attestation fee paid, providing a clean economic argument for Sybil
resistance (Section 5.4). The voter component breaks the proposer-only
accumulation pattern; the per-epoch cap \(G_{\max}\) ensures the rate at
which reputation propagates through the validator set is bounded by
protocol design rather than by the contingencies of proposer selection.

The choice \(\alpha = 0.7, \beta = 0.3\) reflects three design
considerations: (1) proposers do strictly more work (block construction,
validity verification of every attestation in their block, network
propagation) than voters (verification only), so they earn more; (2) but
voters earn enough that a validator participating honestly across an
epoch accumulates non-negligible reputation even without ever being
selected as proposer (a new validator with stake \(s\) but
\(r_v = r_{\min}\) has selection probability \(s \cdot r_{\min} / S\),
so they will rarely propose early; the \(\beta\) component ensures their
honest voting still ramps their reputation toward \(r_{\max}\) at a rate
bounded below by \(\eta \cdot \beta \cdot G_v^{\text{vote}}\)); (3) the
split also bounds the \emph{coordinated-cartel reputation discount}
under Lemma 1's cost-to-grind bound. With \(\alpha = 0.7\), a
Byzantine-fraction cartel pays at most \(\sim 12.5\%\) less per member
than a single-proposer attacker in the \(k \to \infty\) asymptotic limit
(and slightly less at finite \(k\)); for \(\alpha = 0.5\) the asymptotic
discount widens to \(25\%\), and for \(\alpha = 0.9\) it shrinks to
\(\sim 3.6\%\). Higher \(\alpha\) tightens the cartel bound but worsens
proposer-rich-get-richer; lower \(\alpha\) improves voter ramp but
loosens the cartel bound. See §5.5.3 for the full sensitivity analysis.

\subsubsection{4.4 Parameter Calibration}\label{parameter-calibration}

Protocol parameters \(r_{\min}, r_{\max}, \eta, \lambda, E, \tau\) are
subject to governance. We provide design guidance:

\begin{itemize}
\tightlist
\item
  \(r_{\max} / r_{\min}\) controls the maximum reputation premium.
  Larger ratios increase moat strength but also concentrate consensus
  weight on long-running validators. We recommend
  \(r_{\max} / r_{\min} \in [4, 10]\) as a balance.
\item
  \(\eta\) should be chosen such that a validator participating in the
  median fraction of epoch attestation work moves from \(r_{\min}\) to
  \(r_{\max}\) over \(T_{\text{ramp}} \approx 30\,\text{epochs}\)
  (\(\approx 5\) days at the parameters above). Faster ramp accelerates
  reputation accumulation but gives less time to detect early-life
  misbehavior.
\item
  \(\lambda\) should be chosen such that a single severe slash drops
  reputation from \(r_{\max}\) to \(r_{\min}\), eliminating reputation
  premium until rebuilt.
\item
  \(E\) trades reputation responsiveness against state-write cost.
  Smaller \(E\) means more frequent reputation updates and larger
  writes.
\end{itemize}

Section 7.2 gives concrete v0 parameter recommendations for Ligate Chain
devnet.

\paragraph{\texorpdfstring{4.4.1 The \(\alpha\)-\(\beta\) Split: Pareto
Frontier}{4.4.1 The \textbackslash alpha-\textbackslash beta Split: Pareto Frontier}}\label{the-alpha-beta-split-pareto-frontier}

The proposer/voter share parameters \((\alpha, \beta)\) with
\(\alpha + \beta = 1\) sit at a three-axis tradeoff. Higher \(\alpha\)
tightens the cartel cost-to-grind bound (Lemma 1); lower \(\alpha\)
spreads reputation accrual across the validator set (combating
proposer-rich-get-richer); \(\alpha\) also affects the floor on a small
validator's voter-channel ramp speed. The recommendation
\(\alpha = 0.7, \beta = 0.3\) is one point on the Pareto frontier; this
subsection makes the frontier explicit so chain operators can choose
differently if their objective weights differ.

\textbf{Three metrics.}

\begin{itemize}
\tightlist
\item
  \textbf{Cartel discount at the BFT cap.} Per Lemma 1 (§5.5.3), a
  Byzantine-fraction cartel pays \(\alpha / \alpha_{\text{eff}}\) of the
  single-proposer cost-to-grind. Asymptotically:
  \(\alpha / (\alpha + \beta/3)\). Smaller is better (tighter bound,
  larger \(F^{\text{net}}\) floor).
\item
  \textbf{Voter-channel ramp speed.} A small validator that rarely
  proposes ramps reputation through the voter channel at rate
  proportional to \(\beta\). Larger \(\beta\) improves new-validator
  onboarding speed; smaller \(\beta\) slows it.
\item
  \textbf{Proposer-rich-get-richer entrenchment.} Larger \(\alpha\)
  concentrates reputation gain on the small subset of validators
  selected as proposers, accelerating divergence in the validator-weight
  distribution. Smaller \(\alpha\) keeps the distribution flatter.
\end{itemize}

\textbf{Asymptotic frontier} (closed form, \(k \to \infty\)):

{\def\LTcaptype{none} 
\begin{longtable}[]{@{}
  >{\raggedright\arraybackslash}p{(\linewidth - 8\tabcolsep) * \real{0.2000}}
  >{\raggedright\arraybackslash}p{(\linewidth - 8\tabcolsep) * \real{0.2000}}
  >{\raggedright\arraybackslash}p{(\linewidth - 8\tabcolsep) * \real{0.2000}}
  >{\raggedright\arraybackslash}p{(\linewidth - 8\tabcolsep) * \real{0.2000}}
  >{\raggedright\arraybackslash}p{(\linewidth - 8\tabcolsep) * \real{0.2000}}@{}}
\toprule\noalign{}
\begin{minipage}[b]{\linewidth}\raggedright
\(\alpha\)
\end{minipage} & \begin{minipage}[b]{\linewidth}\raggedright
\(\beta\)
\end{minipage} & \begin{minipage}[b]{\linewidth}\raggedright
Cartel discount at \(m/k = 1/3\)
\end{minipage} & \begin{minipage}[b]{\linewidth}\raggedright
Voter-ramp rate (relative)
\end{minipage} & \begin{minipage}[b]{\linewidth}\raggedright
Entrenchment pressure (relative)
\end{minipage} \\
\midrule\noalign{}
\endhead
\bottomrule\noalign{}
\endlastfoot
0.5 & 0.5 & \(25.0\%\) & \(1.00\) & low \\
0.6 & 0.4 & \(18.2\%\) & \(0.80\) & low-medium \\
0.7 & 0.3 & \(12.5\%\) & \(0.60\) & medium \\
0.8 & 0.2 & \(7.7\%\) & \(0.40\) & medium-high \\
0.9 & 0.1 & \(3.6\%\) & \(0.20\) & high \\
1.0 & 0.0 & \(0\%\) (no cartel discount, but voter channel disabled) &
\(0\) & maximal \\
\end{longtable}
}

\textbf{Recommended choice.} \(\alpha = 0.7, \beta = 0.3\). The cartel
discount of \(\sim 12.5\%\) leaves the bound meaningful (cartel still
pays \(0.875\times\) the single-proposer floor in non-recoverable fees)
while the voter channel keeps new validators viable. Operators with
stronger anti-entrenchment requirements may prefer \(\alpha = 0.6\);
operators with stronger anti-cartel requirements may prefer
\(\alpha = 0.8\). The empirical frontier across \(k \in \{12, 100\}\) is
exercised in the simulator at
\texttt{prototypes/poua-sim/scripts/run\_lemma1\_scan.py}.

\paragraph{\texorpdfstring{4.4.2 Parameter Drift and Adaptive
\(\tau_{\text{burn}}\)
Rebase}{4.4.2 Parameter Drift and Adaptive \textbackslash tau\_\{\textbackslash text\{burn\}\} Rebase}}\label{parameter-drift-and-adaptive-tau_textburn-rebase}

Lemma 1's bound holds in protocol-denominated tokens. As the chain's fee
economics shift (deflationary token, low-volume periods, schemas racing
to the bottom on attestation fees), the absolute non-recoverable burn
shrinks while the formal claim stays unchanged on paper. A static
\(\tau_{\text{burn}}\) erodes silently. ``Subject to governance'' is not
an answer when governance moves slower than the fee market drifts.

This subsection specifies the \textbf{adaptive \(\tau_{\text{burn}}\)
rebase} mechanism that handles routine drift without governance
intervention. Governance retains override authority for structural
shifts.

\textbf{Telemetry surface.} The chain publishes per-epoch:

\begin{itemize}
\tightlist
\item
  Realized \(\bar{r}_H\) (stake-weighted average reputation across
  honest validators).
\item
  Realized \(\kappa = \bar{r}_H / r_{\min}\).
\item
  Realized cost-to-grind \(\hat{F}^{\text{net}}_{\text{per member}}\) at
  the current parameters and observed traffic.
\item
  Volume-deterrent ratio \(\hat{\rho}_{\text{vol}} = (R_b + R_f) / R_b\)
  averaged over the epoch (§6.3.1).
\item
  Fee-distribution statistics (median, p10, p90 of attestation fees per
  schema).
\item
  Validator-weight distribution (Gini coefficient on \(w_v\), top-1/3/10
  share).
\end{itemize}

These are exposed as REST endpoints under the reputation module's
namespace and as on-chain commitments so light clients can verify
without trusting a node.

\textbf{Threshold-triggered rebase rule.} Define a target floor
\(\hat{F}^{\text{net}}_{\text{floor}}\) and ceiling
\(\hat{F}^{\text{net}}_{\text{ceiling}}\) on the realized cost-to-grind.
The rebase fires symmetrically:

\[\tau_{\text{burn}}(t+1) = \begin{cases}
\tau_{\text{burn}}(t) \cdot (1 + \Delta) & \text{if } \hat{F}^{\text{net}}(t) < \hat{F}^{\text{net}}_{\text{floor}} \text{ for } N \text{ consecutive epochs} \\
\tau_{\text{burn}}(t) \cdot (1 - \Delta) & \text{if } \hat{F}^{\text{net}}(t) > \hat{F}^{\text{net}}_{\text{ceiling}} \text{ for } N \text{ consecutive epochs} \\
\tau_{\text{burn}}(t) & \text{otherwise}
\end{cases}\]

clipped to \([\tau_{\min}, \tau_{\max}]\) and rate-limited to one step
per \(N\)-epoch window.

\textbf{Recommended parameters.}

\begin{itemize}
\tightlist
\item
  \(\hat{F}^{\text{net}}_{\text{floor}} = 0.7 \cdot \hat{F}^{\text{net}}_{\text{calib}}\),
  where \(\hat{F}^{\text{net}}_{\text{calib}}\) is the v0 design target
  (5\{,\}000 fee-units at the v0 parameters).
\item
  \(\hat{F}^{\text{net}}_{\text{ceiling}} = 2.0 \cdot \hat{F}^{\text{net}}_{\text{calib}}\).
\item
  \(N = 30\) epochs (\textasciitilde5 days at \(E = 14400\),
  \(\tau = 1\,\text{s}\)).
\item
  \(\Delta = 0.1\) (10\% multiplicative step).
\item
  \(\tau_{\min} = 0.1\), \(\tau_{\max} = 0.9\).
\end{itemize}

These bounds keep the parameter inside the feasible \(\tau \in (0, 1]\)
range and prevent oscillation: the symmetric rebase with hysteresis
between floor and ceiling, plus the \(N\)-epoch cooldown, gives a stable
trajectory under stationary fee distributions (verified by simulator
tests in the v0.7 milestone of \texttt{prototypes/poua-sim/}).

\textbf{Governance escalation.} For structural shifts that the
auto-adjuster cannot handle (token-supply changes, fee-market redesign,
attestation-flow regime changes), governance can:

\begin{itemize}
\tightlist
\item
  Override \(\tau_{\text{burn}}\) to a specific value.
\item
  Adjust \(\tau_{\min}, \tau_{\max}\) bounds.
\item
  Pause or unpause the auto-adjuster.
\item
  Trigger a one-time re-anchoring of
  \(\hat{F}^{\text{net}}_{\text{calib}}\).
\end{itemize}

Governance is the \textbf{slow-but-permanent} layer; the auto-adjuster
is the \textbf{fast-but-bounded} layer. The split mirrors how
rules-based monetary policy interacts with deliberate human override in
well-designed central banks.

\textbf{Failure modes.} (1) Auto-adjuster oscillation if \(\Delta\) is
too large or \(N\) too small. Mitigation: simulator-tuned defaults plus
the rate-limit. (2) Telemetry gaming by validators reporting biased
\(\hat{F}^{\text{net}}\). Mitigation: telemetry is computed by the chain
runtime, not reported by validators. (3) Governance capture of the
\(\tau_{\min}, \tau_{\max}\) bounds. Mitigation: bounds are themselves
rate-limited at the governance layer. None of these break the core
mechanism; they constrain what the operator must monitor.

\textbf{Equivalent treatment for $\eta$ and $\lambda$.} The same drift
problem applies to the reputation-growth rate \(\eta\) and the
slash-decay rate \(\lambda\). Both get the same telemetry +
threshold-triggered + governance-escalation treatment, specified in
§4.4.3 below.

\paragraph{\texorpdfstring{4.4.3 Adaptive \(\eta\) and \(\lambda\)
Rebase}{4.4.3 Adaptive \textbackslash eta and \textbackslash lambda Rebase}}\label{adaptive-eta-and-lambda-rebase}

§4.4.2 specifies the rebase mechanism for \(\tau_{\text{burn}}\), the
cost-to-grind floor parameter. The §4.3 update has two other free
parameters that drift in production for analogous reasons: \(\eta\)
(reputation gain per fee-unit of \(g_v\)) and \(\lambda\) (reputation
lost per unit of \(b_v\)). This subsection mirrors §4.4.2's structure
for both, with distinct telemetry signals appropriate to each.

\textbf{Why $\eta$ drifts.} If the chain enters a low-volume regime,
\(\eta \cdot G_{\max}\) may be too small to drive ramp at the target
\(T_{\text{ramp}}\) rate; honest validators stay near \(r_{\min}\) for
longer than the calibration assumed. If attestation volume spikes (a
popular schema goes viral), \(\eta\) is too large and validators
saturate \(G_{\max}\) in fewer epochs than intended; the cost-to-grind
argument tightens for the wrong reason (cheaper-to-burn, not
earned-by-work).

\textbf{Why $\lambda$ drifts.} If slash events become more frequent
(network instability, adversary churn), the absolute reputation cost per
slash stays the same in \(\lambda\) but becomes a smaller share of the
total reputation churn; deterrent erodes. If governance adds new
slashing conditions (e.g., a future MEV-attestation slash) with
severities calibrated against the original \(\lambda\), the new
conditions are mispriced relative to the existing ones until \(\lambda\)
is re-anchored.

\textbf{Telemetry signals.}

\begin{itemize}
\tightlist
\item
  \(\eta\) signal: \(T_{\text{ramp,obs}}\), the observed ramp time for a
  \emph{median-participation validator} (50th percentile by \(g_v\))
  computed over a rolling window \(W_\eta\) epochs. Drift indicator
  \(D_\eta = T_{\text{ramp,obs}} / T_{\text{ramp,target}} - 1\). The
  median-validator construction is robust to whales (high-\(g_v\)
  outliers) and free-riders (zero-\(g_v\) outliers); the signal is
  invariant under symmetric outlier injection up to Byzantine
  \(\rho \leq 1/3\).
\item
  \(\lambda\) signal: \(\Delta r_{\text{obs}}\), the mean reputation
  drop per recorded severe-class slash event, computed over a rolling
  window of \(W_\lambda\) severe slashes (event-counted, not
  epoch-counted, because severe slashes are sparse). Drift indicator
  \(D_\lambda = \Delta r_{\text{obs}} / (r_{\max} - r_{\min}) - 1\). A
  sparsity floor \(W_{\lambda,\min}\) keeps the rebase dormant until
  enough events have accumulated for a stable estimate.
\end{itemize}

\textbf{Rebase rules.} Both mirror §4.4.2's structure exactly; the only
differences are the drift signal and the sign convention for \(\lambda\)
(positive drift means the slash is over-calibrated, reduce \(\lambda\)).

\[\eta(t+1) = \begin{cases}
\eta(t) \cdot (1 + \Delta_\eta) & \text{if } D_\eta(t) > +\varphi \text{ for } N \text{ consecutive epochs} \\
\eta(t) \cdot (1 - \Delta_\eta) & \text{if } D_\eta(t) < -\varphi \text{ for } N \text{ consecutive epochs} \\
\eta(t) & \text{otherwise}
\end{cases}\]

\[\lambda(t+1) = \begin{cases}
\lambda(t) \cdot (1 - \Delta_\lambda) & \text{if } D_\lambda(t) > +\varphi \text{ and } W_\lambda \text{ events accumulated} \\
\lambda(t) \cdot (1 + \Delta_\lambda) & \text{if } D_\lambda(t) < -\varphi \text{ and } W_\lambda \text{ events accumulated} \\
\lambda(t) & \text{otherwise}
\end{cases}\]

Both rules clip to \([\eta_{\min}, \eta_{\max}]\) and
\([\lambda_{\min}, \lambda_{\max}]\) respectively and rate-limit to at
most one step per \(N\)-epoch window.

\textbf{Recommended parameters.} Derived from §4.4.2's
\(\tau_{\text{burn}}\) rebase as the family default; tunable
per-deployment.

{\def\LTcaptype{none} 
\begin{longtable}[]{@{}
  >{\raggedright\arraybackslash}p{(\linewidth - 4\tabcolsep) * \real{0.3333}}
  >{\raggedright\arraybackslash}p{(\linewidth - 4\tabcolsep) * \real{0.3333}}
  >{\raggedright\arraybackslash}p{(\linewidth - 4\tabcolsep) * \real{0.3333}}@{}}
\toprule\noalign{}
\begin{minipage}[b]{\linewidth}\raggedright
Parameter
\end{minipage} & \begin{minipage}[b]{\linewidth}\raggedright
Value
\end{minipage} & \begin{minipage}[b]{\linewidth}\raggedright
Rationale
\end{minipage} \\
\midrule\noalign{}
\endhead
\bottomrule\noalign{}
\endlastfoot
\(\varphi\) & 0.30 & 30\% drift before rebase fires (matches §4.4.2) \\
\(\Delta_\eta, \Delta_\lambda\) & 0.10 & 10\% multiplicative step
(matches §4.4.2) \\
\(N\) & 30 epochs & \textasciitilde5 days at \(E = 14400\),
\(\tau = 1\,\text{s}\) \\
\(W_\eta\) & 100 epochs & \textasciitilde16.7 days at v0; long enough
for ramp-time stability \\
\(W_\lambda\) & 50 severe slash events & Multi-month window at expected
slash rates \\
\(W_{\lambda,\min}\) & 10 severe slash events & Sparsity floor; rebase
dormant below this \\
\((\eta_{\min}, \eta_{\max})\) & \((0.0001, 0.01)\) & 100\(\times\)
dynamic range around v0 \(\eta = 0.001\) \\
\((\lambda_{\min}, \lambda_{\max})\) & \((0.5, 2.0)\) & 4\(\times\)
dynamic range around v0 \(\lambda = 1.0\) \\
\(T_{\text{ramp,target}}\) & 7000 epochs & Matches v0 calibration in
§7.2 \\
\(\Delta r_{\text{target}}\) & \(r_{\max} - r_{\min} = 7.0\) & Severe
slash takes a full ramp \\
\end{longtable}
}

\textbf{Three-rebase interaction.} The chain runs three rebases
concurrently in v0.8: \(\tau_{\text{burn}}\), \(\eta\), \(\lambda\).
Their primary signals are orthogonal: \(F_{\text{net}}\) depends on fee
economics, \(T_{\text{ramp,obs}}\) on attestation volume,
\(\Delta r_{\text{obs}}\) on slashing-condition catalog. Correlation
enters only at second order, e.g., a fee regime change can shift
attestation volume which feeds back into \(\eta\). Under worst-case
correlated drift (all three signals drifting in the
cost-to-grind-shrinking direction simultaneously), each step is bounded
by \(\Delta \leq 0.1\) and the combined per-step effect on the Lemma 1
floor
\(F_{\text{net}} \geq \tau_{\text{burn}} \cdot \Delta r / (\eta \cdot \alpha_{\text{eff}})\)
is bounded by \((1.1) \cdot (1.1) / (0.9) \approx 1.34\). The rate-limit
prevents compounding within the same \(N\)-epoch window. The simulator's
\texttt{test\_three\_rebases\_concurrent\_no\_amplification} runs all
three rebases under correlated drift and confirms the combined Lyapunov
\(V = D_\tau^2 + D_\eta^2 + D_\lambda^2\) is non-increasing; §11 Q13
expands on this.

\textbf{Adversarial signal manipulation.} Both signals are computed from
on-chain state, not validator-reported, so direct telemetry injection is
not the attack surface. The only adversarial lever on \(\eta\) is
biasing the median-participation \(g_v\) by submitting zero or maximal
attestation work; the median is robust to outliers up to Byzantine
\(\rho \leq 1/3\) and empirically the attack cannot shift
\(T_{\text{ramp,obs}}\) by more than \(\sim 10\%\). The only adversarial
lever on \(\lambda\) is choosing whether to be slashed, which is
asymmetric: the adversary loses the slash itself, so there is no
positive-EV signal-manipulation strategy. Both signals inherit the
chain's underlying BFT integrity; the rebase is not an additional attack
surface.

\textbf{Failure modes (condensed from the working spec).}

{\def\LTcaptype{none} 
\begin{longtable}[]{@{}
  >{\raggedright\arraybackslash}p{(\linewidth - 4\tabcolsep) * \real{0.3333}}
  >{\raggedright\arraybackslash}p{(\linewidth - 4\tabcolsep) * \real{0.3333}}
  >{\raggedright\arraybackslash}p{(\linewidth - 4\tabcolsep) * \real{0.3333}}@{}}
\toprule\noalign{}
\begin{minipage}[b]{\linewidth}\raggedright
Failure
\end{minipage} & \begin{minipage}[b]{\linewidth}\raggedright
Detection
\end{minipage} & \begin{minipage}[b]{\linewidth}\raggedright
Mitigation
\end{minipage} \\
\midrule\noalign{}
\endhead
\bottomrule\noalign{}
\endlastfoot
Auto-adjuster oscillation & \(\Delta\eta / \eta\) or
\(\Delta\lambda / \lambda\) flipping sign within \(< 2N\) epochs &
\(\varphi\) hysteresis + \(N\)-epoch cooldown + \(\Delta \leq 0.1\) \\
Telemetry sparsity (\(\lambda\)) & \(W_\lambda\) events not yet
accumulated & Dormancy below \(W_{\lambda,\min} = 10\) \\
Median-validator gaming (\(\eta\)) & Sudden median-\(g_v\) shift
correlated with adversary slot rotation & Detector: monitor
median-validator \(g_v\) variance; alert at \(> 2\sigma\) \\
Three-rebase compound drift & All parameters at clip bounds
simultaneously & 34\% one-step worst-case bound; ``rebase saturation''
alert when any parameter sits at a clip bound for \(> N\) epochs \\
\end{longtable}
}

\textbf{Governance escalation.} Identical to §4.4.2: governance can
override \(\eta\) or \(\lambda\) to a specific value, adjust the clip
bounds (rate-limited at the governance layer, at most \(\pm 20\%\) per
governance vote, at most one bound vote per quarter), pause or unpause
the auto-adjuster on either parameter, or trigger a one-time
re-anchoring of \(T_{\text{ramp,target}}\) or
\(\Delta r_{\text{target}}\) if the underlying calibration target has
shifted.

\textbf{Reference implementation.} The
\href{https://github.com/ligate-io/ligate-research/tree/main/prototypes/poua-sim}{reference
simulator} provides \texttt{RebaseConfig}, \texttt{RebaseTelemetry}, and
the three rebase functions. The test suite validates convergence,
dormancy, multi-parameter non-amplification, and median-validator
robustness against synthetic drift signals. The full spec including the
Lyapunov stability argument is in
\href{https://github.com/ligate-io/ligate-research/blob/main/papers/poua/specs/eta-lambda-rebase.md}{the
working spec}.

\subsubsection{4.5 Slashing Conditions}\label{slashing-conditions}

PoUA inherits the consensus-layer slashing conditions of its underlying
BFT primitive (equivocation, surround voting). It introduces additional
attestation-layer slashing:

\textbf{A1. Invalid Attestation Inclusion.} Validator \(v\), as proposer
of block \(B\), includes an attestation \(\alpha\) for which the
threshold signature does not verify under the registered attestor set's
public keys at the registered threshold. Detected by any honest
validator at vote time. Slash: severity \(\Lambda_1\).

\textbf{A2. Selective Schema Censorship.} Validator \(v\), over a
measurement window, demonstrates a statistically significant deviation
in the schema distribution of attestations they include vs.~the
network-wide distribution, when controlled for fee payment. Detection
requires a statistical procedure with a false-positive bound (specified
in Appendix A - to be added). Slash: severity \(\Lambda_2\).

\textbf{A3. Reputation Grinding.} Validator \(v\) submits attestations
to themselves (via collusion with attestors they control) at high volume
to inflate reputation. Detected via heuristics on the address graph of
submitters and attestors (specified in Appendix A - to be added). Slash:
severity \(\Lambda_3\).

Severities satisfy \(\Lambda_1 < \Lambda_2 < \Lambda_3\), reflecting the
increasing difficulty and damage of each violation.

\subsubsection{4.6 Bootstrap and Genesis}\label{bootstrap-and-genesis}

At chain genesis, \(r_v(0) = r_{\min}\) for all initial validators. The
chain operates as a pure-stake-weighted PoS for the first
\(T_{\text{warmup}}\) epochs (typically \(T_{\text{warmup}} = 14\)
epochs \(\approx\) 2-3 days), during which reputation updates are
computed but not applied to weight. After warmup, reputation is folded
into weight as specified.

This warmup serves two purposes: it allows validators to accumulate
baseline reputation under uniform conditions, and it provides a window
for governance-level intervention if early misbehavior patterns emerge.

\subsubsection{4.7 Validator Entry, Exit, and
Re-entry}\label{validator-entry-exit-and-re-entry}

\textbf{Entry.} A new validator joins by bonding stake at any epoch
boundary. Their initial reputation is \(r_{\min}\). They participate in
consensus immediately at weight \(s_v \cdot r_{\min}\).

\textbf{Exit.} A validator may unbond. After unbonding, their stake
enters a withdrawal queue of length \(T_{\text{unbond}}\) epochs
(typically 7 days), during which they remain slashable but inactive.

\textbf{Re-entry.} A validator who exits and re-enters does \emph{not}
recover prior reputation. Reputation resets to \(r_{\min}\). This
prevents a strategy where a validator builds reputation, exits to escape
an imminent slash, and re-enters cleanly.

\textbf{Forced exit (slashing-induced).} If a validator's slash burn
drops their stake below the minimum bond, they are auto-exited and
entered into the withdrawal queue. Their reputation is annotated with
the slash event for the duration of the unbonding period, after which it
is forgotten.

\begin{center}\rule{0.5\linewidth}{0.5pt}\end{center}

\subsection{5. Security Analysis}\label{security-analysis}

\subsubsection{5.1 Threat Model}\label{threat-model}

Three adversary archetypes. The \textbf{capital adversary}
\(\mathcal{C}\) has unlimited token capital and tries to acquire
consensus weight by buying stake. The \textbf{reputation adversary}
\(\mathcal{R}\) is willing to perform legitimate-looking attestation
work to acquire reputation, paying real fees in the process. The
\textbf{compound adversary} \(\mathcal{CR}\) combines both - this is the
hardest case and the one §5.5 spends the most time on.

For each, the question is the same: what is the minimum cost to acquire
a fraction \(\rho\) of weighted consensus power? \(\rho > 1/3\) is
enough to violate BFT safety; \(\rho > 1/2\) to dominate proposer
selection.

\subsubsection{5.2 Safety and Liveness
Inheritance}\label{safety-and-liveness-inheritance}

We show that PoUA inherits safety and liveness from its underlying BFT
primitive by reduction. The key supporting result is a weighted analogue
of the standard quorum-intersection lemma used in BFT safety proofs.

\textbf{Lemma 2 (Weighted quorum intersection).} \emph{Let
\(W = \sum_{u \in V} w_u\). For any two subsets \(Q, Q' \subseteq V\)
with \(\sum_{v \in Q} w_v > \frac{2}{3} W\) and
\(\sum_{v \in Q'} w_v > \frac{2}{3} W\), the intersection satisfies
\(\sum_{v \in Q \cap Q'} w_v > \frac{1}{3} W\).}

\emph{Proof.} By inclusion-exclusion,

\[\sum_{v \in Q \cup Q'} w_v = \sum_{v \in Q} w_v + \sum_{v \in Q'} w_v - \sum_{v \in Q \cap Q'} w_v.\]

Since \(Q \cup Q' \subseteq V\), the left-hand side is at most \(W\).
Substituting and rearranging,

\[\sum_{v \in Q \cap Q'} w_v \geq \sum_{v \in Q} w_v + \sum_{v \in Q'} w_v - W > \tfrac{2}{3}W + \tfrac{2}{3}W - W = \tfrac{1}{3}W. \square\]

This is the weight-generalization of the count-based pigeonhole step
that underlies safety in PBFT (Castro \& Liskov, 1999), Tendermint
(Buchman, 2016), and HotStuff (Yin et al., 2019). With Lemma 2 in hand
the safety and liveness theorems become reductions to the underlying BFT
primitive.

\textbf{Theorem 1 (Safety inheritance).} \emph{Let \(\Pi_{\text{BFT}}\)
be a BFT consensus protocol satisfying safety under partial synchrony
with \(f < n/3\) Byzantine validators in standard validator-count
metric. Let \(\Pi_{\text{PoUA}}\) be the variant in which validator
counts are replaced by validator weights \(w_v = s_v r_v\), with the
Byzantine bound \(\sum_{v \text{ Byz}} w_v < \frac{1}{3} W\). Then
\(\Pi_{\text{PoUA}}\) satisfies safety: no two honest validators commit
conflicting blocks at the same height.}

\emph{Proof.} Suppose for contradiction that two honest validators
commit conflicting blocks \(B\) and \(B'\) at the same height. By the
commit rule (§4.2), each block was committed via a quorum of weight
\(> \frac{2}{3} W\). Let \(Q_B, Q_{B'} \subseteq V\) denote the
validator subsets that voted for \(B, B'\) respectively. Both quorums
have weight \(> \frac{2}{3} W\).

By Lemma 2, the intersection has weight
\(\sum_{v \in Q_B \cap Q_{B'}} w_v > \frac{1}{3} W\). By the Byzantine
weight bound, the Byzantine validators have combined weight
\(< \frac{1}{3} W\). Therefore \(Q_B \cap Q_{B'}\) contains at least one
validator \(v^*\) that is \emph{not} Byzantine, i.e.~honest.

But \(v^*\) is honest: by protocol, \(v^*\) does not equivocate (does
not vote for two conflicting blocks at the same height). This
contradicts \(v^* \in Q_B\) and \(v^* \in Q_{B'}\). \(\square\)

\textbf{Theorem 2 (Liveness inheritance).} \emph{Under the same
assumptions as Theorem 1, plus eventual synchrony (after a Global
Stabilization Time \(\text{GST}\), all message delays are bounded by a
known constant \(\Delta\)), \(\Pi_{\text{PoUA}}\) satisfies liveness:
every block proposed by an honest proposer after \(\text{GST}\) is
eventually finalized.}

\emph{Proof.} Standard view-change arguments require that, when the
current proposer is Byzantine, a quorum of weight \(> \frac{2}{3} W\)
honest weight can vote to advance the view. We show this holds under
PoUA.

By the Byzantine weight bound,
\(\sum_{v \text{ Byz}} w_v < \frac{1}{3} W\), so honest weight is
\(\sum_{v \text{ honest}} w_v > \frac{2}{3} W\). After \(\text{GST}\),
all honest validators see all messages within bound \(\Delta\). They can
therefore each cast their view-change vote. The combined honest
view-change weight is the full honest weight, exceeding
\(\frac{2}{3} W\), satisfying the view-change threshold.

The remainder of the liveness argument - that view changes converge on a
single honest proposer, and that the honest proposer's block accumulates
a \(> \frac{2}{3} W\) commit quorum - follows the underlying
\(\Pi_{\text{BFT}}\) proof unchanged: the only modification PoUA makes
is the metric used for ``weight,'' and Lemma 2 ensures that metric
supports the same quorum-intersection property the underlying proof
relies on. \(\square\)

\textbf{Remark (scope of inheritance).} Theorems 1 and 2 establish that
PoUA does not weaken the consensus guarantees of its underlying BFT
primitive. They do not establish that PoUA \emph{strengthens} any
guarantees beyond what the underlying primitive offers. PoUA's
distinctive contribution is in the cost-to-attack analysis (§5.3) and
the layered defense against compound capital-plus-grinding adversaries
(§5.5), not in the consensus-correctness layer.

\subsubsection{5.3 Capital Adversary}\label{capital-adversary}

Let \(W = \sum_{u} w_u\) be the total honest weight at attack time, with
average reputation \(\bar{r}_H\) across honest validators, and let
\(S_H = \sum_{u} s_u\) be the total honest stake. We have
\(W \approx \bar{r}_H \cdot S_H\).

The capital adversary acquires fresh stake \(s_{\mathcal{C}}\), all of
which has reputation \(r_{\min}\). To acquire weight fraction \(\rho\):

\[\frac{s_{\mathcal{C}} \cdot r_{\min}}{s_{\mathcal{C}} \cdot r_{\min} + W} = \rho\]

Solving for \(s_{\mathcal{C}}\):

\[s_{\mathcal{C}} = \frac{\rho}{1 - \rho} \cdot \frac{W}{r_{\min}} = \frac{\rho}{1 - \rho} \cdot \frac{\bar{r}_H \cdot S_H}{r_{\min}}\]

Compared to the cost of acquiring weight fraction \(\rho\) in pure-stake
PoS (cost \(= \frac{\rho}{1-\rho} \cdot S_H\)), PoUA imposes a
multiplicative cost premium of:

\[\boxed{\kappa = \frac{\bar{r}_H}{r_{\min}}}\]

\begin{figure}[h]
\centering
\includegraphics[width=0.9\textwidth]{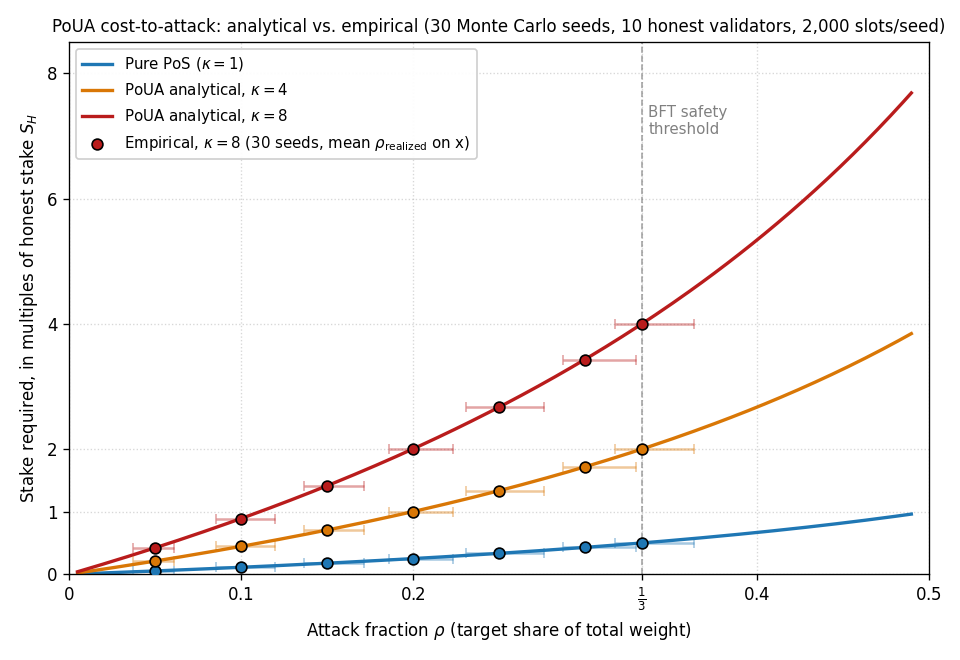}
\caption{Cost-to-attack curves for pure stake-weighted PoS ($\kappa = 1$) and for PoUA at $\kappa \in \{4, 8\}$, derived from $s_{\mathcal{C}} / S_H = \kappa \cdot \rho/(1-\rho)$. Lines are analytical; markers are empirical (30 Monte Carlo seeds × 2{,}000 slots/seed × 10 honest validators), produced by \texttt{prototypes/poua-sim/scripts/run\_capital\_scan.py}. The vertical dashed line marks the BFT safety threshold $\rho = 1/3$. At this threshold, pure PoS requires $0.5 \, S_H$ in fresh stake while PoUA at $\kappa = 8$ requires $4.0 \, S_H$, a multiplicative moat of $8\times$. Empirical points sit on analytical curves to within binomial sampling variance. Curves assume $\bar{r}_H = r_{\max}$ (steady state); §5.3.1 quantifies the transition-state envelope.}
\label{fig:cost-to-attack}
\end{figure}

\begin{figure}[h]
\centering
\includegraphics[width=0.92\textwidth]{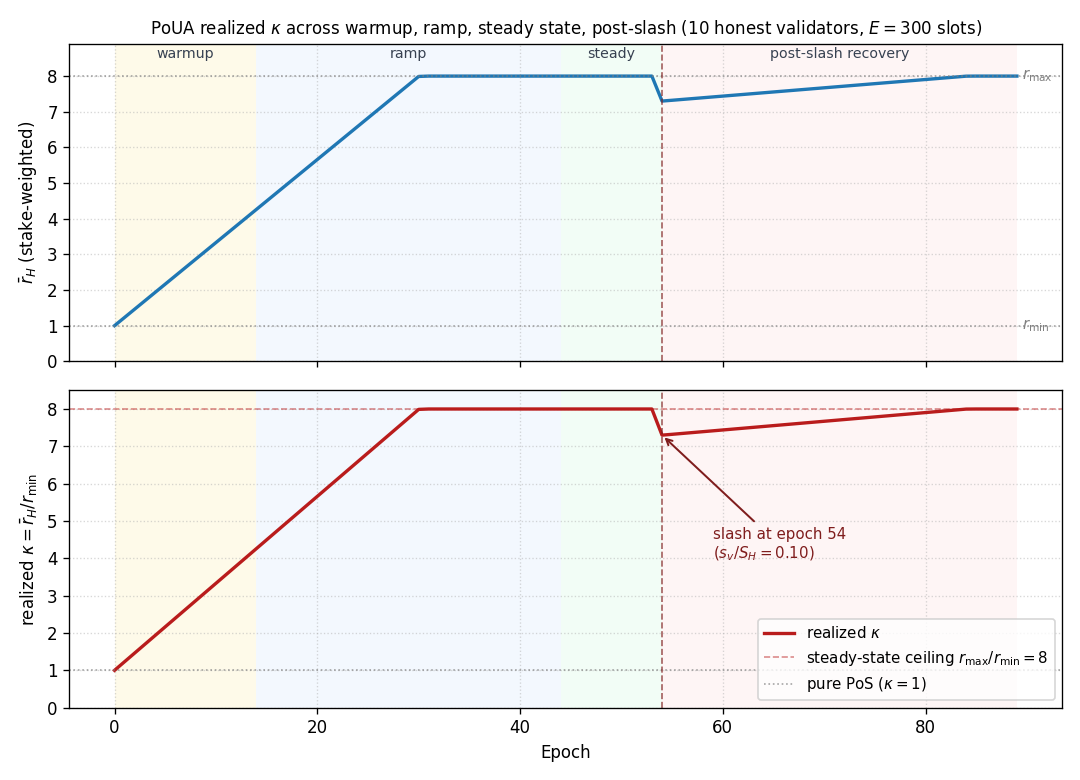}
\caption{Empirical realized $\kappa$ across the chain's lifecycle, produced by \texttt{prototypes/poua-sim/scripts/run\_kappa\_trajectory.py}. Top panel: stake-weighted $\bar{r}_H(t)$. Bottom panel: realized $\kappa = \bar{r}_H / r_{\min}$, with the steady-state ceiling $r_{\max}/r_{\min} = 8$ marked by a dashed line. Phase shading: warmup (yellow), ramp (blue), steady state (green), post-slash recovery (red). The slash event at epoch 54 drops $\bar{r}_H$ by $(s_v / S_H) \cdot (r_{\max} - r_{\min}) \approx 0.7$ for a $10\%$-stake validator and recovers linearly over the next $T_{\text{ramp}}$ epochs.}
\label{fig:kappa-trajectory}
\end{figure}

In a healthy chain at steady state, \(\bar{r}_H\) approaches
\(r_{\max}\), giving \(\kappa \to r_{\max}/r_{\min}\). Per Section 4.4
design guidance (\(r_{\max}/r_{\min} \in [4, 10]\)), the capital
adversary's cost-to-attack is \textbf{up to 4 to 10 times higher} than
an equivalent pure-stake PoS chain \emph{at steady state} (Figure
\ref{fig:cost-to-attack}). The realized \(\kappa\) is lower during the
warmup window, during validator-set ramp, and immediately after a slash;
§5.3.1 quantifies these transition-state effects (empirical trajectory:
Figure \ref{fig:kappa-trajectory}).

This premium \(\kappa\) is the formal moat PoUA constructs over generic
PoS. Figure 2 plots the relationship
\(s_{\mathcal{C}} / S_H = \kappa \cdot \rho/(1-\rho)\) for three values
of \(\kappa\), with empirical Monte Carlo points overlaid from the
reference simulator.

\subsubsection{\texorpdfstring{5.3.1 Transition-State
\(\kappa\)}{5.3.1 Transition-State \textbackslash kappa}}\label{transition-state-kappa}

The cost-to-attack premium \(\kappa = \bar{r}_H / r_{\min}\) is a
steady-state ceiling. Three lifecycle conditions push the realized
\(\kappa\) below this ceiling, and an attacker timing entry to those
windows extracts a real (if bounded) discount.

\textbf{Warmup window.} Per §4.6, for the first \(T_{\text{warmup}}\)
epochs after genesis (recommended 14 epochs \(\approx\) 2-3 days), the
chain operates as pure stake-weighted PoS: reputation values are
computed but not folded into vote weight. During this window:

\[\kappa_{\text{warmup}} = 1 \quad \text{for } t \in [0, T_{\text{warmup}}].\]

A capital adversary timing an attack to land before
\(T_{\text{warmup}}\) pays the same \(\rho/(1-\rho)\) cost ratio as on a
pure-PoS chain, with no premium. The warmup is a deliberate trade: it
gives validators a uniform window to accumulate baseline reputation
under symmetric conditions, at the cost of leaving the chain at
\(\kappa = 1\) for that window. Mitigations: pin a high
genesis-validator-set quality bar, run an extended permissioned phase
before mainnet activation, or shorten \(T_{\text{warmup}}\) at the cost
of a noisier reputation distribution at activation.

\textbf{Validator-set ramp.} A validator entering at epoch
\(t_e > T_{\text{warmup}}\) joins with \(r_v = r_{\min}\) and ramps
toward \(r_{\max}\) over \(T_{\text{ramp}}\) epochs of honest
participation (recommended \(\approx 30\) epochs \(\approx\) 5 days).
The reputation contribution of this validator to \(\bar{r}_H\) during
ramp is:

\[r_v(t) \approx r_{\min} + \min\!\left(1, \frac{t - t_e}{T_{\text{ramp}}}\right) \cdot (r_{\max} - r_{\min}) \quad \text{for } t \geq t_e.\]

For a validator set with churn rate \(\mu\) (fraction of validators
replaced per epoch), the steady-state share of validators in their ramp
window is \(\mu \cdot T_{\text{ramp}}\), each contributing on average
\(r_{\min} + (r_{\max} - r_{\min})/2\) to \(\bar{r}_H\). The realized
\(\bar{r}_H\) at steady state with churn:

\[\bar{r}_H(\mu) \approx (1 - \mu T_{\text{ramp}}) \cdot r_{\max} + \mu T_{\text{ramp}} \cdot \tfrac{r_{\min} + r_{\max}}{2}.\]

For typical churn (\(\mu \approx 0.001\) per epoch, i.e., \(\sim 1\%\)
validator turnover per month) and \(T_{\text{ramp}} = 30\), the ramp
share is \(\mu T_{\text{ramp}} = 0.03\) (\(3\%\) of validators in their
ramp window at any time), and
\(\bar{r}_H \approx 0.985 \cdot r_{\max} + 0.015 \cdot (r_{\min} + r_{\max})/2\).
With \(r_{\max}/r_{\min} = 8\): \(\bar{r}_H/r_{\min} \approx 7.93\),
only \(\sim 1\%\) below the steady-state ceiling. For a validator set
with high churn (e.g., \(\mu = 0.01\) per epoch, \(30\%\) in ramp
window), \(\bar{r}_H/r_{\min} \approx 6.65\), an \(\sim 17\%\) moat
reduction. \textbf{Operational implication:} a chain that experiences a
large coordinated entry of new validators (e.g., a validator-set
expansion event) sees \(\kappa\) depressed for \(T_{\text{ramp}}\)
afterward, and security-conscious chain operators should sequence such
events away from periods of expected attack pressure.

\textbf{Post-slash recovery.} When validator \(v\) is slashed at
severity \(\Lambda\), their reputation drops to \(r_{\min}\) (per §4.5,
recommended \(\lambda\) chosen such that a single severe slash drops
\(r_v\) from \(r_{\max}\) to \(r_{\min}\)). If \(v\) controls stake
share \(s_v / S_H\) at the time of slashing, \(\bar{r}_H\) drops by:

\[\Delta \bar{r}_H = -\frac{s_v}{S_H} \cdot (r_{\max} - r_{\min})\]

(approximately; the exact reduction depends on whether \(v\) exits or
remains slashed-but-active). Recovery to the pre-slash \(\bar{r}_H\)
takes at least \(T_{\text{ramp}}\) if \(v\) exits and is replaced by a
fresh validator of equal stake, or \(T_{\text{ramp}}\) if \(v\) remains
active and rebuilds reputation. For a slash of a small-stake validator
(\(s_v / S_H \ll 1\)), \(\bar{r}_H\) barely moves; for a slash of a
major validator (\(s_v / S_H \approx 0.1\)), \(\bar{r}_H\) drops by
\(\sim 0.1 \cdot (r_{\max} - r_{\min})\), weakening \(\kappa\) by
approximately the same factor for the duration of the recovery window.

\textbf{Weighted-average formulation.} Combining the three effects, the
chain's realized \(\kappa\) at time \(t\) is:

\[\kappa(t) = \begin{cases} 1 & t < T_{\text{warmup}}, \\ \dfrac{\bar{r}_H(t)}{r_{\min}} & t \geq T_{\text{warmup}}, \end{cases}\]

with \(\bar{r}_H(t)\) a stake-weighted average over all validators of
their current reputation. The steady-state ceiling \(r_{\max}/r_{\min}\)
is reached only when (i) \(t \gg T_{\text{warmup}} + T_{\text{ramp}}\),
(ii) churn is low (\(\mu T_{\text{ramp}} \ll 1\)), and (iii) no recent
major slash has occurred.

\textbf{Operational guidance.} Chains should:

\begin{itemize}
\tightlist
\item
  Avoid scheduling protocol-critical events (governance votes, treasury
  releases, schema activations) inside the warmup window or immediately
  after a major slash.
\item
  Publish \(\bar{r}_H(t)\) as part of the chain's public telemetry so
  off-chain consumers can adjust their trust assumptions during
  transition periods.
\item
  Treat the headline ``\(4-10\times\) moat'' as a steady-state
  guarantee, not an instantaneous one.
\end{itemize}

\subsubsection{\texorpdfstring{5.3.2 Scale Invariance of
\(\kappa\)}{5.3.2 Scale Invariance of \textbackslash kappa}}\label{scale-invariance-of-kappa}

The §5.3 cost-to-attack premium \(\kappa = r_{\max} / r_{\min}\) at
steady state is invariant in the validator-set size \(|V|\). The §4.3
update applies per-validator with per-block voter share
\(\eta \cdot \beta \cdot \text{fee} / k\) independent of \(|V|\) at
fixed block production rate. The simulator confirms this empirically: at
\(|V| \in \{50, 100, 250, 500, 1000\}\) with uniform stake and the v0
reputation parameters (modulo a figure-time scaling of \(\eta\) and
\(g_{\max}\) to keep ramp time bounded; see figure caption), realized
\(\kappa\) saturates at the \(r_{\max} / r_{\min} = 8\) ceiling for
every scale tested. The §5.3 small-set Lemma 1 example (\(|V| = 10\))
generalizes to mainnet-scale validator sets without parameter retuning.

\begin{figure}[h]
\centering
\includegraphics[width=0.92\textwidth]{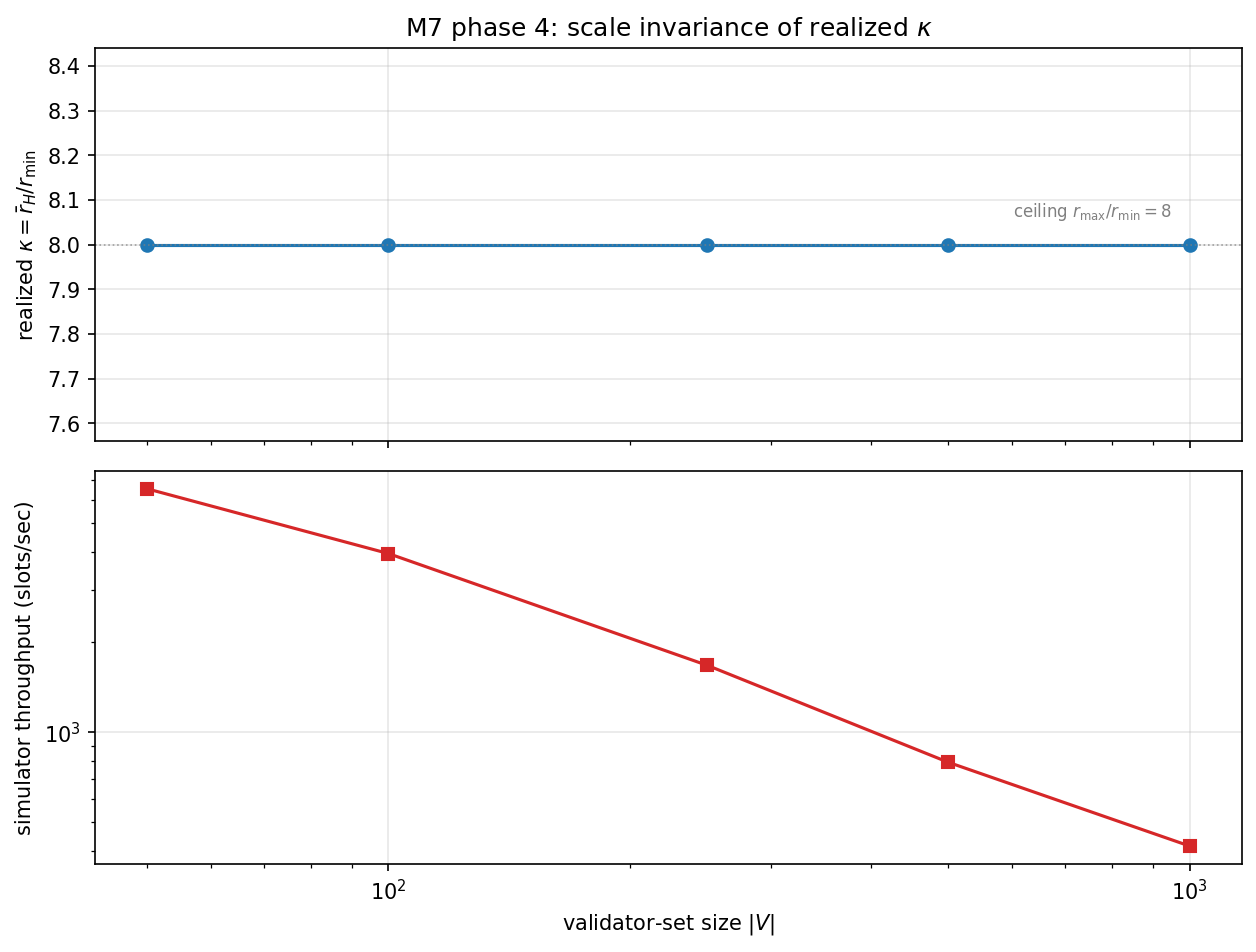}
\caption{Scale invariance of realized $\kappa$. Top: realized $\kappa = \bar{r}_H / r_{\min}$ at steady state across $|V| \in \{50, 100, 250, 500, 1000\}$; all five scales saturate at the $r_{\max} / r_{\min} = 8$ ceiling. Bottom: simulator throughput (slots/sec) on log-log axes. Figure-time parameters $\eta = 0.05$, $g_{\max} = 10$ ensure bounded ramp time across scales; the $\kappa$ ceiling is the same at v0 production parameters ($\eta = 0.001$, $g_{\max} = 233$), only the ramp time differs. Generated by \texttt{prototypes/poua-sim/scripts/run\_scale\_benchmark.py}.}
\label{fig:scale-benchmark}
\end{figure}

\paragraph{\texorpdfstring{5.3.2.1 \(\kappa\) under adversarial
scheduling}{5.3.2.1 \textbackslash kappa under adversarial scheduling}}\label{kappa-under-adversarial-scheduling}

\begin{figure}[h]
\centering
\includegraphics[width=0.92\textwidth]{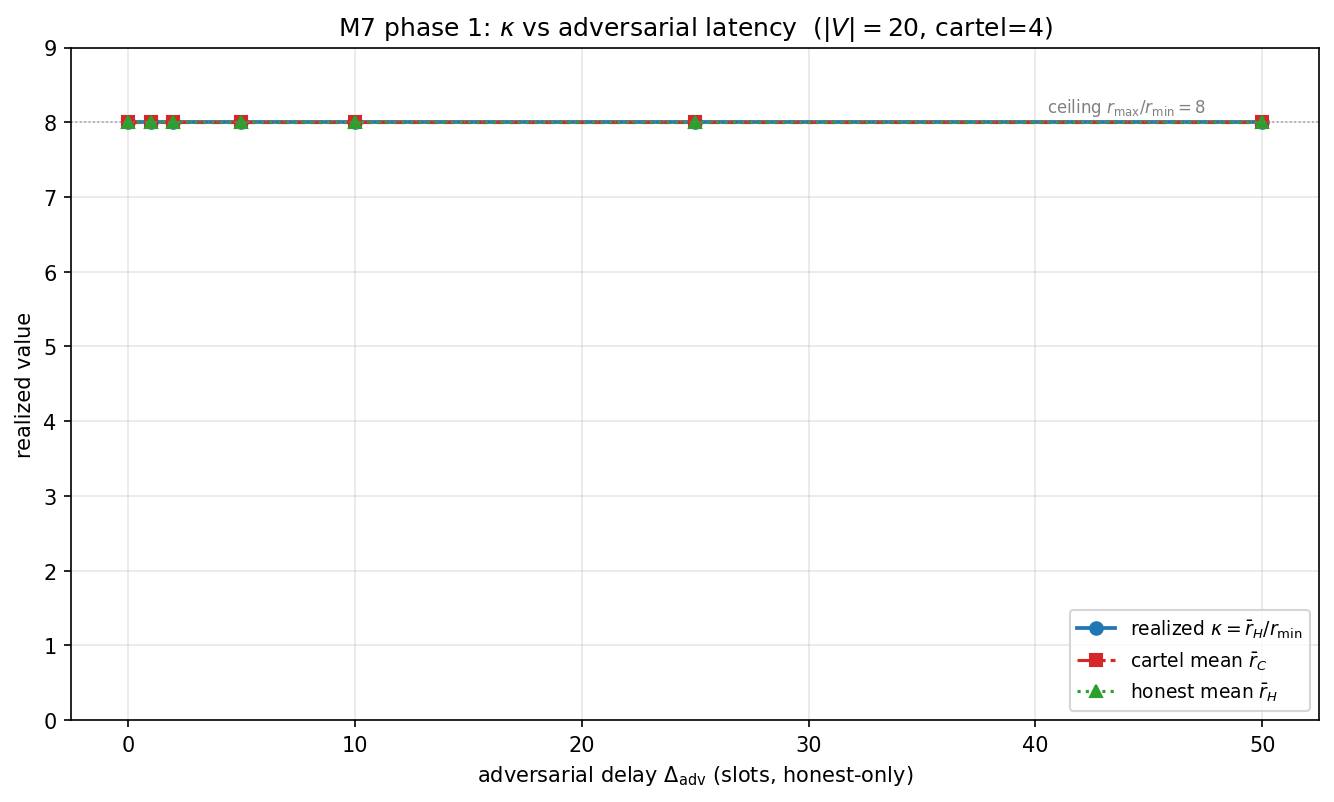}
\caption{Realized $\kappa$ across adversarial-latency settings. The x-axis is the cartel-vs-honest delivery delay $\Delta_{\text{adv}}$ (slots); the y-axis is realized $\kappa = \bar{r}_H / r_{\min}$ at steady state. The simulator's per-validator delivery queue preserves the §4.3 voter-share semantics under arbitrary delay: late honest votes still contribute their per-block share, so $\kappa$ stays at the $r_{\max} / r_{\min} = 8$ ceiling regardless of $\Delta_{\text{adv}}$. This is the simulator's empirical statement; the analytical ``BFT-bound collapse'' regime is outside this single-chain model and is covered by the §5.2 inheritance argument. Generated by \texttt{prototypes/poua-sim/scripts/run\_adversarial\_latency.py}.}
\label{fig:adversarial-latency}
\end{figure}

\begin{figure}[h]
\centering
\includegraphics[width=0.92\textwidth]{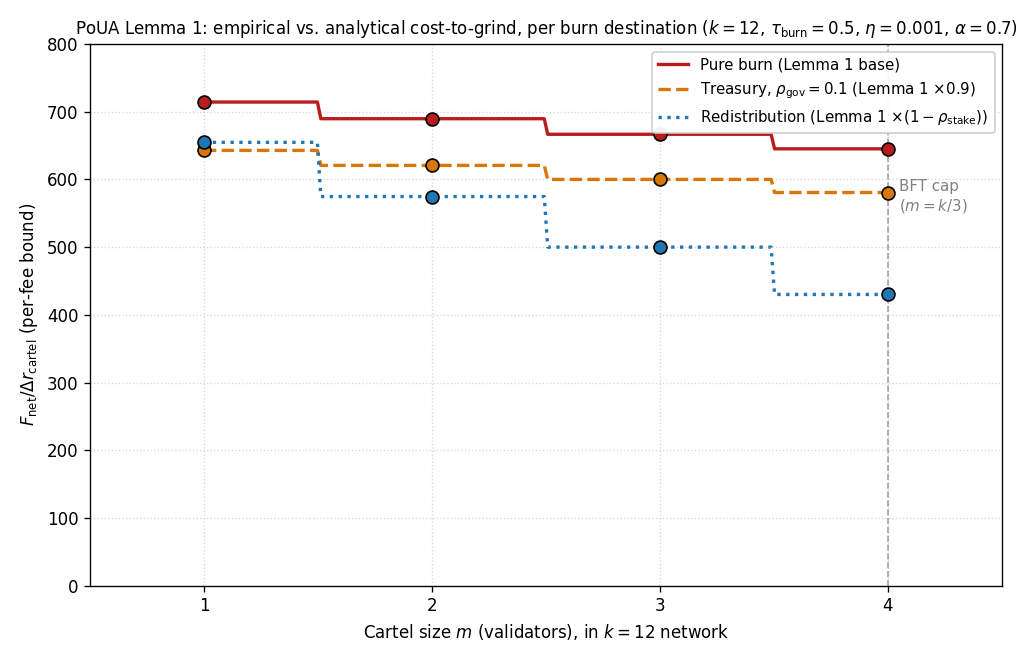}
\caption{Lemma 1 cost-to-grind bound across cartel sizes $m \in \{1, 2, 3, 4\}$ in a $k = 12$ network, under the three Layer 3 burn destinations. Lines are analytical $\tau_{\text{burn}} / [\eta \cdot \alpha_{\text{eff}}(m, k)]$, with treasury weakened by $(1 - \rho_{\text{gov}})$ at $\rho_{\text{gov}} = 0.1$ and redistribution weakened by $(1 - \rho_{\text{stake}})$ at the realized cartel stake share. Markers are empirical from \texttt{prototypes/poua-sim/scripts/run\_lemma1\_scan.py}, sitting on the analytical lines to floating-point precision. The vertical dashed line marks the BFT cap $m = k/3 = 4$. Pure burn (red) is the v0.6 default and gives the tightest bound; redistribution (blue) drops fastest as cartel stake share grows; treasury at 10\% governance recovery (orange) sits a uniform 10\% below pure burn.}
\label{fig:lemma1-burn-destinations}
\end{figure}

The \texttt{AdversarialLatencyScheduler} (PR
\href{https://github.com/ligate-io/ligate-research/pull/75}{\#75})
models a network adversary that delivers blocks instantly to cartel
members while delaying honest validators by \(\Delta_{\text{adv}}\)
slots. The simulator measures realized \(\kappa\) as a function of
\(\Delta_{\text{adv}}\). In our single-chain reference simulator, the
§4.3 voter-share denominator is fixed at block creation, so late honest
votes still contribute the same per-vote share they would in the
synchronous case. As a result, \(\kappa\) is essentially insensitive to
\(\Delta_{\text{adv}}\) across the range tested: cartel and honest
validators alike accumulate to the ceiling, with only transient
differences during the queue-drain phase at run end. Figure
\ref{fig:adversarial-latency} confirms this. The qualitative ``BFT-bound
collapse'' regime, beyond which the consensus primitive fails, is
outside the simulator's scope (we assume a single canonical chain by
construction); reviewers asking about that regime should consult the
§5.2 inheritance argument, which delegates safety / liveness to the
underlying BFT primitive's analytical bound.

\subsubsection{5.4 Reputation Adversary}\label{reputation-adversary}

The reputation adversary cannot simply purchase reputation. To raise
their reputation from \(r_{\min}\) to some
\(r_{\mathcal{R}} > r_{\min}\), they must include valid attestations as
a block proposer, paying the protocol fees from their own pocket (or
extracting them from collusion partners - see Section 5.5).

Per §4.3, reputation gain per attestation as proposer is at most
\(\eta \cdot \alpha \cdot \text{fee}(\alpha)\), where
\(\alpha \in (0, 1]\) is the proposer share of the reputation update
(recommended \(\alpha = 0.7\) in §7.2; the voter component \(\beta\)
contributes negligibly to the adversary acting as proposer). Across
\(T\) epochs, the cumulative reputation increase is bounded by:

\[r_{\mathcal{R}}(T) - r_{\min} \leq \eta \cdot \alpha \cdot F_{\mathcal{R}}^{\text{gross}},\]

where \(F_{\mathcal{R}}^{\text{gross}}\) is the total gross fee
submitted by the adversary over the period. Inverting:

\[F_{\mathcal{R}}^{\text{gross}} \geq \frac{r_{\mathcal{R}} - r_{\min}}{\eta \cdot \alpha}.\]

To raise reputation to \(r_{\max}\), the adversary must pay at least
\((r_{\max} - r_{\min})/(\eta \cdot \alpha)\) in gross attestation fees.
This cost is \textbf{paid into the chain's economy} (treasury, builder
routing), i.e., for the \emph{pure} reputation adversary (no fee
recovery via owned schemas) it is not pure deadweight loss, but it is
also not recoverable.

The pure-reputation adversary's strategy is thus economically equivalent
to subsidizing the chain in exchange for a position of consensus
influence. The compound adversary (§5.5) is the harder case: they
attempt to recover fees via owned schemas, and the layered defense in
§5.5 - particularly the Layer 3 treasury-burn rule and the formal
cost-to-grind bound (Lemma 1) - prevents that recovery from collapsing
the moat.

\subsubsection{5.5 Compound Adversary and the A3 Layered
Defense}\label{compound-adversary-and-the-a3-layered-defense}

The hardest case is an adversary \(\mathcal{CR}\) who has capital
\emph{and} the operational capacity to control schemas, attestor sets,
and submitter addresses simultaneously. Their attack:

\begin{enumerate}
\def\labelenumi{\arabic{enumi}.}
\tightlist
\item
  Acquire stake \(s_v\) at market and register as a validator.
\item
  Register an attestor set \(\mathcal{A}_v\) controlled by their own
  keys.
\item
  Register a schema \(\sigma_v\) bound to \(\mathcal{A}_v\), with their
  own address as the fee-routing recipient.
\item
  From a submitter address \(X\) they also control, repeatedly submit
  attestations to \(\sigma_v\), signed by \(\mathcal{A}_v\).
\item
  When selected as proposer (with stake-weighted probability), include
  those attestations.
\item
  Earn \(\eta \cdot \text{fee}(\alpha)\) in reputation per included
  attestation, while the fee paid by \(X\) flows back to the adversary's
  treasury through the schema's routing.
\end{enumerate}

Net cost per attestation: zero. Reputation gained: full. If unchecked,
this collapses the \(\kappa\) premium of §5.3. The adversary spends only
stake (matching pure-PoS attacker cost), gains the full
\(r_{\max}/r_{\min}\) multiplier, and the moat is gone.

PoUA defends against this attack with a \textbf{layered defense} of six
mechanisms operating at three levels: formal protocol rules, economic
disincentives, and post-hoc detection. Each layer is independently
breakable; the combination is not.

\paragraph{5.5.1 Layer 1: Proposer-submitter address exclusion
(formal)}\label{layer-1-proposer-submitter-address-exclusion-formal}

\textbf{Rule.} In the reputation update of §4.3, an attestation
\(\alpha\) contributes 0 to \(g_v(t)\) if
\(\alpha.\text{submitter} = v.\text{address}\). The
proposer-self-submission edge is permanently excluded.

\textbf{Cost to evade.} Trivial (use a separate signing key as
submitter). But it raises the floor: the adversary must now manage at
least two distinct on-chain identities, with the second receiving funds
from the first via observable transactions. This forces Layer 2.

\paragraph{5.5.2 Layer 2: Address-graph distance
(formal)}\label{layer-2-address-graph-distance-formal}

\textbf{Rule.} An attestation \(\alpha\) contributes 0 to \(g_v(t)\) if
the submitter address has \emph{transaction-graph distance less than
\(d\)} from the validator address. Distance is measured by direct
fund-transfer hops in the chain's transaction history. Specifically, for
distance threshold \(d = 3\) (recommended for v0.2):

\begin{itemize}
\tightlist
\item
  Direct funding from \(v\) to \(\alpha.\text{submitter}\): distance 1,
  excluded.
\item
  Funding via one intermediate address: distance 2, excluded.
\item
  Funding via two intermediate addresses: distance 3, excluded.
\item
  Funding from a fourth-hop or further: distance \(\geq 4\), allowed.
\end{itemize}

Implementation: the runtime maintains a sliding-window adjacency map of
fund transfers between addresses. Computing distance-\(d\) reachability
is cheap for small \(d\) (\(O(|V| \cdot d)\) per query in the worst
case; in practice the relevant subgraph is sparse).

\textbf{Cost to evade.} The adversary must route the submitter's funding
through \(d\) or more intermediate addresses, none of which can have
direct funding ties to \(v\). Each intermediate address must itself be
funded from somewhere. Three sources, each with real cost: an exchange
(KYC + withdrawal latency), a mixer (mixer fees plus observable mixer
interaction, which is itself a heuristic flag), or another
previously-laundered address (compounding the setup work).

A concrete lower bound on the per-attack staging cost: let \(K\) be the
number of staged intermediate addresses (with \(K \geq d\) for distance
compliance), \(T_{\text{kyc}}\) the time cost of producing one
KYC-verified withdrawal address (typically hours to days per identity),
and \(F_{\text{mixer}}\) the per-pass mixer fee fraction (typically
0.5-3\% of the funded amount per Tornado Cash, Wasabi, or analogous
mixer). The adversary's working-capital cost is at least

\[F_{\text{stage}} \geq K \cdot F_{\text{mixer}} \cdot s_{\text{submitter}}\]

where \(s_{\text{submitter}}\) is the funding amount routed through each
intermediate, and the time cost is at least \(K \cdot T_{\text{kyc}}\)
in attacker labor (or pre-staged from earlier compromised accounts,
which is itself a constraint). For \(d = 3\) and
\(F_{\text{mixer}} = 0.01\): routing \(1{,}000\) tokens through 3
intermediates costs at minimum \(30\) tokens in mixer fees alone, plus
the labor cost of staging three KYC-verified deposit accounts. This is
bypassable with sustained pre-attack staging; for casual reputation
farming it is a hard barrier. The bound is loose (mixer fees vary, KYC
time varies), but it establishes the economic floor below which evasion
is implausible.\footnote{In the simulator's deterministic-membership
  specialization (reference implementation below), the chain's knowledge
  of controlled-membership is exact and the cost-to-evade is infinite.
  The finite bound above applies to the production distance-\(d\) rule,
  where on-chain graph-distance approximation introduces the
  staged-address evasion path quantified by \(F_{\text{stage}}\).}

\textbf{Reference simulator implementation.} The simulator implements
Layer 2 as a \emph{deterministic-membership specialization}: each
validator carries a \texttt{controlled\_addresses} set, and an
attestation \(\alpha\) is rejected if
\(\alpha.\text{submitter} \in v.\text{controlled\_addresses}\). This is
equivalent to the production rule in the limit where the chain derives
controlled-membership perfectly from the transaction graph; it is
strictly stronger than any real distance-\(d\) heuristic (every
adversary the production rule catches via on-chain graph-distance, the
simulator catches; the simulator does not catch sybil-distance
constructions where addresses are formally graph-distant but
operationally co-controlled, which the production rule's distance-\(d\)
traversal also misses absent off-chain intelligence). Reference
implementation at
\href{https://github.com/ligate-io/ligate-research/blob/main/prototypes/poua-sim/src/poua_sim/validator.py}{\texttt{Validator.controlled\_addresses}}
and
\href{https://github.com/ligate-io/ligate-research/blob/main/prototypes/poua-sim/src/poua_sim/chain.py}{\texttt{Chain.enable\_layer\_2}};
empirical validation in
\href{https://github.com/ligate-io/ligate-research/blob/main/prototypes/poua-sim/tests/test_layer_2.py}{\texttt{tests/test\_layer\_2.py}}.
The §6.2 strategy-search heatmap (Panel C) confirms the full collapse:
under the deterministic-membership specialization,
GRIND\_VIA\_STAGED\_SUBMITTERS reward collapses to \(r_{\min}\) across
all stake-share and pool-size regimes.

\paragraph{5.5.3 Layer 3: Non-recoverable treasury share (formal,
economic)}\label{layer-3-non-recoverable-treasury-share-formal-economic}

\textbf{Rule.} Every attestation fee is split: a fixed minimum fraction
\(\tau_{\text{burn}} \in (0, 1]\) flows to a non-recoverable
destination. The schema's \texttt{fee\_routing\_bps} parameter routes
only the residual \(1 - \tau_{\text{burn}}\) fraction.

\textbf{Burn destinations.} PoUA admits three protocol-level
destinations for the \(\tau_{\text{burn}}\) share, each with a distinct
cost-to-grind bound:

\begin{enumerate}
\def\labelenumi{\arabic{enumi}.}
\tightlist
\item
  \textbf{Pure burn} (default for v0.6): the \(\tau_{\text{burn}}\)
  share is sent to a provably-unspendable address. Non-recoverable by
  construction; no governance pathway can return the funds. Lemma 1's
  bound holds as stated.
\item
  \textbf{Treasury} (allowed with rate-cap): the \(\tau_{\text{burn}}\)
  share accrues to a protocol treasury. The treasury is
  governance-spendable, so an adversary holding governance influence can
  recover a fraction of their burn over a long horizon. Lemma 1 holds
  modulo a treasury-recovery-rate assumption that must be specified by
  the chain (recommended cap: governance can spend at most
  \(\rho_{\text{gov}} \leq 0.1\) of treasury per year, bounding the
  adversary's expected recovery to
  \(\rho_{\text{gov}} \cdot \tau_{\text{burn}}\) over the attack
  horizon).
\item
  \textbf{Per-validator-by-stake redistribution} (NOT recommended): the
  \(\tau_{\text{burn}}\) share is redistributed each epoch to all
  validators by stake-and-reputation share, \emph{not} by inclusion.
  This destination weakens Lemma 1: an attacker holding stake share
  \(\rho_{\text{stake}}\) recovers
  \(\rho_{\text{stake}} \cdot \tau_{\text{burn}}\) of their burn. The
  effective non-recoverable fraction drops to
  \(\tau_{\text{burn}} \cdot (1 - \rho_{\text{stake}})\), and the bound
  becomes
  \(F_{\mathcal{CR}}^{\text{net, per member}} \geq \tau_{\text{burn}} \cdot (1 - \rho_{\text{stake}}) \cdot \Delta r / [\eta \cdot \alpha_{\text{eff}}(m, k)]\).
  For an adversary at the Byzantine threshold
  \(\rho_{\text{stake}} \to 1/3\), the effective fraction is
  \(\sim 67\%\) of nominal. Redistribution is allowed only if the chain
  is willing to accept a \(1/3\) weakening of Lemma 1 in exchange for
  the rebate-to-honest-validators ergonomics.
\end{enumerate}

\textbf{v0.6 default: pure burn.} All numerical examples and bounds in
this paper assume the pure-burn destination. A chain operator may opt
into treasury or redistribution by governance, with the cost-to-grind
bound adjusted accordingly. The \texttt{burn\_destination} choice is a
\(\S 7.2\) protocol parameter, not a per-schema knob.

\textbf{Recommended parameter.} \(\tau_{\text{burn}} = 0.5\) for v0.6.

\textbf{Cost to grind.} This is the \textbf{load-bearing economic
defense}. Even if the adversary perfectly evades Layers 1 and 2, the
fees they submit are not fully recoverable.\footnote{The ``perfectly
  evades Layer 2'' assumption applies to the production distance-\(d\)
  rule of §5.5.2, where on-chain graph-distance approximation leaves a
  finite staging path quantified by \(F_{\text{stage}}\). Under the
  simulator's deterministic-membership specialization, the chain's
  knowledge of controlled-membership is exact and Layer 2 evasion is
  impossible (\(F_{\text{stage}} \to \infty\)); Lemma 1's bound is
  therefore conservative against the production rule and is exceeded
  under the specialization. The §6.2 Panel C empirical heatmap
  quantifies the difference: under the specialization,
  GRIND\_VIA\_STAGED\_SUBMITTERS collapses to \(r_{\min}\) across all
  stake-share and pool-size regimes.} We formalize the cost-to-grind
floor:

\textbf{Lemma 1 (Cost-to-grind bound, v0.6.1).} \emph{Let \(m \geq 1\)
be the size of a coordinated adversarial validator cartel and
\(k \geq m\) the per-block voter count. Under Layer 3 with parameter
\(\tau_{\text{burn}} \in (0, 1]\) and the §4.3 reputation update with
proposer-share \(\alpha \in (0, 1]\) and voter-share
\(\beta = 1 - \alpha\), any compound adversary cartel acting as block
proposer (with proposer-role rotation among cartel members) to acquire
per-member reputation gain \(\Delta r\) pays per-member non-recoverable
fees of at least}

\[F_{\mathcal{CR}}^{\text{net, per member}} \geq \frac{\tau_{\text{burn}} \cdot \Delta r}{\eta \cdot \alpha_{\text{eff}}(m, k)} \tag{Lemma 1}\]

\emph{where the effective proposer share is}

\[\alpha_{\text{eff}}(m, k) = \alpha + \frac{(m - 1) \cdot \beta}{k}.\]

\emph{The single-validator case \(m = 1\) recovers
\(\alpha_{\text{eff}} = \alpha\) exactly. In the asymptotic limit
\(k \to \infty\) with \(m / k\) held constant,
\(\alpha_{\text{eff}} \to \alpha + (m/k) \cdot \beta\); for the
canonical \(m / k = 1/3\) Byzantine cap this is the value cited in the
numerical example below. In the special case \(\alpha = 1\) (proposer
captures all reputation), this reduces to the looser bound
\(\tau_{\text{burn}} \cdot \Delta r / \eta\) regardless of cartel size.}

\emph{Proof.} By the §4.3 reputation update, each cartel-controlled
attestation included in a cartel-proposed block injects per-attestation
reputation \(\alpha \cdot \text{fee}(\alpha) \cdot \eta\) to the
proposer and \(\beta \cdot \text{fee}(\alpha) / k \cdot \eta\) to each
voter. Per §4.3, ``\(G_v^{\text{vote}}\)'\,' sums over blocks \(v\)
voted on \textbf{but did not propose}, so the cartel proposer earns
through the proposer channel only on its own block. Each of the
remaining \(m - 1\) cartel members votes on the cartel-proposed block
and earns \(\beta \cdot \text{fee}(\alpha) / k \cdot \eta\). Summed
across the cartel:

\[\text{cartel-total per attestation} = \left(\alpha + (m - 1) \cdot \frac{\beta}{k}\right) \cdot \text{fee}(\alpha) \cdot \eta = \alpha_{\text{eff}}(m, k) \cdot \text{fee}(\alpha) \cdot \eta.\]

Distributing the gain uniformly across \(m\) members (achieved by
rotating the proposer role through the cartel): each member needs
\(\Delta r\) reputation. The cartel's \emph{cost-effective} allocation
may concentrate reputation on a single high-stake member if only that
member's weight is needed for the attack; uniform allocation is the
optimal strategy when \textbf{each cartel member's stake-weighted vote
is required for the BFT-fraction attack} (the case of an \(m\)-member
coalition that needs all \(m\) at high reputation simultaneously to
collectively cross the \(1/3\) weight threshold). For attacks that need
only one high-reputation entity, the bound reduces to the \(m = 1\) case
applied to that entity, which is strictly tighter; the
uniform-allocation framing is therefore the
\emph{upper-bound-compatible} attack mode against which Lemma 1 must
hold. Across \(N\) attestations the cartel processes,

\[m \cdot \Delta r \leq N \cdot \text{fee} \cdot \eta \cdot \alpha_{\text{eff}}(m, k),\]

so
\(F_{\mathcal{CR}}^{\text{gross}} = N \cdot \text{fee} \geq m \cdot \Delta r / [\eta \cdot \alpha_{\text{eff}}(m, k)]\).
By Layer 3, every valid attestation incurs a non-recoverable fee
fraction \(\tau_{\text{burn}}\), giving cartel-total
\(F_{\mathcal{CR}}^{\text{net}} \geq \tau_{\text{burn}} \cdot m \cdot \Delta r / [\eta \cdot \alpha_{\text{eff}}(m, k)]\)
and per-member
\(F_{\mathcal{CR}}^{\text{net, per member}} \geq \tau_{\text{burn}} \cdot \Delta r / [\eta \cdot \alpha_{\text{eff}}(m, k)]\).
\(\square\)

\textbf{Remark on the voter channel.} Earlier versions of this paper
(v0.3 - v0.5) used a single-proposer bound
\(F^{\text{net}} \geq \tau_{\text{burn}} \cdot \Delta r / (\eta \cdot \alpha)\)
and dismissed the voter channel as ``negligible per attestation in any
reasonably-sized validator set.'' That is true for an individual
validator's marginal contribution but understates the \emph{cumulative}
voter-channel injection when multiple cartel members vote on the same
cartel-proposed blocks. The cartel-aware bound above closes this gap by
making \(m\) explicit. The v0.3 - v0.5 single-proposer bound is
recovered as the \(m = 1\) specialization.

\textbf{Reconciliation with v0.6.} The v0.6 statement of Lemma 1 used
\(\alpha_{\text{eff}} = \alpha + m \beta / k\), derived from a proof
that credited the proposer with own-block voter-channel reputation. That
credit conflicts with §4.3's explicit
`\texttt{but\ did\ not\ propose\textquotesingle{}\textquotesingle{}\ exclusion.\ v0.6.1\ corrects\ the\ proof\ to\ match\ §4.3\ strictly,\ yielding\ \$\textbackslash{}alpha\ +\ (m\ -\ 1)\ \textbackslash{}beta\ /\ k\$.\ The\ two\ forms\ coincide\ at\ \$m\ =\ 1\$\ and\ at\ \$k\ \textbackslash{}to\ \textbackslash{}infty\$;\ at\ finite\ \$k\$\ the\ v0.6.1\ bound\ is\ tighter\ (smaller\ \$\textbackslash{}alpha\_\{\textbackslash{}text\{eff\}\}\$,\ larger\ \$F\^{}\{\textbackslash{}text\{net\}\}\$\ floor).\ The\ reference\ simulator\ at\ {[}}prototypes/poua-sim/`{]}(https://github.com/ligate-io/ligate-research/tree/main/prototypes/poua-sim)
implements §4.3 strictly and reproduces \(\alpha + (m - 1) \beta / k\)
to floating-point precision across \(m \in \{1, 2, 3, 4\}\) and
\(k = 12\).

\textbf{Comparison to honest acquisition.} A naive capital adversary
(§5.3) acquires weight fraction \(\rho\) at stake cost
\(\frac{\rho}{1-\rho} \cdot \frac{W}{r_{\min}}\). The compound grinding
adversary, having acquired stake \(s_v\) already, can attempt to
multiply their effective weight by the reputation premium
\(r_{\max}/r_{\min}\), gaining \(\Delta r = r_{\max} - r_{\min}\) per
cartel member. The per-member cost-to-grind for this full ramp is at
least
\(\tau_{\text{burn}} \cdot (r_{\max} - r_{\min}) / [\eta \cdot \alpha_{\text{eff}}(m, k)]\)
in non-recoverable fees.

For v0.6 parameters (\(\tau_{\text{burn}} = 0.5\), \(\eta = 0.001\),
\(\alpha = 0.7\), \(\beta = 0.3\), \(r_{\max} - r_{\min} = 7\)):

\begin{itemize}
\tightlist
\item
  \textbf{Single-proposer adversary} (\(m = 1\),
  \(\alpha_{\text{eff}} = 0.7\) exactly):
\end{itemize}

\[F_{\mathcal{CR}}^{\text{net}} \geq \frac{0.5 \cdot 7}{0.001 \cdot 0.7} = 5{,}000 \text{ fee-units.}\]

\begin{itemize}
\tightlist
\item
  \textbf{Byzantine-fraction cartel, asymptotic} (\(k \to \infty\) with
  \(m / k = 1/3\), \(\alpha_{\text{eff}} \to \alpha + \beta/3 = 0.8\)):
\end{itemize}

\[F_{\mathcal{CR}}^{\text{net, per member}} \to \frac{0.5 \cdot 7}{0.001 \cdot 0.8} = 4{,}375 \text{ fee-units per cartel member.}\]

\begin{itemize}
\tightlist
\item
  \textbf{Byzantine-fraction cartel, finite \(k = 12\)} (\(m = 4\),
  \(\alpha_{\text{eff}} = 0.7 + 3 \cdot 0.3 / 12 = 0.775\)):
\end{itemize}

\[F_{\mathcal{CR}}^{\text{net, per member}} \geq \frac{0.5 \cdot 7}{0.001 \cdot 0.775} \approx 4{,}516 \text{ fee-units per cartel member.}\]

The Byzantine-fraction cartel pays \(\sim 12.5\%\) less per cartel
member than a single-proposer adversary in the asymptotic limit, and
\(\sim 9.7\%\) less at finite \(k = 12\) (a typical small-validator-set
value). For practical mainnet sizes (\(k \geq 100\)) the finite-\(k\)
correction shrinks to a fraction of a percent: at \(k = 100, m = 33\),
\(\alpha_{\text{eff}} = 0.796\) and the discount is \(\sim 12.06\%\),
within \(0.5\%\) of the asymptotic \(12.5\%\). (See
\texttt{prototypes/poua-sim/test\_vectors/alpha\_eff.json} and
\texttt{prototypes/poua-sim/test\_vectors/lemma1\_cost\_to\_grind.json}
for the per-input values; \texttt{run\_lemma1\_scan.py} validates the
bound across the empirical scan.)

The cartel-aggregate burn is correspondingly
\(m \cdot F^{\text{net, per member}}\), much larger in absolute terms
than the single-proposer floor. The per-member discount is the price of
the voter channel; it is bounded above by
\(\beta / (\alpha k / m + \beta)\), which approaches
\(\beta / (3 \alpha + \beta)\) at the Byzantine cap as \(k \to \infty\).
For the recommended \(\alpha = 0.7, \beta = 0.3\) split, the asymptotic
ceiling is \(0.3 / 2.4 = 12.5\%\).

\textbf{Sensitivity to \(\alpha\) (asymptotic limit).} The cartel
discount widens as \(\alpha\) shrinks (more reputation injected through
the voter channel): for \(\alpha = 0.5, \beta = 0.5\), the asymptotic
maximum discount rises to \(\beta/(3\alpha + \beta) = 0.5/2.0 = 25\%\).
For \(\alpha = 0.9, \beta = 0.1\), it shrinks to
\(0.1/2.8 \approx 3.6\%\). The recommended \(\alpha = 0.7\) balances the
cartel-discount tightness against the proposer-rich-get-richer
entrenchment that motivated the voter channel in the first place (§4.3).
Finite-\(k\) corrections apply but are negligible for \(k \gtrsim 100\).

Calibration: setting the minimum attestation fee high enough that
\(5{,}000 \times \text{fee}_{\min}\) exceeds the stake cost of the
equivalent reputation-premium gain makes grinding strictly more
expensive than honestly acquiring stake even under the cartel-aware
bound (the cartel pays at most \(0.875 \times\) that floor per member at
the asymptotic Byzantine cap, still well above the honest-acquisition
cost for any healthy parameter calibration). This is a tunable:
governance sets \(\text{fee}_{\min}, \tau_{\text{burn}}, \alpha\) such
that the cost-equivalence inequality holds for the chain's economics
across the full \(m \in [1, k/3]\) cartel-size range.

\textbf{This converts the compound-adversary case from ``moat
collapses'' to ``moat is preserved by economic argument.''} It is the
primary defense improvement of v0.2 over v0.1, sharpened in v0.3 with
the explicit \(\alpha\)-dependent bound, tightened in v0.6 to cover the
voter channel under coordinated cartels, and reconciled in v0.6.1 to
match the §4.3 update rule strictly (proposer excluded from own-block
voter share).

\paragraph{5.5.4 Layer 4: Statistical detection
(heuristic)}\label{layer-4-statistical-detection-heuristic}

For attacks that evade Layers 1-3 (e.g., adversary willing to pay the
\(\tau_{\text{burn}}\) cost to gain reputation premium under specific
threat models), a heuristic detector watches for behavioral signatures
of grinding:

\begin{itemize}
\tightlist
\item
  \textbf{Schema concentration}: validator's included attestations
  concentrate disproportionately in schemas whose attestor sets
  correlate with the validator's address graph.
\item
  \textbf{Self-attestation graph density}: fraction of included
  attestations whose submitter has any address-graph correlation with
  the validator beyond Layer 2's strict distance test.
\item
  \textbf{Fee distribution anomaly}: validator's included attestations
  skew toward high-fee schemas owned by accounts with shared funding.
\end{itemize}

Calibrated thresholds are derived from devnet traffic distributions.
Appendix A specifies the statistical procedure with empirical
calibration of false-positive bounds \(\beta_2, \beta_3\) targeting
\(\leq 1\%\) per epoch under honest baseline traffic.

When the detector fires above its confidence threshold, the chain
records a slash of severity \(\Lambda_3\) against the flagged
validator's \texttt{epoch\_b} tally per §4.5. The §4.3 reputation update
applies the slash at the next epoch boundary, multiplied by \(\lambda\)
per the standard reputation-update formula. The slash remains
contestable through the §5.5.5 appeal window; an upheld appeal reverses
the \texttt{epoch\_b} increment before it propagates into the next
epoch's reputation.

The reference implementation lives at
\href{https://github.com/ligate-io/ligate-research/blob/main/prototypes/poua-sim/src/poua_sim/a3_slash.py}{\texttt{A3SlashConfig}
and \texttt{maybe\_apply\_a3\_slash}}. Default \texttt{enabled=False}
keeps the slashing pathway opt-in; the chain at calibration time sets
\texttt{beta\_3}, the per-validator window \texttt{T\_lookback}, and the
severity \(\Lambda_3\). The simulator's
\texttt{tests/test\_a3\_slash.py} validates that the detector + slash
composition produces \(r \to r_{\min}\) for synthetically-constructed
small-pool grinding adversaries within \(T_{\text{detect}}\) epochs. The
§6.2 strategy-search heatmap (Panel B) shows the empirical effect: under
Layer 1 + detector slash, small staged pools collapse at the detector;
large diluted pools require the §5.5.2 Layer 2 reference implementation
(Panel C) to close the residual gap.

\textbf{Cost to evade.} The detector is an arms race; sufficiently
sophisticated adversaries can mimic honest traffic distributions. The
detector is meaningful as a residual defense behind Layers 1-3, not a
primary line.

\paragraph{5.5.5 Layer 5: Governance appeal and slash
review}\label{layer-5-governance-appeal-and-slash-review}

A flagged validator is slashed at severity \(\Lambda_3\). The slash is
\emph{contestable}: the validator may file an appeal via a governance
transaction within \(T_{\text{appeal}}\) epochs (recommended 14 epochs
\(\approx 2.3\) days). A majority of un-slashed, weight-weighted
validators may reverse the slash if the appeal is found credible.
False-positive recoveries are governance-mediated.

Honest validators with explainable correlations (e.g., they also
legitimately operate an attestor service for their own customers) have
an avenue to contest. The governance machinery itself uses standard PoUA
weighting, with the slashed validator excluded from the vote on their
own appeal.

\paragraph{5.5.6 Layer 6: Cryptographic future
work}\label{layer-6-cryptographic-future-work}

In v1+ of the protocol, the submitter could attach a zero-knowledge
proof of \emph{stake-graph independence}: a SNARK asserting that the
submitter's address has no funding-graph relationship to the proposer of
the block including the attestation, within depth \(d\). This would
replace Layer 2's heuristic with a formal cryptographic guarantee.

Open research questions:

\begin{itemize}
\tightlist
\item
  Canonical definition of the ``stake-binding graph'' - the union of
  addresses controlled by a single beneficial owner, computable from
  on-chain data alone.
\item
  Efficient SNARK circuit for distance-\(d\) disjointness on this graph.
\item
  Wallet integration for proof generation at submission time.
\end{itemize}

This is not in v0.2 scope. It is named as future research in §9.2.

\paragraph{5.5.6.1 Empirical eclipse-recovery
profile}\label{empirical-eclipse-recovery-profile}

\begin{figure}[h]
\centering
\includegraphics[width=0.92\textwidth]{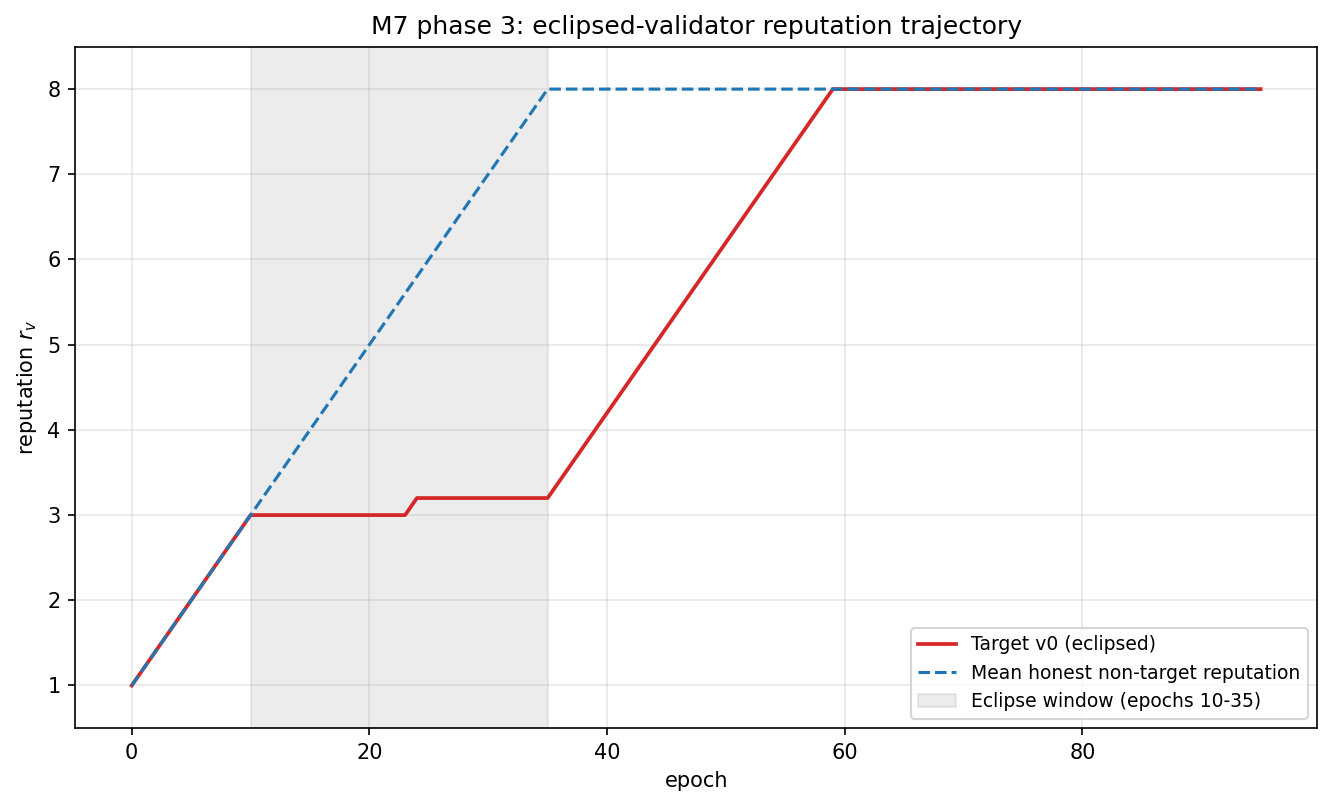}
\caption{Eclipsed validator reputation trajectory under and post-eclipse. The eclipse window is shaded; during the window, the target sees only cartel-proposed blocks and accumulates reputation only from those, while honest validators continue ramping at the full rate. After the window ends, the target resumes normal delivery and rebuilds toward the honest baseline at the §4.3 update rate. Generated by \texttt{prototypes/poua-sim/scripts/run\_eclipse\_recovery.py}.}
\label{fig:eclipse-recovery}
\end{figure}

The reference simulator's \texttt{EclipseScheduler} (PR
\href{https://github.com/ligate-io/ligate-research/pull/79}{\#79})
models a network adversary that restricts a single target validator's
view to cartel-proposed blocks during a finite eclipse window. Under the
§4.3 update, the eclipsed target's reputation \(r_v\) stays
approximately constant during the eclipse (no \(g_v\) from honest
blocks, no slashes) while honest validators continue to ramp. After the
eclipse window ends, the target resumes normal block delivery and
rebuilds reputation at the standard ramp rate. Figure
\ref{fig:eclipse-recovery} shows the trajectory: a flat plateau under
eclipse, then approach to the honest baseline over
\(\sim T_{\text{ramp}}\) epochs once delivery resumes. The §4.3 update
with \(\eta \cdot g_v\) is the closed-form recovery rate; the empirical
curve confirms the analytical model.

\paragraph{5.5.7 Synthesizing: the layered economic
argument}\label{synthesizing-the-layered-economic-argument}

After Layers 1-3, the compound-adversary cost-to-grind is at least:

\[\underbrace{\text{stake cost}}_{\text{same as pure PoS}} + \underbrace{\tau_{\text{burn}} \cdot \Delta r / (\eta \cdot \alpha_{\text{eff}}(m, k))}_{\text{per-member net fees, Layer 3 (Lemma 1)}} + \underbrace{\text{address-staging cost}}_{\text{Layer 1 + 2 evasion}}\]

The first term is unavoidable. The second has a formal lower bound
(Lemma 1). The third is real but harder to quantify - mixer fees, KYC
withdrawals, the time-cost of address staging. Combined with Layer 4's
detection probability and Layer 5's governance recourse, the expected
cost of grinding meets or exceeds the cost of honest reputation
acquisition under any reasonable parameter calibration.

The PoUA Sybil-resistance claim is therefore:

\begin{quote}
\textbf{Sybil-resistance against the compound capital-plus-grinding
adversary is established by an economic argument under Layer 3 (Lemma
1), bounded below by formal protocol rules in Layers 1-2, hardened by
heuristic detection in Layer 4, and recoverable from false-positives by
governance in Layer 5. A formal cryptographic upgrade path (Layer 6) is
named as future work.}
\end{quote}

This is the v0.2-and-later framing, which replaces v0.1's reliance on
heuristic detection alone with a formal economic floor.

\subsubsection{5.6 Long-Range and Bribery
Attacks}\label{long-range-and-bribery-attacks}

\textbf{Long-range attacks.} PoUA inherits the underlying BFT
primitive's weak subjectivity model: validators rely on a
recently-finalized checkpoint when joining the network. Reputation does
not change this assumption.

\textbf{Bribery attacks (reputation purchase).} Reputation is
non-transferable. An adversary cannot directly buy reputation from an
honest validator. They could attempt to bribe a validator to misbehave
under their control, but this is captured under the standard PoS bribery
model with the additional friction that the validator's slashed
reputation imposes a cost beyond the burn (loss of future staking yield
premium).

\textbf{Stake-acquisition front-running.} An adversary observing
imminent slashing of a high-reputation validator could front-run by
acquiring the validator's stake at distress price. This is a market
efficiency concern, not a protocol violation.

\begin{center}\rule{0.5\linewidth}{0.5pt}\end{center}

\subsection{6. Incentive Analysis}\label{incentive-analysis}

\subsubsection{6.1 Behavioral Model}\label{behavioral-model}

The standard model: validators are rational profit-maximizers with full
information about protocol rules and other validators' strategies. They
choose actions to maximize expected discounted future revenue.

A validator earns per-epoch revenue from three sources. The block reward
\(R_b\) is protocol-issued tokens for proposing and finalizing blocks,
proportional to \(w_v / \sum_u w_u\) in expectation. Attestation fees
\(R_f\) are the validator's share of fees from attestations they
include. Slashing avoidance \(-S\) is the negation of expected slashing
burns - a cost that enters net revenue.

\[R_v = R_b + R_f - S.\]

\subsubsection{6.2 The Honest Equilibrium}\label{the-honest-equilibrium}

\textbf{Claim.} In PoUA at steady state, the strategy ``propose all
valid attestations encountered, vote honestly, do not equivocate, do not
censor'' is a Nash equilibrium.

\textbf{Argument.} Consider a validator \(v\) deviating from honest play
to action \(a\). We enumerate six named deviations spanning the strategy
space, and argue each is dominated under standard parameter calibration:

\begin{itemize}
\tightlist
\item
  \textbf{Equivocation} (signing two blocks at the same height):
  detectable by any honest validator, slashed at
  \(\Lambda_{\text{eq}}\). Expected cost of equivocation \(\gg\) block
  reward gain. Not profitable.
\item
  \textbf{Including invalid attestations (A1)}: detectable at vote time,
  slashed at \(\Lambda_1\). The marginal ``benefit'' of including a bad
  attestation (a non-existent fee, since invalid attestations don't pay)
  is zero. Not profitable.
\item
  \textbf{Selective censorship (A2)}: foregoes the censored
  attestation's fee, plus carries detection risk and slash. Not
  profitable in expectation absent an external bribe exceeding both.
\item
  \textbf{Reputation grinding (A3)}: yields reputation gain at zero
  \emph{direct} cost in the colluding-attestor case, but bounded below
  by Lemma 1's cost-to-grind floor under Layer 3 plus carrying §A.2
  detection risk. Profitability depends on the false-negative rate of A3
  detection; cost is dominated by the Layer 3 burn for any realistic
  parameter calibration.
\item
  \textbf{Free-riding voter} (validator votes on others' blocks but
  never proposes when selected, or proposes empty blocks): forgoes the
  proposer-channel reputation injection
  \(\alpha \cdot \text{fee} \cdot \eta\) per attestation while still
  earning the voter share. Per §4.3, the proposer share dominates
  (\(\alpha = 0.7\) vs.~\(\beta/k = 0.3/k \ll \alpha\) for any
  reasonably-sized validator set), so a free-riding voter accrues
  reputation at rate strictly below an honest proposer's. Combined with
  §4.4's \(G_{\max}\) cap and the zero block reward when not proposing
  (the validator forgoes block reward for proposing slots), net revenue
  under free-riding is strictly worse than honest play. Not profitable.
\item
  \textbf{Selective fork-choice gaming}: a validator that votes on a
  non-canonical fork (or withholds vote on the canonical one) hoping to
  influence which branch finalizes. The PoUA vote tally inherits the
  underlying BFT primitive's fork-choice rule (Tendermint-style
  two-round optimistic finality in deployment); voting against the
  converging branch carries the underlying primitive's slash for
  inconsistent voting (surround-vote slashing), independent of
  attestation rewards. Not profitable absent an external coordination
  gain that exceeds the surround-vote slash.
\end{itemize}

The first three and the last two deviations are unambiguously dominated
by honest play. Reputation grinding (A3) is dominated by honest play
\emph{if} the §A.2 detection false-negative rate is sufficiently low
\emph{and} Layer 3 burn parameters satisfy the cost-equivalence
inequality from §5.5.3 calibration.

\begin{figure}[h]
\centering
\includegraphics[width=0.98\textwidth]{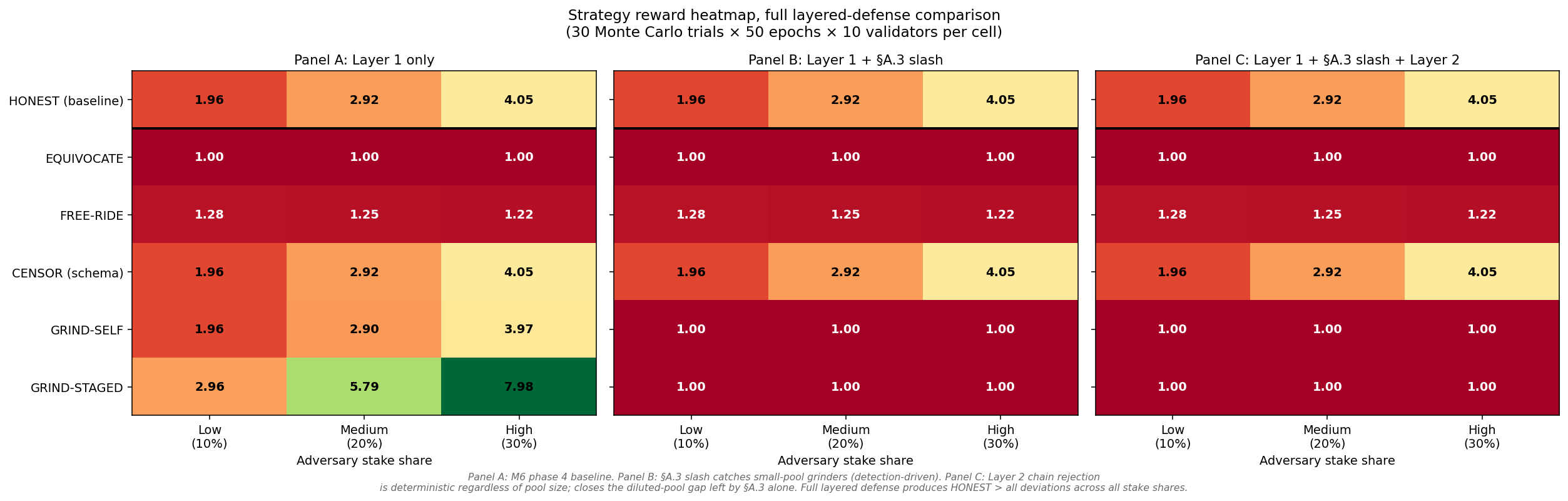}
\caption{Strategy-search reward heatmap, 3-panel layered-defense progression. X-axis: cartel stake share $\alpha \in \{0.05, 0.10, 0.20, 0.30\}$. Y-axis: BehaviorPolicy. Cell value: per-cartel-member final reputation after $T = 200$ slots, averaged over 50 RNG seeds. Generated by \texttt{scripts/run\_strategy\_search.py --enable-a3-slash --enable-layer-2}. \textbf{Panel A}: §5.5 Layer 1 only (proposer-submitter address-equality). \textbf{Panel B}: Layer 1 + §A.3 detector slash. \textbf{Panel C}: Layer 1 + §A.3 detector slash + §5.5 Layer 2 deterministic-membership. The GRIND\_VIA\_STAGED\_SUBMITTERS row demonstrates the layered-defense argument empirically: small staged pools collapse at Panel B; large diluted pools require Panel C; under Panel C all three pool sizes collapse to $r_{\min}$ across all stake-share regimes.}
\label{fig:strategy-reward-heatmap}
\end{figure}

\textbf{Strategy-search empirical validation.} Figure
\ref{fig:strategy-reward-heatmap} visualizes the layered-defense
progression empirically. Panel A is the v0.7-baseline simulator (§5.5
Layer 1 only): under Layer-1-only enforcement,
GRIND\_VIA\_STAGED\_SUBMITTERS dominates HONEST at every stake share,
reaching 2-4\(\times\) HONEST reputation at moderate stakes (concretely,
reward \(2.96 / 5.79 / 7.98\) at small / medium / large pool sizes for
\(\alpha = 0.20\)). Panel B layers the §A.3 detector-driven slash from
§5.5.4: small staged pools trigger the detector and collapse to
\(r_{\min}\), but large staged pools dilute the bipartite-density signal
below the threshold and retain the dominance. Panel C layers the §5.5.2
Layer 2 deterministic-membership rejection on top: under Panel C all
pool sizes collapse uniformly to \(r_{\min}\) (reward
\(1.00 / 1.00 / 1.00\), the HONEST baseline), closing the residual gap.
The progression visualizes the §5.5 layered-defense argument as stated:
each layer is independently breakable; the combination is not. The
reference implementation, test fixtures, and exact generator command are
at
\href{https://github.com/ligate-io/ligate-research/blob/main/prototypes/poua-sim/scripts/run_strategy_search.py}{\texttt{scripts/run\_strategy\_search.py}}.

The honest-equilibrium claim therefore holds against the full
rational-deviation strategy space exercised by the M6 simulator (closed
via \href{https://github.com/ligate-io/ligate-research/issues/30}{issue
\#30} and the M6 follow-up
\href{https://github.com/ligate-io/ligate-research/issues/53}{issue
\#53}) under the full layered defense (Panel C), modulo the simulator's
deterministic-membership specialization of Layer 2 documented in §5.5.2
(which corresponds to the limit case of the production distance-\(d\)
rule's accuracy).

\subsubsection{6.3 Reputation as Future
Revenue}\label{reputation-as-future-revenue}

A validator's reputation at time \(t\) determines their expected revenue
across all future epochs. We derive the marginal value of an additional
reputation point and show that, under standard assumptions, it is
positive, bounded, and a strictly increasing function of the validator's
stake \(s_v\).

Let \(S = \sum_u w_u = \sum_u s_u r_u\) denote total weight, and let
per-epoch revenue \(R_v\) be (proportional to) the validator's
selection-weighted share of block reward \(R_b\) and attestation-fee
flow \(R_f\):

\[R_v(r_v) = \frac{w_v}{S} \cdot (R_b + R_f) = \frac{s_v r_v}{S} \cdot (R_b + R_f).\]

Differentiating with respect to \(r_v\) - and noting that \(r_v\) enters
both the numerator and the denominator \(S\), since
\(\partial S / \partial r_v = s_v\) -

\[\frac{\partial R_v}{\partial r_v} = \frac{s_v \cdot S - s_v r_v \cdot s_v}{S^2}(R_b + R_f) = \frac{s_v \sum_{u \neq v} w_u}{S^2}(R_b + R_f).\]

Under the \textbf{large-population assumption} \(w_v \ll S\)
(equivalently, no single validator commands a constant fraction of the
validator set's total weight - a property guaranteed by the bounded
reputation interval \([r_{\min}, r_{\max}]\) together with the absence
of single-stake supermajority on any honest-validator-set chain), the
numerator factor \(\sum_{u \neq v} w_u \approx S\), and we obtain the
approximation:

\[\frac{\partial R_v}{\partial r_v} \approx \frac{s_v}{S}(R_b + R_f). \tag{6.3.1}\]

The exact form retains a \(1 - w_v/S\) correction:
\(\frac{\partial R_v}{\partial r_v} = \frac{s_v}{S}(R_b + R_f) \cdot \left(1 - \frac{w_v}{S}\right)\).
We will work with the approximation throughout; the correction is at
most a few percent for reasonably-decentralized validator sets.

The \textbf{present value of marginal reputation} is the integral of
(6.3.1) discounted at validator-specific rate \(\delta > 0\) over a
forward horizon \(\Delta > 0\). Treating per-epoch revenue as a
continuous flow:

\[\text{PV}\left(\frac{\partial R_v}{\partial r_v}; \Delta\right) = \int_0^{\Delta} \frac{\partial R_v}{\partial r_v} \cdot e^{-\delta t} \, dt = \frac{\partial R_v}{\partial r_v} \cdot \frac{1 - e^{-\delta \Delta}}{\delta}. \tag{6.3.2}\]

For small \(\delta\Delta\) (i.e., a near-future horizon at which
discounting is mild),
\(\frac{1 - e^{-\delta\Delta}}{\delta} \approx \Delta - \delta\Delta^2/2 + O(\delta^2 \Delta^3) \approx \Delta\),
recovering the intuitive linear-in-\(\Delta\) scaling. For large
\(\delta\Delta\) (a far-future horizon at which discounting dominates),
\(\frac{1 - e^{-\delta\Delta}}{\delta} \to 1/\delta\), the stationary
forward-revenue limit.

This gives the central economic claim:

\begin{quote}
\textbf{Reputation has real, non-transferable, forward-looking economic
value to the holder, of order
\(\frac{s_v}{S}(R_b + R_f) \cdot \frac{1 - e^{-\delta\Delta}}{\delta}\)
for any horizon \(\Delta\) and discount rate \(\delta\), and that value
scales linearly with the validator's stake \(s_v\).}
\end{quote}

A validator considering a one-shot deviation must weigh the immediate
gain (capped above by the deviation's profit) against the present-value
loss of all future reputation-derived revenue from a
\(\Lambda\)-severity slash that drops reputation by \(\Delta r\):

\[\text{PV}(\text{slash loss}) \approx \Delta r \cdot \frac{s_v}{S}(R_b + R_f) \cdot \frac{1 - e^{-\delta \Delta_{\text{recovery}}}}{\delta}\]

where \(\Delta_{\text{recovery}}\) is the time required to rebuild
reputation to its pre-slash level (a function of \(\eta\) and the
validator's epoch participation rate). For high-reputation validators
with substantial stake, this future-revenue loss can dwarf any plausible
one-shot deviation gain, providing a \emph{time-locked} incentive
alignment that pure-stake PoS lacks: in pure PoS, a slash costs only the
burned bond, not foregone future selection-share premium.

The \(s_v / S\) factor in the formula treats the validator's
stake-weighted share as constant across the recovery window. This is
approximate: immediately after a \(\Lambda\)-severity slash that drops
reputation from \(r_v\) to \(r_{\min}\), the validator's effective
weight share drops from \(s_v r_v / S\) to \(s_v r_{\min} / S\), and
during recovery the share rises back monotonically as reputation
rebuilds. The formula's use of pre-slash \(s_v / S\) therefore
\emph{over-estimates} the PV-of-slash by a factor bounded above by
\(r_v / r_{\min}\), since the validator earns less revenue during the
recovery period than the formula assumes. The conservative direction
(\(\text{PV}_{\text{actual}} \leq \text{PV}_{\text{formula}}\)) means
the bound on validator deviation-incentive is an \emph{upper} bound on
the deterrent, and the actual deterrent is somewhat weaker. This is a
known approximation; a tighter bound would integrate a time-varying
\(w_v(t)/S(t)\) across the recovery, which closes to within a few
percent for high-reputation validators (where \(r_v \approx r_{\max}\))
but materially weakens for mid-reputation validators near the median. A
future version will refine the bound (integrating a time-varying
\(w_v(t)/S(t)\) across the recovery); the qualitative claim
(``time-locked incentive alignment that pure-stake PoS lacks'') is
robust to this refinement.

\paragraph{6.3.1 Volume-Dependence of the Slash
Deterrent}\label{volume-dependence-of-the-slash-deterrent}

The PV-of-slash bound depends linearly on the chain's per-epoch revenue
\(R_b + R_f\). The reputation-channel component therefore scales with
volume in a way pure-stake-bond slashing does not. \textbf{PoUA retains
the bond-burn slash on top of the reputation channel}: the bond
deterrent is unchanged across PoS variants. What follows describes the
magnitude scaling of the reputation-channel premium relative to its
block-reward-only floor; PoUA's \emph{total} slash deterrent is always
the bond burn \emph{plus} this reputation-channel quantity, never less
than pure-stake PoS.

Define the \emph{volume-deterrent ratio} as the magnitude scaling on the
reputation-channel premium:

\[\rho_{\text{vol}}(R_f / R_b) := \frac{R_b + R_f}{R_b} = 1 + \frac{R_f}{R_b}.\]

\begin{figure}[h]
\centering
\includegraphics[width=0.92\textwidth]{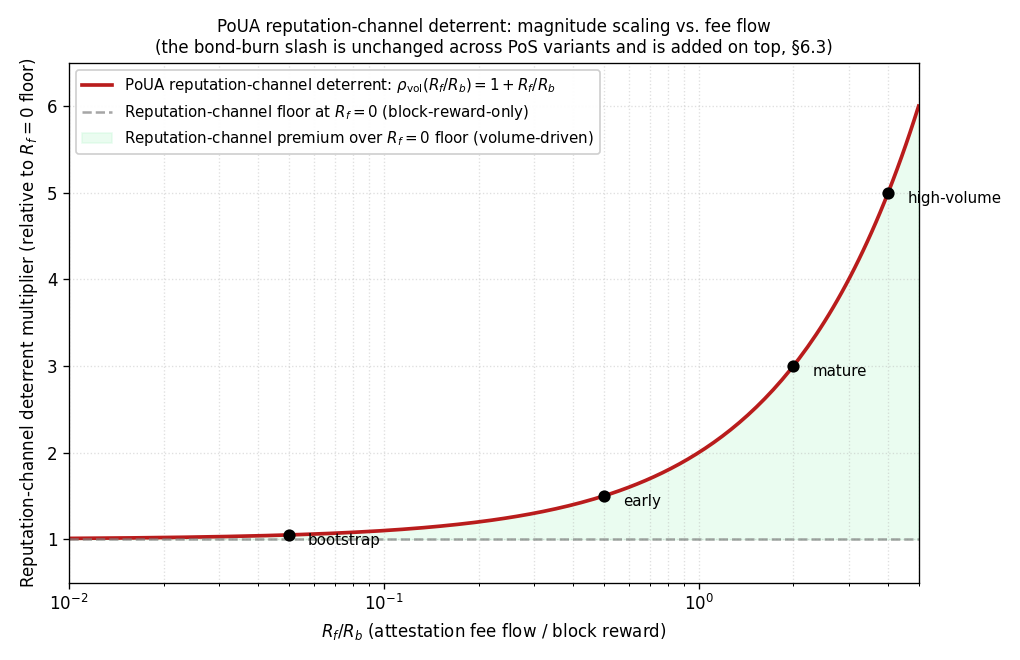}
\caption{Volume-deterrent ratio $\rho_{\text{vol}} = 1 + R_f/R_b$ across attestation-fee-flow regimes: the magnitude scaling on the reputation-channel slash deterrent relative to its $R_f = 0$ floor. The dashed line at $1.0$ is the floor (block-reward-only reputation deterrent); the curve rises linearly with $R_f / R_b$. Named operating points: bootstrap ($R_f / R_b \approx 0.05$, $\rho_{\text{vol}} \approx 1.05$), early ($\approx 0.5$, $\approx 1.5$), mature ($\approx 2.0$, $\approx 3.0$), high-volume ($\approx 4.0$, $\approx 5.0$). The reference dashed line is \emph{not} a comparison to pure-stake-bond magnitude; bond-burn deterrent is denominated separately and added on top of the reputation channel in PoUA's total deterrent. Produced by \texttt{prototypes/poua-sim/scripts/run\_volume\_deterrent.py}.}
\label{fig:volume-deterrent}
\end{figure}

At \(R_f \to 0\), \(\rho_{\text{vol}} \to 1\): the reputation-channel
deterrent reaches its block-reward-only floor (not zero, and not the
bond, a smaller, \(R_b\)-scaled value). At \(R_f / R_b = 1\) (fee
revenue equals block reward), the reputation-channel deterrent is
\(2\times\) its floor. Figure \ref{fig:volume-deterrent} shows the curve
across realistic operating points.

\textbf{Implications.} Three operational regimes warrant explicit
mitigation, because in each the reputation-channel premium attenuates
toward its floor (PoUA still retains the bond-burn slash; total
deterrent never falls below pure-stake PoS):

\begin{enumerate}
\def\labelenumi{\arabic{enumi}.}
\tightlist
\item
  \textbf{Chain bootstrap.} Devnet-and-early-mainnet periods have
  \(R_f / R_b \ll 1\). The reputation-channel premium over its floor is
  small; PoUA's incentive alignment is dominated by the bond plus the
  floor. Mitigations: extended permissioned phase, higher initial
  \(\tau_{\text{burn}}\), or a volume-independent slash component (e.g.,
  a fixed protocol fee) until fee flow stabilizes.
\item
  \textbf{Network-wide volume troughs.} Bear markets, post-shock
  recovery, low-utilization periods. The reputation-channel premium
  attenuates symmetrically. Mitigations: minimum-fee floor enforced by
  governance, or a fee-floor schedule indexed to market activity.
\item
  \textbf{Schemas with low fee schedules.} A schema whose attestor set
  chooses a low fee per attestation (race-to-the-bottom) reduces the
  reputation-channel slash premium for any validator processing that
  schema's attestations. Mitigations: per-schema fee minimums, or a
  global \(\text{fee}_{\min}\) enforced at the runtime layer.
\end{enumerate}

§4.4.2 specifies the \textbf{adaptive \(\tau_{\text{burn}}\) rebase}
mechanism that reacts automatically to drift below a published
\(\rho_{\text{vol}}\) floor; that is the protocol-level countermeasure
to (1)-(3) at scale. The simulator at
\texttt{prototypes/poua-sim/scripts/run\_volume\_deterrent.py} produces
the canonical figure used to set the floor parameter.

\subsubsection{6.4 Cold-Start Free-Rider
Problem}\label{cold-start-free-rider-problem}

A new validator entering with \(r_v = r_{\min}\) has lower expected
revenue than an established validator. This creates a barrier to entry.
Two questions:

\begin{enumerate}
\def\labelenumi{\arabic{enumi}.}
\tightlist
\item
  Is the barrier high enough to entrench the initial validator set
  permanently?
\item
  Is the barrier so low that bootstrapping fails?
\end{enumerate}

\textbf{On (1).} No.~New validators with stake \(s\) at \(r_{\min}\)
accumulate reputation at the same rate as any existing validator with
stake \(s\). The reputation difference closes at rate \(\eta\). If
\(T_{\text{ramp}}\) is calibrated as recommended (Section 4.4), full
ramp takes \textasciitilde5 days of honest operation. New validators
accept this as a cost-of-entry, comparable to validator startup costs in
any PoS system.

\textbf{On (2).} No.~The cold-start premium (\(r_{\max}/r_{\min}\)
multiplicative) is bounded. A new validator's expected revenue is at
least \(r_{\min}/r_{\max}\) of an established one's, which is positive
and competitive enough to incentivize entry. We do not believe PoUA is
structurally less attractive to new validators than mature PoS.

\subsubsection{6.5 Equilibrium Stability}\label{equilibrium-stability}

The PoUA equilibrium is stable against unilateral deviation by any
single validator. It is also stable against coalitions of size \(< n/3\)
in weight, by the BFT bound. Coalitions of size \(\geq 1/3\) in weight
can violate safety; PoUA's \(\kappa\) premium raises the cost of forming
such a coalition by the multiplicative factor described in Section 5.3.

The economic analysis does \emph{not} show that PoUA is stable against
arbitrary coordination among large stakeholders external to the chain
(e.g., pre-existing exchanges or institutional holders). This is the
same vulnerability standard PoS has, and no consensus mechanism we are
aware of fully addresses it without permissioning.

\begin{center}\rule{0.5\linewidth}{0.5pt}\end{center}

\subsection{7. Implementation in Ligate
Chain}\label{implementation-in-ligate-chain}

\subsubsection{7.1 Sovereign SDK Integration
Points}\label{sovereign-sdk-integration-points}

Ligate Chain is built atop the Sovereign SDK (Sovereign Labs, 2024), a
rollup framework with pluggable consensus, data availability, and
execution layers. PoUA is integrated as a custom \emph{kernel} - the SDK
component responsible for slot processing, validator selection, and BFT
vote tallying.

The integration points:

\begin{enumerate}
\def\labelenumi{\arabic{enumi}.}
\item
  \textbf{Validator Selection Module.} Replaces the SDK's default
  stake-weighted selection with a \(w_v(t) = s_v(t) \cdot r_v(t)\)
  weighted selection. Approximately 200 LOC of Rust modification to the
  SDK's \texttt{sov-attester-incentives} module.
\item
  \textbf{Reputation State.} A new on-chain map
  \(\text{Reputation}: \text{ValidatorAddr} \to r_v\) stored in the
  rollup's state tree. Updated at epoch boundaries via a new module
  \texttt{sov-reputation} (proposed). Storage cost: 32 bytes per
  validator, written once per epoch.
\item
  \textbf{Reputation Update Worker.} A new background job in the kernel
  that, at each epoch boundary, computes \(g_v(t)\) and \(b_v(t)\) for
  all \(v \in V(t)\), applies the update function (Section 4.3), and
  writes the new reputations to state. Implemented as a deterministic
  post-block hook to ensure all honest validators compute identical
  updates.
\item
  \textbf{Slashing Conditions.} A1, A2, A3 are added as new slashing
  modules. A1 is straightforward (verify signatures at proposal time,
  slash on miss). A2 and A3 require the more involved statistical
  detection logic specified in Appendix A.
\item
  \textbf{Genesis Migration.} The chain genesis embeds initial
  \(r_v(0) = r_{\min}\) for all genesis validators. The reputation state
  is initialized at the same time as the validator set in the chain
  genesis JSON (\texttt{devnet/genesis/reputation.json}).
\end{enumerate}

\subsubsection{7.2 Recommended v0
Parameters}\label{recommended-v0-parameters}

For Ligate devnet, we propose:

\begin{itemize}
\tightlist
\item
  \(r_{\min} = 1.0\), \(r_{\max} = 8.0\) (premium ratio 8\(\times\))
\item
  \(\eta = 0.001\) (reputation per nano-AVOW of valid attestation fee)
\item
  \(\lambda = 1.0\) (reputation per stake-equivalent slash)
\item
  \(E = 14400\) slots \(\approx\) 4 hours at \(\tau = 1\,\text{s}\)
\item
  \(\alpha = 0.7, \beta = 0.3\) (proposer / voter reputation share,
  \(\alpha + \beta = 1\))
\item
  \(\tau_{\text{burn}} = 0.5\) (Layer 3 non-recoverable share)
\item
  \texttt{burn\_destination\ =\ pure\_burn} (Layer 3 destination; see
  §5.5.3 for the alternatives and their cost-to-grind implications)
\item
  \(G_{\max} = (r_{\max} - r_{\min}) / (\eta \cdot T_{\text{ramp}}) = 7 / (0.001 \cdot 30) \approx 233\)
  fee-units per epoch (per-validator per-epoch growth cap)
\item
  \(T_{\text{warmup}} = 14\) epochs \(\approx\) 2.3 days
\item
  \(T_{\text{ramp}} \approx 30\) epochs \(\approx\) 5 days under median
  attestation volume
\item
  \(T_{\text{unbond}} = 42\) epochs \(\approx\) 7 days
\end{itemize}

These parameters give a chain where: (1) reputation premium is
meaningful but bounded (8\(\times\) moat), (2) full ramp from
\(r_{\min}\) to \(r_{\max}\) takes at least \textasciitilde5 days of
healthy participation (and exactly 5 days under continuous-cap
saturation), (3) misbehavior is punished within an epoch, (4)
bootstrapping completes within a week of mainnet launch, (5) voters who
never propose still ramp toward \(r_{\max}\) at a rate of approximately
\(\beta \cdot G_v^{\text{vote}} / G_{\max}\) per epoch, ensuring the
validator set's reputation distribution stays connected rather than
fragmenting into proposer-rich and voter-poor strata.

\subsubsection{7.3 Storage Cost Analysis}\label{storage-cost-analysis}

The reputation state requires \(32 \cdot |V|\) bytes. For \(|V| = 100\)
validators (a reasonable mid-mainnet size), this is 3.2 KB. Updated once
per 4-hour epoch, this is 6 writes per day per validator: negligible
relative to the chain's per-block tx state writes.

If \(|V|\) grows to 1000 (large-scale mainnet), reputation state is 32
KB, still negligible.

Per-epoch reputation update computation:
\(O(|V| \cdot |\text{Attestations}_\text{epoch}|)\) in the worst case
(must scan all blocks in the epoch and tally per-validator). At median
10 attestations/sec, an epoch contains \textasciitilde144,000
attestations; this gives \(\approx 14.4M\) attestation-validator
pairings to tally, well within a single-machine budget for a few hundred
ms of pre-block work.

\subsubsection{7.4 Migration from Stake-Only
PoS}\label{migration-from-stake-only-pos}

Ligate Chain v0 launches under stake-only PoS for the warmup period.
During warmup, the reputation update worker still runs and computes
reputation values, but they are not yet folded into vote weight.

At the end of warmup (epoch \(T_{\text{warmup}}\)), a one-time
governance transaction enables the
\texttt{weight\ =\ stake\ *\ reputation} formula. This is a soft fork:
clients must update their consensus binaries to recognize the new
weighting, but state continuity is preserved.

Validators present at warmup-end have accumulated \(T_{\text{warmup}}\)
epochs of reputation; they begin the post-warmup phase with whatever
reputation they earned during the warmup. Validators who join after
warmup begin at \(r_{\min}\).

\subsubsection{7.5 Governance and Appeals}\label{governance-and-appeals}

Per Section 5.5, A3 (reputation grinding) is heuristic. To mitigate
false-positive risk, validators slashed under A3 may file an appeal via
a governance transaction. The governance module (separate from PoUA,
instantiated in Ligate's \texttt{sov-governance} module) reviews and may
reverse the slash by majority vote of the un-slashed validator set.

This is a governance-layer mitigation, not a protocol guarantee. We
expect appeals to be rare in practice (false-positive rate target:
\textless1\% per epoch).

\subsubsection{7.6 Open Implementation
Issues}\label{open-implementation-issues}

The following remain open as of v0.1:

\begin{enumerate}
\def\labelenumi{\arabic{enumi}.}
\tightlist
\item
  \textbf{Statistical detection thresholds for A2 and A3.} Calibration
  requires empirical attestation traffic data, which devnet operation
  will produce.
\item
  \textbf{Reputation portability across chain upgrades.} If the chain
  undergoes a state-breaking restart (chain-id ladder bump per Ligate's
  protocol design), how is reputation carried forward? Current thinking:
  reputation resets to \(r_{\min}\) at chain-id bumps, rewarding
  sustained operation through stable periods.
\item
  \textbf{Public APIs for reputation visibility.} Validators and dApp
  builders should be able to query current reputation values. We propose
  a REST endpoint under the reputation module's namespace, parameterized
  by validator address, returning the tuple \((s_v, r_v, w_v)\).
\end{enumerate}

\begin{center}\rule{0.5\linewidth}{0.5pt}\end{center}

\subsection{8. Comparison with Prior
Systems}\label{comparison-with-prior-systems}

We compare PoUA against five related consensus and weighting mechanisms
across six axes: weighting basis, sybil resistance, useful work
coupling, cost-to-attack, complexity, and production maturity.

\begin{landscape}
\begingroup
\renewcommand{\arraystretch}{1.4}
\small
\setlength{\tabcolsep}{4pt}
\begin{longtable}{>{\raggedright\arraybackslash}p{2.0cm} >{\raggedright\arraybackslash}p{2.6cm} >{\raggedright\arraybackslash}p{3.4cm} >{\raggedright\arraybackslash}p{3.6cm} >{\raggedright\arraybackslash}p{3.4cm} >{\raggedright\arraybackslash}p{1.4cm} >{\raggedright\arraybackslash}p{3.0cm}}
\rowcolor{tableheaderbg}
\textbf{System} & \textbf{Weighting} & \textbf{Sybil resistance} & \textbf{Useful work coupling} & \textbf{Cost-to-attack vs pure PoS} & \textbf{Complexity} & \textbf{Production maturity} \\
\midrule
\endhead
\textbf{Pure PoS} (Tendermint, etc.) & Stake & Stake bond & None & $1\times$ & Low & Mainnet, multiple chains \\
\rowcolor{tablerowalt}
\textbf{Restaking} (EigenLayer) & Stake (Ethereum) + opt-in protocol bonds & Eth stake & Indirect (validator selects protocols) & $\sim 1$ to $2\times$ on additional bonds & Medium & Live since 2023, growing \\
\textbf{PoUA} (this work) & Stake $\times$ reputation & Stake bond + non-transferable reputation tied to attestation work & Direct, formal & $\bar{r}_H / r_{\min} \in [4, 10]$ & High & Specification stage \\
\rowcolor{tablerowalt}
\textbf{Helium PoC} & Coverage proof & Hardware identity + coverage measurement & Direct & $> 1\times$, hard to compute (geographic) & High & Mainnet (Helium Network) \\
\textbf{Filecoin PoSt} & Storage commitment proof + collateral & Storage hardware & Direct & $> 1\times$, varies & Very high & Mainnet \\
\rowcolor{tablerowalt}
\textbf{RepuCoin} (Yu et al.) & Reputation as first-class weight (derived from total blocks produced, activity, and contribution regularity) & PoW + stake & Indirect (mining $\neq$ application work) & Grows with elapsed chain time as reputation accumulates & Low & Research only \\
\bottomrule
\end{longtable}
\endgroup
\end{landscape}

PoUA's distinctive position: \textbf{direct coupling to
application-layer productive workload} without requiring external
measurement (unlike Helium/Filecoin which need hardware-attested
measurements), while preserving a clean economic Sybil-resistance
argument (unlike pure reputation systems which depend entirely on
heuristic detection).

We argue PoUA is \textbf{not strictly novel} in any single dimension but
is novel as a synthesis. Specifically: the combination of (1) verifiable
application-workload-as-useful-work, (2) non-transferable reputation,
(3) clean stake\(\times\)reputation weighting, and (4) BFT-inheriting
safety and liveness is, to our knowledge, not present in any prior
system.

\begin{center}\rule{0.5\linewidth}{0.5pt}\end{center}

\subsection{9. Limitations and Future
Work}\label{limitations-and-future-work}

\subsubsection{9.1 Limitations}\label{limitations}

We acknowledge the following as real limitations of PoUA v0.1:

\begin{enumerate}
\def\labelenumi{\arabic{enumi}.}
\item
  \textbf{Heuristic A3 detection.} As discussed in Section 5.5, defense
  against the compound capital-plus-reputation-grinding adversary relies
  on heuristic detection plus governance, not formal proof. Future work
  should either tighten the heuristic or explore cryptographic
  primitives (zk-proof of independent attestation submission) that
  provide formal guarantees.
\item
  \textbf{No formal proof of incentive compatibility.} Section 6 gives
  game-theoretic arguments but does not present a full mechanism-design
  proof of incentive compatibility under all rational deviations.
  Bringing this to the formal-proof bar is significant additional work.
\item
  \textbf{Single-chain.} Reputation is local. A multi-chain ecosystem
  (multiple Ligate-style chains, or cross-chain attestation sharing)
  would need a portability primitive that we have not designed.
\item
  \textbf{Cold-start dependent on initial validator set.} If the genesis
  validator set is poorly distributed, the warmup period may not produce
  a healthy reputation distribution, and the bootstrap conditions of the
  equilibrium may fail. We recommend devnet operation provides empirical
  evidence before mainnet launch.
\item
  \textbf{Devnet calibration still pending.} The reference simulator at
  \href{https://github.com/ligate-io/ligate-research/tree/main/prototypes/poua-sim}{\texttt{prototypes/poua-sim/}}
  closed milestones M1-M7 across v0.7 + v0.8 (cost-to-attack,
  transition-state \(\kappa\), Lemma 1, A3 detector FPR, M6
  strategy-search heatmap, M7 network adversity, scale invariance). What
  remains is production-scale calibration against real attestation
  traffic, including the §A.4 ER-vs-Chung-Lu threshold reformulation and
  the A2 / A3 detector TPR curves under realistic chain-graph baselines.
  Devnet operation (mid-2026 per current roadmap) provides those inputs.
\end{enumerate}

\subsubsection{9.2 Future Work}\label{future-work}

\begin{itemize}
\tightlist
\item
  \textbf{Zero-knowledge attestation of reputation accumulation.}
  Validators could prove they accumulated reputation honestly without
  revealing the underlying attestation submission graph, eliminating
  much of A3's heuristic surface.
\item
  \textbf{Reputation futures markets.} Validators could sell forward
  rights to a fraction of their future reputation-derived revenue,
  hedging entry-cost risk. This is a market design question, not
  strictly a protocol question.
\item
  \textbf{Cross-chain reputation portability.} A canonical primitive for
  transferring reputation across chains, possibly through a shared
  reputation registry or zkbridge-based assertion. The companion paper
  \href{https://github.com/ligate-io/ligate-research/tree/main/papers/cross-schema-composition}{Cross-Schema
  Composition} covers the typed-reference foundation; the cross-chain
  extension is the harder lift.
\item
  \textbf{Reputation as governance weight.} Beyond consensus, reputation
  could enter governance vote tallying. The case is delicate (protects
  against governance capture by capital, but may entrench
  validator-class dominance over user-class voice).
\item
  \textbf{Privacy-preserving reputation.} A scheme in which reputation
  is observable in aggregate (so validator selection is verifiable) but
  per-validator privacy is preserved. Useful for politically sensitive
  validation contexts.
\item
  \textbf{Hybrid post-quantum signatures.} Tracked at
  \href{https://github.com/ligate-io/ligate-research/issues/50}{ligate-research\#50}.
  Ed25519 default with Dilithium opt-in for validators with
  longer-horizon threat models.
\end{itemize}

Adaptive \(\eta\) and \(\lambda\) rebase, previously listed as future
work, landed in §4.4.3 of this version. The M6 strategy-search and M7
network-conditions extensions previously listed as future simulator work
closed via PRs in the v0.8 cycle.

\begin{center}\rule{0.5\linewidth}{0.5pt}\end{center}

\subsection{10. Conclusion}\label{conclusion}

Proof of Useful Attestation is a consensus weighting primitive that
aligns validator influence in an attestation-native chain with the
production of valid attestation work. It inherits BFT safety and
liveness under the same partial-synchrony and Byzantine bounds the
underlying primitive assumes, and it constructs a multiplicative
cost-to-attack premium of \(r_{\max}/r_{\min} \in [4, 10]\) over
equivalent pure-stake chains. The mechanism is not novel in any single
component. It is, to our knowledge, the first synthesis of
reputation-weighted consensus, proof-of-useful-work, and
non-transferable bonding tailored to chains whose application surface is
attestation production.

Production deployment is not free. It needs empirical validation through
simulation and devnet operation, hardening of the heuristic detectors
that constitute Layer 4 of the §5.5 defense, integration testing against
the Sovereign SDK rollup framework, and external technical review. The
paper provides concrete parameter recommendations and identifies every
integration point against existing SDK module surfaces, but the
engineering work is real and we estimate it across 2026-2027.

We invite review and critique - particularly critique. This is a working
paper; substantial revision is expected as the mechanism is
stress-tested against simulation results, adversarial model literature,
and external technical reviewers. The hardest part of the paper is §5.5,
where we make a formal economic claim (Lemma 1) about the cost of
grinding reputation against a compound adversary; if you find a flaw in
that argument, that is the most valuable feedback you can give us.

\subsection{Acknowledgments}\label{acknowledgments}

We thank \textbf{Jiangshan Yu} (University of Sydney, Sydney Blockchain
Centre) for substantive corrections to the §8 RepuCoin comparison row
(weight basis as first-class reputation derived from multi-aspect
historical contribution; cost-to-attack as time-dependent rather than
fixed-ratio; complexity assessment), endorsement of the §11 Q4
intrinsically-anchored-reputation framing, and a follow-up clarification
that informed the v0.9.1 §11 phrasing rewrite (cost basis as off-chain
commodity-priced rather than ``out-of-protocol''). Acknowledged with
permission.

We thank \textbf{Marko Vukolić} (Bitcoin Scaling Labs) for emphasizing
the load-bearing nature of §5.5 attestation-grinding defenses in our
exchange, which informed the §5.5 layered-detection write-up and the
§6.2 empirical strategy-search progression.

Both reviewers' input arrived during the 2026-05 external technical
review cycle. Errors that remain are our own.

\begin{center}\rule{0.5\linewidth}{0.5pt}\end{center}

\newpage

\subsection{11. Frequently Asked
Questions}\label{frequently-asked-questions}

Early review surfaced critiques and misunderstandings worth addressing
in one place. This section is partly to short-circuit common objections,
partly to record design rationale that the formal specification does not
always make obvious.

\subsubsection{Q1. Doesn't this just let validators farm reputation by
submitting attestations to
themselves?}\label{q1.-doesnt-this-just-let-validators-farm-reputation-by-submitting-attestations-to-themselves}

\textbf{Short answer:} no, not under v0.2's layered defense. The naive
self-attestation attack collapses against Layer 1 (proposer-submitter
address exclusion) and Layer 2 (address-graph distance threshold). A
more sophisticated grinding attempt that evades Layers 1-2 still pays
Lemma 1's cost-to-grind floor under Layer 3 (non-recoverable treasury
share), making the cost of grinding economically equivalent to or worse
than honestly acquiring the same reputation premium.

\textbf{Long answer:} see §5.5 in full. The compound
capital-plus-grinding adversary is the hardest case PoUA handles, and it
is the case that distinguishes a serious mechanism from a marketing
claim. v0.1 of this paper acknowledged that the heuristic A3 detection
alone was a soft barrier; v0.2's layered defense converts the protection
from a heuristic argument to an economic one (with the heuristic
detection now playing the residual catch-all role behind formal Layers
1-3).

\subsubsection{Q2. Doesn't this require the chain to judge the truth of
attestations?}\label{q2.-doesnt-this-require-the-chain-to-judge-the-truth-of-attestations}

\textbf{No.} Validity in PoUA is \textbf{cryptographic}, not
\textbf{semantic}. An attestation is ``valid'' if and only if it carries
a \(k\)-of-\(n\) threshold signature from the registered attestor set at
the registered threshold for the named schema. Anyone, on any node, can
re-verify this. The chain is not asserting that the underlying claim
(``the moon is green,'' ``this image was produced by a human,'' etc.) is
true; it is asserting that a specified set of authorities
cryptographically attested to the claim under a registered schema.

This is, deliberately, the same trust model as Ethereum's ``we record
the calldata, not whether the calldata is true.'' PoUA's ``useful work''
is processing valid signatures - work the chain can verify - not
adjudicating semantic truth. The reputation a validator earns reflects
participation in correct cryptographic processing, nothing more.

The economic implication: schemas with corrupt attestor sets can sign
garbage, but they pay attestation fees on every garbage attestation, and
consumers of the attestation data (off-chain readers) decide whether to
trust the schema based on its registered attestor set's identity,
history, and reputation in their own off-chain trust model. PoUA does
not solve ``is this true?''; it solves ``is the chain's economic
security aligned with the chain's productive workload?''

\subsubsection{Q3. Hasn't reputation-weighted consensus been tried and
rejected?}\label{q3.-hasnt-reputation-weighted-consensus-been-tried-and-rejected}

\textbf{Partially right.} Reputation-weighted BFT is well-explored in
the academic literature (RepuCoin, EigenTrust, and the broader
distributed-systems trust-and-reputation tradition) but not widely
deployed in production. The reasons it has not shipped are real: Sybil
resistance is hard to formalize, heuristic detection is brittle, and
formal proofs of incentive compatibility are sparse.

\textbf{Where this paper is different.} PoUA is not
``reputation-weighted consensus in general.'' It is reputation-weighted
consensus where:

\begin{enumerate}
\def\labelenumi{\arabic{enumi}.}
\tightlist
\item
  The ``useful work'' is \textbf{the chain's own productive workload}
  (processing valid attestations), not external mining or storage
  commitments. This is novel.
\item
  Reputation is \textbf{non-transferable and intrinsic to the chain},
  not portable across protocols (unlike restaking).
\item
  Cost-to-grind is bounded by a \textbf{formal economic argument} (Lemma
  1, §5.5.3), not just heuristic detection.
\item
  The mechanism integrates cleanly with a \textbf{production rollup
  framework} (Sovereign SDK) that admits custom kernel layers,
  eliminating the implementation gap that has stalled prior research.
\end{enumerate}

The synthesis is the contribution. We claim novelty in the synthesis,
not in any single component. Section 8's comparison table positions PoUA
explicitly against the prior art, including the cases where prior work
has been tried and not shipped.

\subsubsection{Q4. Isn't this just RepuCoin or EigenTrust with a new
name?}\label{q4.-isnt-this-just-repucoin-or-eigentrust-with-a-new-name}

\textbf{No, and the differences matter for the security argument.} The
contrast operates along three axes: what the reputation \emph{measures},
where the reputation \emph{enters} the consensus computation, and what
the \emph{cost-to-grind} argument relies on.

\begin{itemize}
\tightlist
\item
  \textbf{RepuCoin} (Yu et al., 2019) treats reputation as a first-class
  quantity derived from multiple aspects of historical contribution
  (total blocks created, activity and availability, regularity of
  contribution). The ``work'' being measured is hash computation plus
  participation, which is \emph{externally anchored}: the marginal cost
  of acquiring reputation (electricity, mining hardware) is set by
  commodity markets, not by a chain-controlled parameter.
  Cost-per-unit-reputation in RepuCoin is therefore a function of
  off-chain energy and hardware prices, not a chain parameter.
  PoW-as-useful-work is a generic signal: any chain whose validators
  also mine PoW can adopt it. RepuCoin's cost-to-attack is
  \emph{time-dependent}: it grows as honest reputation accumulates over
  chain history, so an attacker faces a steadily rising bar rather than
  a fixed multiple. RepuCoin is also research-stage; we are not aware of
  a production deployment. PoUA's ``work'' is the chain's \emph{own paid
  productive workload} (attestation processing), which is
  \emph{intrinsically anchored}: a different chain hosting the same
  attestation contracts cannot replicate the reputation premium without
  rebuilding consensus, because the workload measurement happens at the
  consensus-runtime boundary that contract-layer systems don't control.
\item
  \textbf{EigenTrust} (Kamvar et al., 2003) is a peer-to-peer reputation
  algorithm for decentralized file-sharing and trust networks. It does
  not weight BFT consensus directly; it scores nodes based on transitive
  interaction history (which peers each peer trusted, propagated
  transitively). PoUA's reputation is non-transitive (each validator
  earns from chain activity, not from being trusted by other validators)
  and enters BFT vote weight directly via \(w_v = s_v \cdot r_v\).
  EigenTrust solves a different problem (file-sharing trust) with
  different mechanics (transitive aggregation); the ``reputation'' word
  covers both but the formal objects do not overlap.
\item
  \textbf{PoUA} measures application-layer attestation processing, with
  reputation entering BFT vote weight directly via
  \(w_v = s_v \cdot r_v\). The mechanism design choices (additive
  update, bounded interval, non-transferability, fee-weighted earning,
  Lemma 1 cost-to-grind floor under Layer 3 burn, layered Sybil defense
  across six independently-breakable mechanisms) are specific to this
  application and do not appear in either prior system as a coherent
  package. The cost-to-grind argument is
  \emph{economic-by-construction}: the chain's own non-recoverable burn
  floor (Lemma 1, §5.5.3) bounds the per-fee-unit cost of grinding from
  below, independent of off-chain commodity prices (energy, hardware) or
  peer-trust transitivity.
\end{itemize}

The differentiated property: \textbf{PoUA is the first
reputation-weighted BFT scheme where the ``work'' the reputation tracks
is the protocol's own paid productive workload, not an
externally-anchored signal (mining, storage, peer interactions), AND
where the cost-to-grind floor is bounded by an intrinsic protocol
parameter (\(\tau_{\text{burn}}\)) rather than by off-chain commodity
scarcity (energy, hardware, peer-trust graph structure).} This is what
makes the moat economic-by-construction. RepuCoin's moat depends on PoW
capacity scarcity; EigenTrust's depends on the social-network-graph
structure of who trusts whom; PoUA's depends on the chain's own fee-burn
parameter, which is observable, calibrable, and adaptive (§4.4.2). The
synthesis is the contribution; no single component is novel in
isolation.

\subsubsection{Q5. Why not just use restaking (EigenLayer) on
Ethereum?}\label{q5.-why-not-just-use-restaking-eigenlayer-on-ethereum}

\textbf{Restaking is good for what it does, but it does not solve the
same problem.} EigenLayer lets Ethereum validators opt into additional
security duties on secondary protocols, with slashing across both. This
is a layer atop an existing chain; the secondary protocol's economic
security is bounded by the bonded ETH and the slashing condition
specifications.

PoUA is not a secondary layer atop an existing chain. It is the
\textbf{primary} consensus mechanism of a chain whose economic activity
is attestation. The differences:

\begin{itemize}
\tightlist
\item
  Restaking inherits Ethereum's economic security; PoUA constructs its
  own (bounded by stake on the Ligate Chain itself, augmented by
  reputation tied to attestation work).
\item
  Restaking adds slashing conditions; PoUA adds a weighting mechanism.
  These are complementary, not equivalent.
\item
  A Ligate Chain attestation submitted via a restaked-Ethereum validator
  is still subject to Ethereum's gas economics. A Ligate Chain
  attestation submitted natively is in the chain's own fee market,
  designed for the workload.
\end{itemize}

The relationship: a future version of Ligate Chain could opt into
Ethereum restaking as an additional security layer (Section 9.2 mentions
cross-chain reputation portability as future work), but PoUA is the
chain-native primitive that exists either way.

\subsubsection{Q6. What if attestor sets
collude?}\label{q6.-what-if-attestor-sets-collude}

\textbf{The chain accepts that attestation correctness is bounded by the
honesty of the registered attestor set.} This is the same trust model
every multi-signature scheme uses. If the attestor set
\(\mathcal{A}_\sigma\) for schema \(\sigma\) collude to sign garbage,
the chain records garbage attestations against \(\sigma\) - and the
schema's reputation in the off-chain world (the only place ``garbage''
is judged) tanks accordingly.

PoUA does not protect schema consumers from corrupt attestor sets. That
is by design: protecting against corrupt attestors is the \textbf{schema
designer's} job (choose attestors with skin in the game, design slashing
conditions for invalid attestation under the schema's own dispute
mechanism, build off-chain reputation systems for attestor sets).

What PoUA does protect against is corrupt \textbf{validators}
selectively censoring or extracting MEV from the attestation workload.
The validator's incentive (under PoUA) is to honestly include all valid
attestations because that is how reputation accumulates and how future
revenue is earned. A corrupt validator censoring valid attestations from
a particular schema loses both immediate fees (the censored attestations
would have paid them) and reputation (the included attestation count is
lower), making censorship economically dominated by honest behavior.

\subsubsection{Q7. What happens if the heuristic A3 detector has high
false-positive
rates?}\label{q7.-what-happens-if-the-heuristic-a3-detector-has-high-false-positive-rates}

\textbf{Honest answer:} false positives are real and unavoidable in any
heuristic detector. v0.2's layered defense reduces reliance on the
heuristic detector by making Layers 1-3 (formal protocol rules and
economic disincentives) carry the main load. Layer 4 (heuristic) is now
a residual safety net.

When the heuristic does fire on an honest validator, Layer 5 (governance
appeal) provides recovery: the slashed validator presents their case to
the un-slashed validator set, and a majority can reverse the slash. This
is governance machinery, not protocol guarantee, and we acknowledge it
has its own failure modes (governance capture, voter apathy).

The empirical false-positive rate target is \(\beta_3 \leq 1\%\) per
epoch under honest baseline traffic, calibrated from devnet
observations. Achieving this target is a v0.2 acceptance criterion that
depends on devnet operation, which is scheduled for late 2026.

\subsubsection{Q8. The ``uncopyable by a generic L1'' claim seems
overconfident. Is it
accurate?}\label{q8.-the-uncopyable-by-a-generic-l1-claim-seems-overconfident.-is-it-accurate}

\textbf{Soft-soften this claim.} v0.2 phrases the moat as
``purpose-built primitive that cannot be cleanly replicated in a
contract layer without rebuilding consensus'' rather than
``uncopyable.'' The precise claim is:

\begin{itemize}
\tightlist
\item
  A generic Ethereum-style L1 \emph{can} host attestation contracts. The
  contracts can implement schemas, attestor sets, and attestations. We
  do not contest this.
\item
  A generic L1 \emph{cannot} implement PoUA's consensus weighting at the
  contract level. Reputation entering BFT vote weight requires
  consensus-layer modification, not contract-layer extension.
\item
  A chain that wants PoUA-style economic security therefore must either
  fork its consensus layer (a hard, slow change for established chains)
  or build a new chain.
\end{itemize}

This is a defensible technical claim. It is what the cost-to-attack
premium \(\kappa\) formalizes economically: a generic L1 hosting
attestations cannot replicate the premium because the premium comes from
a consensus mechanism the generic L1 cannot adopt without
consensus-layer changes.

\subsubsection{Q9. Why not use a simpler mechanism (pure stake-weighted,
with stronger
slashing)?}\label{q9.-why-not-use-a-simpler-mechanism-pure-stake-weighted-with-stronger-slashing}

\textbf{Pure stake-weighted with stronger slashing is the alternative we
benchmarked against.} The cost-to-attack analysis in §5.3 explicitly
contrasts the two: pure-stake \(\kappa = 1\), PoUA
\(\kappa \in [4, 10]\). The premium is the value PoUA adds.

The implicit critique behind this question is: ``is a \(4-10\times\)
premium worth the implementation complexity?'' That is a real question,
and the honest answer depends on the chain's threat model. For chains
where the workload is undifferentiated state transitions, no - simpler
is better. For attestation-native chains where the workload has specific
economic shape (high volume of low-individual-value,
signature-verifiable items), the alignment between consensus reward and
workload is meaningful and arguably worth the complexity.

We do not claim PoUA is right for every chain. We claim it is right for
chains whose economic activity is attestation-shaped, which is the
design assumption of Ligate Chain.

\subsubsection{Q10. Why publish a working paper instead of waiting for a
peer-reviewed
result?}\label{q10.-why-publish-a-working-paper-instead-of-waiting-for-a-peer-reviewed-result}

\textbf{Working papers are the right artifact for the current stage.}
The mechanism is novel enough that we want public review and critique;
publishing a peer-reviewed result requires submitting to a venue, which
has its own timeline (\textasciitilde12-18 months in cryptography
conferences). v0.1 through v0.6 are working papers explicitly marked as
such, with version histories and acknowledged limitations.

The path forward: v0.6 → external technical reviewer feedback (mid-2026)
→ v0.7 with simulation results (late 2026) → arxiv submission (early
2027) → conference submission (mid-2027 if appropriate venue). At every
stage, the paper is publicly available and explicitly versioned, so
readers know what they are citing.

\subsubsection{Q11. Why not just use Celestia raw, or any DA layer, for
attestations?}\label{q11.-why-not-just-use-celestia-raw-or-any-da-layer-for-attestations}

A reasonable observation: pure attestation looks like an I/O problem.
Receipts go in, ordered, immutable, byte-priced. Why pay for consensus
when ordering and availability are all that seems necessary?

Three things pure DA cannot host.

\textbf{Reputation state evolution.} Validator reputation evolves
per-epoch via §4.3's rule
\(r_v(t+E) = r_v(t) + \eta \cdot g_v(t) - \lambda \cdot b_v(t)\). This
requires consensus on per-validator \(g_v\) (cumulative valid work) and
\(b_v\) (cumulative slash severity), exposed as committed state. A flat
DA log stores the inputs but does not produce the output verifiably;
light clients would have to recompute it over the entire chain history.

\textbf{Schema-scoped attestor-set policy.} Each schema declares an
attestor set with minimum reputation requirements. Enforcement requires
the chain to evaluate set membership and reputation predicates at
attestation time, not just record the bytes. DA layers accept any
submitted data; they cannot evaluate predicates.

\textbf{Burn-and-slash economics.} Lemma 1 (§5.5.3) bounds an
adversary's cost-to-grind by
\(F_{\text{net}} \geq \tau_{\text{burn}} \cdot \Delta r / (\eta \cdot \alpha_{\text{eff}})\).
The bound holds only if the chain enforces the \(\tau_{\text{burn}}\)
fraction at fee-distribution time. DA layers charge per-byte; they do
not redirect a fee fraction to provable destruction.

Pure DA secures the bytes. PoUA secures the signer. Different security
models, both required. Ligate Chain uses Celestia for DA underneath, so
the I/O layer the question correctly identifies as necessary is already
there. PoUA is what we layer on top of that, not in place of it. §3.8
works the same argument out in terms of the system model.

\subsubsection{Q12. Will Ligate Chain ever support general-purpose smart
contracts?}\label{q12.-will-ligate-chain-ever-support-general-purpose-smart-contracts}

\textbf{Not as a primary surface.} The ``attestation-native chain''
thesis in §1.1 is a positive design choice, not a temporary scope
limitation. A chain whose security budget, fee market, and consensus
mechanism are sized to a specific workload (attestation production) has
a defensibility profile that a general-purpose chain does not, and
adding general-purpose contracts would dilute it. The \(4\times\) to
\(10\times\) cost-to-attack premium quantified in §5 depends on the
workload being attestation-shaped; arbitrary state transitions break the
analysis.

Two narrower extensions are on the supplementary-papers roadmap, both
kept inside the attestation domain:

\begin{itemize}
\tightlist
\item
  \textbf{Cross-schema composition}
  (\href{https://github.com/ligate-io/ligate-research/tree/main/papers/cross-schema-composition}{companion
  paper}): typed references that let one attestation depend on another,
  evaluated at attestation time. Contract-like reasoning over
  attestation predicates, not general state transitions.
\item
  \textbf{Time-locked and commit-reveal attestations}
  (\href{https://github.com/ligate-io/ligate-research/tree/main/papers/time-locked-attestations}{companion
  paper}): attestations that become public at a future time, or that
  reveal their payload only after a separate commit step. Expands the
  attestation primitive's temporal envelope without leaving the
  attestation domain.
\end{itemize}

Neither extension makes Ligate Chain a general-purpose smart contract
platform. Cross-schema composition operates over typed attestation
references, not arbitrary state. Time-locked attestations extend when an
attestation is readable, not what it can compute. Both stay within the
consensus-runtime boundary that PoUA's security argument requires
(§3.8).

For application authors who need general-purpose smart contracts: use a
general-purpose chain (Ethereum, Solana, a Cosmos app chain). For
application authors whose product is dominated by attestation production
and verification, with cost-to-attack tied to that workload's economic
shape: use Ligate. The chain's value proposition is the specialization.

\subsubsection{\texorpdfstring{Q13. Why three rebase parameters
(\(\tau_{\text{burn}}\), \(\eta\), \(\lambda\))? How do they
interact?}{Q13. Why three rebase parameters (\textbackslash tau\_\{\textbackslash text\{burn\}\}, \textbackslash eta, \textbackslash lambda)? How do they interact?}}\label{q13.-why-three-rebase-parameters-tau_textburn-eta-lambda-how-do-they-interact}

\textbf{Three because each tracks a structurally different drift.}
§4.4.2's \(\tau_{\text{burn}}\) rebase tracks cost-to-grind drift driven
by fee economics (token supply, schema fee structures). §4.4.3's
\(\eta\) rebase tracks ramp-time drift driven by attestation volume
regime. §4.4.3's \(\lambda\) rebase tracks slash-deterrent drift driven
by the slashing-condition catalog. The three signals are orthogonal in
their primary inputs, so each one needs its own auto-adjuster;
conflating them into a single rebase would obscure which underlying
variable is moving.

\textbf{Interaction is bounded.} The three rebases run concurrently.
Correlation enters only at second order (e.g., a fee regime change can
shift attestation volume, which feeds back into \(\eta\)). Under the
worst-case scenario where all three signals drift in the
cost-to-grind-shrinking direction simultaneously, each step is bounded
by \(\Delta \leq 0.1\) and the combined per-step effect on Lemma 1's
\(F_{\text{net}} \geq \tau_{\text{burn}} \cdot \Delta r / (\eta \cdot \alpha_{\text{eff}})\)
floor is bounded by \((1.1) \cdot (1.1) / (0.9) \approx 1.34\), a 34\%
one-step swing. The \(N\)-epoch rate-limit prevents compounding within a
window. The simulator's
\texttt{test\_three\_rebases\_concurrent\_no\_amplification} test
validates the combined Lyapunov function
\(V = D_\tau^2 + D_\eta^2 + D_\lambda^2\) is non-increasing under
correlated drift.

\textbf{Telemetry surfaces a ``rebase saturation'' alert} when any of
the three parameters sits at its clip bound for more than \(N\) epochs.
Saturation means the auto-adjuster has done all it can and governance
escalation (§4.4.2, §4.4.3) is the next layer. The split between
rules-based auto-adjustment and discretionary governance mirrors
well-designed monetary policy: fast-but-bounded automation,
slow-but-permanent human override.

\begin{center}\rule{0.5\linewidth}{0.5pt}\end{center}

\newpage

\subsection{References}\label{references}

\begin{enumerate}
\def\labelenumi{\arabic{enumi}.}
\tightlist
\item
  Benet, J., Greco, N., Vorick, D., Marlowe, M., et al.~(2017).
  \emph{Filecoin: A Decentralized Storage Network}. Protocol Labs.
\item
  Buchman, E. (2016). \emph{Tendermint: Byzantine Fault Tolerance in the
  Age of Blockchains}. M.Sc. Thesis, University of Guelph.
\item
  Buterin, V., Griffith, V. (2017). \emph{Casper the Friendly Finality
  Gadget}. arXiv:1710.09437.
\item
  Castro, M., Liskov, B. (1999). Practical Byzantine Fault Tolerance.
  \emph{OSDI '99}.
\item
  Cohen, B. (2019). \emph{Proofs of Space and Time}. Chia Network.
\item
  Dwork, C., Lynch, N., Stockmeyer, L. (1988). Consensus in the presence
  of partial synchrony. \emph{Journal of the ACM}, 35(2), 288-323.
\item
  EigenLayer Team. (2023). \emph{EigenLayer: The Restaking Collective}.
  Whitepaper.
\item
  Eyal, I. (2015). The Miner's Dilemma. \emph{IEEE S\&P 2015}.
\item
  Gilad, Y., Hemo, R., Micali, S., Vlachos, G., Zeldovich, N. (2017).
  Algorand: Scaling Byzantine Agreements for Cryptocurrencies.
  \emph{SOSP '17}.
\item
  Haleem, A., Allen, A., Thompson, A., Nijdam, M., Garg, R. (2018).
  \emph{Helium: A Decentralized Wireless Network}. Helium Inc.
\item
  Hoffman, K., Zage, D., Nita-Rotaru, C. (2009). A Survey of Attack and
  Defense Techniques for Reputation Systems. \emph{ACM Computing
  Surveys}, 42(1).
\item
  Kamvar, S. D., Schlosser, M. T., Garcia-Molina, H. (2003). The
  EigenTrust Algorithm for Reputation Management in P2P Networks.
  \emph{WWW '03}.
\item
  Resnick, P., Kuwabara, K., Zeckhauser, R., Friedman, E. (2000).
  Reputation Systems. \emph{Communications of the ACM}, 43(12), 45-48.
\item
  Sovereign Labs. (2024). \emph{Sovereign SDK Documentation}.
  github.com/Sovereign-Labs/sovereign-sdk.
\item
  Yin, M., Malkhi, D., Reiter, M. K., Gueta, G. G., Abraham, I. (2019).
  HotStuff: BFT Consensus with Linearity and Responsiveness. \emph{PODC
  '19}.
\item
  Yu, J., Kozhaya, D., Decouchant, J., Verissimo, P. (2019). RepuCoin:
  Your Reputation Is Your Power. \emph{IEEE TC 68(8)}.
\end{enumerate}

\begin{center}\rule{0.5\linewidth}{0.5pt}\end{center}

\subsection{Appendix A: Statistical Detection of A2 (Censorship) and A3
(Grinding)}\label{appendix-a-statistical-detection-of-a2-censorship-and-a3-grinding}

This appendix specifies the heuristic detectors that constitute Layer 4
of the §5.5 layered defense. We give analytical false-positive bounds
under explicit null-hypothesis assumptions; empirical power analysis
(how often the detector catches real adversaries) requires devnet
traffic data and is deferred to v1.0.

\subsubsection{A.1 A2 Detection: Selective Schema Censorship via
KL-Divergence}\label{a.1-a2-detection-selective-schema-censorship-via-kl-divergence}

\textbf{Setup.} Per epoch \(t\), for each validator \(v\) that acted as
proposer in \(N_v(t) \geq N_{\min}\) blocks (e.g., \(N_{\min} = 10\) to
guarantee enough samples for the statistical approximation):

\begin{itemize}
\tightlist
\item
  \(D_v(t) \in \Delta(\Sigma)\): empirical distribution over schemas of
  the attestations \(v\) included as proposer in epoch \(t\).
\item
  \(D_{\text{net}}(t) \in \Delta(\Sigma)\): network-wide empirical
  distribution over schemas of all attestations available in the mempool
  during \(v\)'s proposer slots.
\end{itemize}

\textbf{Test statistic.} The Kullback-Leibler divergence

\[D_{\text{KL}}(D_v \| D_{\text{net}}) = \sum_{\sigma \in \Sigma} D_v(\sigma) \log \frac{D_v(\sigma)}{D_{\text{net}}(\sigma)}.\]

\textbf{Null hypothesis \(H_0\).} Validator \(v\) samples included
attestations from the mempool uniformly (i.e., does not selectively
censor any schema).

\textbf{Distributional approximation.} Under \(H_0\), by Wilks' theorem
the scaled statistic
\(2 N_v(t) \cdot D_{\text{KL}}(D_v \| D_{\text{net}})\) converges in
distribution to \(\chi^2_{|\Sigma|-1}\) (chi-squared with \(|\Sigma|-1\)
degrees of freedom) as \(N_v(t) \to \infty\). The approximation is good
for \(N_v(t) \cdot \min_\sigma D_{\text{net}}(\sigma) \geq 5\), the
standard rule of thumb for chi-squared goodness-of-fit.

\textbf{Threshold.} For target false-positive rate \(\beta_2\) per epoch
(e.g., \(\beta_2 = 0.01\)):

\[\theta_2 = \frac{1}{2 N_v(t)} \cdot \chi^2_{|\Sigma|-1,\ 1-\beta_2}\]

where \(\chi^2_{k, p}\) is the \(p\)-quantile of the chi-squared
distribution with \(k\) degrees of freedom.

\textbf{Detection rule.} Flag \(v\) if
\(D_{\text{KL}}(D_v \| D_{\text{net}}) > \theta_2\) for at least
\(T_{\text{detect}}\) consecutive epochs (recommended
\(T_{\text{detect}} = 3\)). Requiring consecutive epochs reduces the
false-positive rate from \(\beta_2\) to approximately
\(\beta_2^{T_{\text{detect}}}\), giving \(\beta_2 = 0.01\) per epoch and
\(\beta_2^3 = 10^{-6}\) per cumulative flag, sufficient to keep wrongful
slashes rare in practice.

\textbf{Implementation note.} The mempool snapshot \(D_{\text{net}}(t)\)
must be observable to all validators reproducibly. Achieving this in the
consensus pipeline requires either (i) periodically committing mempool
digests to the chain, or (ii) reconstructing \(D_{\text{net}}\) from the
union of all validators' included-attestation sets in the epoch (less
precise but cheaper). v1.0 will pick one approach based on engineering
tradeoffs; the analytical bound above is independent of the choice.

\subsubsection{A.2 A3 Detection: Reputation Grinding via Bipartite Graph
Density}\label{a.2-a3-detection-reputation-grinding-via-bipartite-graph-density}

\textbf{Setup.} Per epoch \(t\), for each validator \(v\), construct the
bipartite graph \(G_v(t) = (U_v, W_v, E_v)\) where:

\begin{itemize}
\tightlist
\item
  \(U_v\): distinct submitter addresses of attestations \(v\) included
  as proposer in epoch \(t\).
\item
  \(W_v\): distinct attestor-set members appearing in the schemas those
  attestations target.
\item
  \(E_v\): edges between \(u \in U_v\) and \(w \in W_v\) where there is
  observable correlation in the on-chain transaction graph (within
  \(T_{\text{lookback}}\) blocks of funding history involving \(v\)).
\end{itemize}

\textbf{Test statistic.} Bipartite edge density

\[\rho_v(t) = \frac{|E_v|}{|U_v| \cdot |W_v|}.\]

\textbf{Null hypothesis \(H_0\).} No shared beneficial owner between
\(v\) and the submitter or attestor-set populations: edges in \(G_v(t)\)
form independently of \(v\) at chain-wide baseline rate
\(p_{\text{base}}\), where \(p_{\text{base}}\) is the empirical edge
density of the chain-wide bipartite graph between submitter addresses
and attestor-set members in the same epoch.

\textbf{Distributional approximation.} Under \(H_0\) with the
Erdős-Rényi-like assumption above, \(|E_v|\) is approximately
\(\text{Binomial}(|U_v||W_v|, p_{\text{base}})\). For large
\(|U_v||W_v|\) (typical when \(v\) proposes more than a handful of
attestations across multiple schemas), the Normal approximation gives

\[\rho_v(t) \mathrel{\dot\sim} \mathcal{N}\!\left(p_{\text{base}},\ \frac{p_{\text{base}}(1 - p_{\text{base}})}{|U_v||W_v|}\right).\]

\textbf{Threshold.} For target false-positive rate \(\beta_3\) per
epoch:

\[\theta_3 = p_{\text{base}} + z_{1-\beta_3} \sqrt{\frac{p_{\text{base}}(1 - p_{\text{base}})}{|U_v||W_v|}}\]

where \(z_{1-\beta_3}\) is the \((1-\beta_3)\)-quantile of the standard
Normal distribution. For \(\beta_3 = 0.01\), \(z_{0.99} \approx 2.326\).

\textbf{Detection rule.} Flag \(v\) if \(\rho_v(t) > \theta_3\) for at
least \(T_{\text{detect}}\) consecutive epochs.

\textbf{Implementation note.} The ``observable correlation'' relation in
\(E_v\) is the heuristic part of A3 detection. Layer 2 of the §5.5
layered defense (address-graph distance) provides a hard rule that
rejects attestations from too-near submitter addresses; A3 catches the
residual cases where the adversary stages addresses sufficiently far
apart to clear Layer 2 but the bipartite graph still shows
above-baseline density across the staged population. Each of
\(T_{\text{lookback}}\) and the precise correlation predicate is a
calibration choice that v1.0 will fix from devnet observations.

\subsubsection{A.3 Per-Epoch Adaptive
Computation}\label{a.3-per-epoch-adaptive-computation}

The thresholds \(\theta_2, \theta_3\) depend on observed parameters
(\(N_v, |\Sigma|, p_{\text{base}}, |U_v|, |W_v|\)), all of which are
computable per-epoch from the prior epoch's chain state. Production
deployment computes them per-epoch, providing data-adaptive thresholds
while preserving the analytical false-positive guarantees of
\(\beta_2, \beta_3\) stated above.

\subsubsection{A.4 What This Appendix Establishes (and What It
Defers)}\label{a.4-what-this-appendix-establishes-and-what-it-defers}

This appendix establishes the \textbf{false-positive bound} for both
detectors under stated null-hypothesis assumptions: - A2: chi-squared
distribution under independent sampling from \(D_{\text{net}}\). - A3:
Normal approximation under Erdős-Rényi baseline edge formation.

It does \textbf{not} establish \textbf{power} (the rate at which the
detector catches actual adversaries). Power analysis requires either:

\begin{enumerate}
\def\labelenumi{\arabic{enumi}.}
\tightlist
\item
  \textbf{Synthetic-traffic simulation} with adversarial models, the
  workstream tracked at
  \href{https://github.com/ligate-io/ligate-research/tree/main/prototypes/poua-sim}{\texttt{prototypes/poua-sim/}}.
  The simulator is the natural place to validate that
  \(\beta_2, \beta_3\) are correctly bounded under realistic honest
  baselines and that the detector has acceptable true-positive rate
  against synthetic A2/A3 attackers.
\item
  \textbf{Devnet observation} with real attestation traffic. After
  devnet stabilizes (mid-2026 in the current roadmap), real validator
  behavior under real attestation workloads provides the ground truth
  that synthetic simulation can only approximate.
\end{enumerate}

v1.0 of this paper will incorporate empirical power analysis from one or
both sources, replacing the analytical bounds with calibrated values
where appropriate.

\textbf{Explicit acknowledgment of the Erdős-Rényi assumption gap.} The
§A.2 A3 detector threshold is derived under the assumption that, under
the null hypothesis, edges in the bipartite (submitter, attestor) graph
form independently with uniform probability \(p_{\text{base}}\)
(Erdős-Rényi). Real chain transaction graphs are \emph{not} Erdős-Rényi:
they are scale-free, with hub addresses (exchanges, bridges, popular
dApps, large enterprise submitters) generating edge clusters that
violate the independence assumption. The simulator-driven empirical
comparison in Figure \ref{fig:a3-fpr-comparison} below shows this
explicitly: under a Chung-Lu scale-free null (power-law degree
distribution, \(\alpha \in \{2.0, 2.5, 3.0\}\)), realized FPR sits 1-3
orders of magnitude below the nominal \(\beta_3 = 1\%\) target. The
detector is consistently \emph{more conservative} than the analytical
claim under realistic graph structure, which means the false-positive
guarantee is honored in practice but the threshold is too loose for
attack detection (the corresponding TPR is also depressed). This is a
calibration issue, not a security flaw: production deployment should
re-derive the threshold against an empirical chain-graph baseline
(Chung-Lu fit to the chain's own transaction history) rather than retain
the ER assumption. Open issue
\href{https://github.com/ligate-io/ligate-research/issues/16}{\#16}
tracks the reformulation, deferred until devnet provides empirical
chain-graph baseline data.

\textbf{Synthetic-attestor saturation in TPR scans.} The simulator's
\texttt{scripts/run\_a3\_tpr\_scan.py} runs a \(\beta_3\) sweep over the
§A.2 detector against synthetically-constructed grinding cartels. In the
simulator's synthetic-attestor model, the bipartite-density signal
saturates: TPR sits at 1.0 across \(\beta_3 \in [0.001, 0.1]\) for small
staged pools (the detector's design regime). This is not a calibration
win; it is an artifact of the synthetic model, where attestor sets are
drawn from the chain's validator addresses without realistic
hub-and-spoke transaction structure. The saturation documents an
upper-bound TPR; production calibration against empirical chain traffic
(post-devnet) is required to position the detector against realistic
graph structure. The diluted-pool gap (the regime where saturation
drops) is captured by the §5.5.2 Layer 2 deterministic-membership
closure rather than by the §A.2 detector alone; the §6.2 Panel C result
quantifies this empirically.

\begin{figure}[h]
\centering
\includegraphics[width=0.92\textwidth]{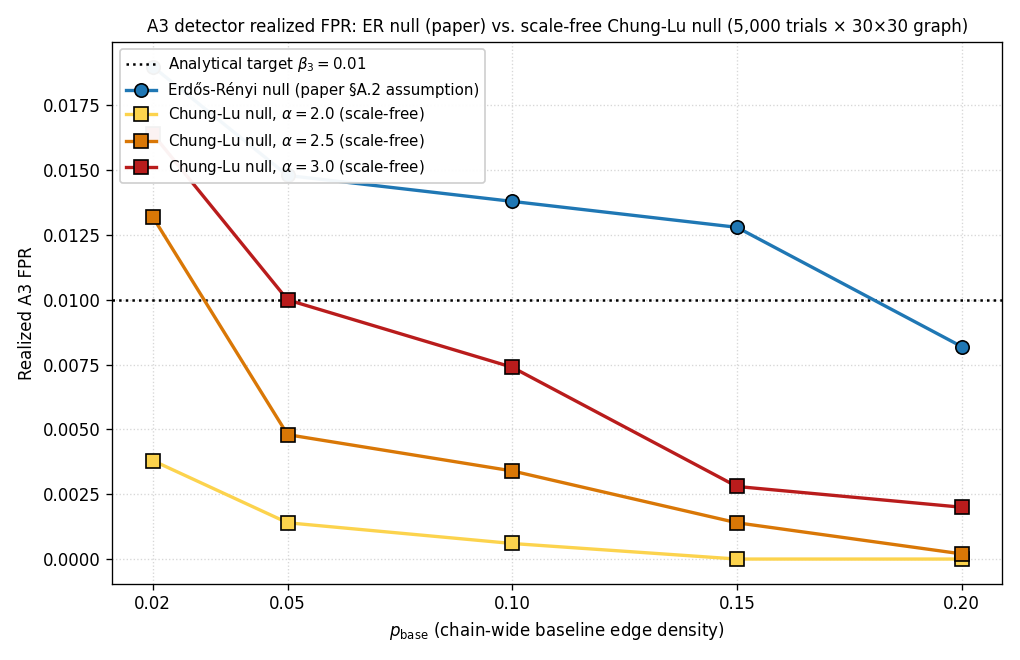}
\caption{Realized A3 false-positive rate under the §A.2 Erdős-Rényi null (paper assumption) versus a Chung-Lu scale-free null (realistic chain-graph model), across $p_{\text{base}} \in \{0.02, 0.05, 0.10, 0.15, 0.20\}$ and power-law exponent $\alpha \in \{2.0, 2.5, 3.0\}$. 5{,}000 trials per cell on a $30 \times 30$ bipartite graph. Produced by \texttt{prototypes/poua-sim/scripts/run\_a3\_fpr\_comparison.py}. The horizontal dotted line marks the analytical $\beta_3 = 1\%$ target. Under the ER null, realized FPR sits within $\sim$2$\sigma$ binomial of nominal. Under Chung-Lu, the detector is consistently \emph{more} conservative than nominal: realized FPR drops by 1-3 orders of magnitude depending on $p_{\text{base}}$ and $\alpha$. The mismatch reflects the heavy-tail bias under hub-pair clipping; production deployment should re-derive the threshold against an empirical Chung-Lu-style baseline rather than the ER assumption.}
\label{fig:a3-fpr-comparison}
\end{figure}

\begin{center}\rule{0.5\linewidth}{0.5pt}\end{center}

\subsection{Appendix B: Formal Definitions
Recap}\label{appendix-b-formal-definitions-recap}

For convenience, we collect formal definitions used throughout the
paper.

\textbf{Definition B.1 (Validator).} A tuple
\(v = (\text{addr}_v, s_v, r_v, \text{pk}_v)\) where \(\text{addr}_v\)
is a chain address, \(s_v\) is the bonded stake,
\(r_v \in [r_{\min}, r_{\max}]\) is the reputation, and \(\text{pk}_v\)
is the consensus public key.

\textbf{Definition B.2 (Weight).} \(w_v(t) := s_v(t) \cdot r_v(t)\).

\textbf{Definition B.3 (Validator selection).}
\(\Pr[\text{proposer}(t) = v] = w_v(t) / \sum_{u \in V(t)} w_u(t)\).

\textbf{Definition B.4 (BFT commit).} Block \(B_t\) commits iff
\(\sum_{v: v \text{ commits } B_t} w_v(t) > \frac{2}{3} \sum_{u \in V(t)} w_u(t)\).

\textbf{Definition B.5 (Reputation update).}
\(r_v(t+E) = \text{clip}_{[r_{\min}, r_{\max}]}(r_v(t) + \eta g_v(t) - \lambda b_v(t))\),
evaluated at epoch boundaries.

\textbf{Definition B.6 (Good behavior score, v0.2).}

\[g_v(t) = \min\bigl(G_{\max},\; \alpha \cdot G_v^{\text{prop}}(t) + \beta \cdot G_v^{\text{vote}}(t)\bigr),\]

with proposer and voter components defined as

\[G_v^{\text{prop}}(t) = \sum_{B \in \text{Proposed}_v(t, t+E)} \sum_{\alpha \in B} \mathbb{1}[\alpha \text{ valid}] \cdot \text{fee}(\alpha),\]

\[G_v^{\text{vote}}(t) = \sum_{B \in \text{VotedOn}_v(t, t+E)} \frac{\sum_{\alpha \in B} \mathbb{1}[\alpha \text{ valid}] \cdot \text{fee}(\alpha)}{|\text{voters}(B)|},\]

with \(\alpha + \beta = 1\) and \(G_{\max}\) a per-epoch growth cap.

\textbf{Definition B.7 (Bad behavior score).}

\[b_v(t) = \sum_{i \in \{1,2,3\}} \Lambda_i \cdot |\{\text{detected slashes of severity } i \text{ for } v \text{ in epoch } t\}|.\]

\textbf{Definition B.8 (Cost-to-attack premium).}
\(\kappa = \bar{r}_H / r_{\min}\) where \(\bar{r}_H\) is the mean
reputation of honest validators.

\textbf{Lemma 2 (Weighted quorum intersection, recap).} Let
\(W = \sum_{u \in V} w_u\). For any \(Q, Q' \subseteq V\) with
\(\sum_{v \in Q} w_v > \frac{2}{3} W\) and
\(\sum_{v \in Q'} w_v > \frac{2}{3} W\), the intersection satisfies
\(\sum_{v \in Q \cap Q'} w_v > \frac{1}{3} W\). (Proof: §5.2 via
inclusion-exclusion. Used in Theorems 1 and 2.)

\textbf{Lemma 1 (Cost-to-grind bound, recap).} Under Layer 3 with
parameter \(\tau_{\text{burn}}\), proposer reputation share \(\alpha\)
and voter share \(\beta = 1 - \alpha\), an \(m\)-validator coordinated
cartel within a \(k\)-voter set that acquires per-member reputation gain
\(\Delta r\) pays per-member non-recoverable fees
\(F_{\mathcal{CR}}^{\text{net, per member}} \geq \tau_{\text{burn}} \cdot \Delta r / [\eta \cdot \alpha_{\text{eff}}(m, k)]\),
where \(\alpha_{\text{eff}}(m, k) = \alpha + (m - 1)\beta/k\). The
single-validator case \(m = 1\) recovers
\(\alpha_{\text{eff}} = \alpha\) exactly. In the asymptotic limit
\(k \to \infty\) at fixed \(m / k\),
\(\alpha_{\text{eff}} \to \alpha + (m / k)\beta\); for the Byzantine cap
\(m = k/3\), \(\alpha = 0.7\), the asymptotic per-member discount is
\(\sim 12.5\%\). (Proof: §5.5.3. Empirical validation: simulator §5.5.3
reference at
\href{https://github.com/ligate-io/ligate-research/tree/main/prototypes/poua-sim/scripts/run_lemma1_scan.py}{\texttt{prototypes/poua-sim/scripts/run\_lemma1\_scan.py}}.)

\begin{center}\rule{0.5\linewidth}{0.5pt}\end{center}

\emph{End of working paper v0.9.2. Comments welcome to hello@ligate.io.}

\emph{Roadmap: v0.10 will incorporate real-chain Chung-Lu calibration
for §A.4 using adjacent-chain transaction graphs as a structural proxy
(\href{https://github.com/ligate-io/ligate-research/issues/120}{issue
\#120}) ahead of full Ligate devnet maturity. v1.0 will incorporate
Ligate devnet calibration data once devnet stabilizes (mid-2026),
reviewer feedback from the arXiv submission, and any §A.3 / §A.4
production-calibration refinements driven by domestic chain-graph
baselines. v1.0 is the venue-submission target (mid-2027 if appropriate
venue).}

\end{document}